\documentclass{JAC2003}

\usepackage{graphicx}

\newcommand{\MeV} {\mathrm{MeV}}
\newcommand{\GeV} {\mathrm{GeV}}

\newcommand{\fbi} {\mathrm{fb}^{-1}}

\newcommand{\cL } {{\cal L}}
\newcommand{\cP } {{\cal P}}
\def\ee    {e^+e^-}
\def\pp    {\gamma\gamma}
\def\ti    {\tilde}
\def\sf    {{\ti f}}
\def\sq    {{\ti q}}

\def\st    {{\ti t}}
\def\sb    {{\ti b}}

\def\stau  {{\ti\tau}}

\def\snu   {{\ti\nu}}
\def\sell  {{\ti\ell}}

\def\sl    {{\ti\ell}}
\def\cx    {\ti {\chi}}
\def\ch    {\ti {\chi}}
\def\cp    {\ti {\chi}^+}
\def\cm    {\ti {\chi}^-}
\def\cpm    {\ti {\chi}^\pm}

\def\nt    {\ti {\chi}^0}
\def\none  {\ti \chi^0_1}
\def\ntwo  {\ti \chi^0_2}

\def\sg    {\ti g}
\def\sG    {\ti G}
\def\sq    {\ti q}

\def\smu   {{\ti\mu}}
\def\smul  {{\ti\mu}_L}
\def\smur  {{\ti\mu}_R}

\def\smurm {{\ti\mu}^-_R}

\def\smurp {{\ti\mu}^+_R}

\def\se    {{\ti e}}
\def\sel   {{\ti e}_L}
\def\ser   {{\ti e}_R}

\def\serm  {{\ti e}^-_R}

\def\serp  {{\ti e}^+_R}
\def\sne  {{\ti\nu}_e}

\def\snu  {{\ti\nu}}

\def\snt      {{\ti\nu}_\tau}

\def\tst   {\theta_{\ti t}}

\def\cstau {\cos\theta_{\ti\tau}}

\def \Eslash {E \kern-.6em\slash }
\def \Mslash {M \kern-.5em\slash }
 
\newcommand{\rpv}{\slash\hspace{-2.5mm}{R}_{p}}

\newcommand{\gsim}{\;\raisebox{-0.9ex}
           {$\textstyle\stackrel{\textstyle>}{\sim}$}\;}

\newcommand{\lsim}{\;\raisebox{-0.9ex}{$\textstyle\stackrel{\textstyle<}
          {\sim}$}\;}

\newcommand{\nn}{\nonumber}
\newcommand{\beq}{\begin{equation}}
\newcommand{\eeq}{\end{equation}}
\newcommand{\bea}{\begin{eqnarray}}
\newcommand{\eea}{\end{eqnarray}}
\newcommand{\eq}[1]{eq.~(\ref{#1})}
\newcommand{\fig}[1]{fig.~\ref{#1}}
\newcommand{\tab}[1]{table~\ref{#1}}

\newcommand{\imag}{\Im {\rm m}}
\newcommand{\real}{\Re {\rm e}}

\begin{document}
\title{SUPERSYMMETRY WORKING GROUP: SUMMARY REPORT\thanks{Much
of the work reported in this talk was done by members of the SUSY Working
Group of the Extended ECFA/DESY Study: B.C.Allanach$^a$, 
M.Ball$^b$, 
A.Bartl$^c$,  
S.Berge$^b$, 
G.Blair$^d$, 
C.Bl\"ochinger$^e$, 
E.Boos$^f$, 
A.Brandenburg$^g$, 
P.Checchia$^h$, 
S.Y.Choi $^i$ 
A.Datta$^j$, 
K.Desch$^b$, 
A.Djouadi$^j$, 
H.Dreiner$^\lambda$ (co-convener),   
H.Eberl$^k$, 
A.Finch$^\zeta$,
H.Fraas$^e$, 
A.Freitas$^l$, 
T.Fritzsche$^m$, 
B.Gaissmaier$^n$, 
N.Ghodbane$^b$, 
D.K.Ghosh$^o$, 
J.Guasch$^p$, 
S.Heinemeyer$^r$,  
C.Hensel$^g$, 
S.Hesselbach$^c$, 
K.Hidaka$^s$, 
M.Hirsch$^t$, 
W.Hollik$^u$, 
T.Kernreiter$^c$, 
M.Kincel$^v$, 
O.Kittel$^e$, 
M.Klasen$^b$, 
S.Kraml$^w$, 
J.L.Kneur$^j$, 
W.Majerotto$^k$, 
M.Maniatis$^g$, 
A.v.Manteuffel$^g$,
H.U.Martyn$^x$ (co-convener), 
M.Melles $^p$, 
D.J.Miller$^w$, 
K.M\"onig$^y$, 
G.Moortgat-Pick$^z$,  
S.Moretti$^\alpha$, 
G.Moultaka$^j$, 
M.M\"uhlleitner$^p$, 
U.Nauenberg$^\beta$, 
H.Nieto-Chaupis$^y$, 
H.Nowak$^y$, 
V.\"Oller$^k$, 
E.Piotto$^w$,
G.Polesello$^\kappa$,  
W.Porod$^\gamma$,
F.Richard$^\rho$, 
J.C.Rom\~ao$^\delta$, 
S.Rosier-Lees$^\mu$ (co-convener),  
H.Rzehak$^u$, 
A.Stahl$^y$, 
J.Sol\`a$^\epsilon$, 
A.Sopczak$^\zeta$, 
C.Tevlin$^\eta$, 
J.W.F.Valle$^t$, 
C.Verzegnassi$^\nu$, 
R.Walczak$^\eta$, 
C.Weber$^k$, 
M.M.Weber$^p$, 
G.Weiglein$^z$, 
Y.Yamada$^\theta$, 
P.M.Zerwas$^g$ 
($^a$~LAPTH Annecy, $^b$~U.Hamburg, $^c$~U.Vienna,  
$^d$~U.London, $^e$~U.W\"urzburg, $^f$~Moscow State~U., 
$^g$~DESY Hamburg, $^h$~U.Padova, $^i$~Chonbuk National~U., 
$^j$~U.Montpellier II, $^k$~\"OAW Vienna, $^l$~Fermilab, 
$^m$~U.Karlsruhe, $^n$~TU Munich, $^o$~Oregon~U.\ Eugene, $^p$~PSI
Villigen, $^r$~LMU Munich, $^s$~Gakugei~U.\ Tokyo, $^t$~U.Val\`encia,
$^u$~MPI Munich, $^v$~Comenius~U.\ Bratislava, 
$^w$~CERN, $^x$~RWTH Aachen,  $^y$~DESY Zeuthen, $^z$~U.Durham, 
$^\alpha$~U.Southampton, $^\beta$~U.Colorado Boulder, 
$^\gamma$~U.Z\"urich, $^\delta$~IST Lisboa,  $^\epsilon$~U.Barcelona,
$^\zeta$~U.Lancaster, $^\eta$~U.Oxford, $^\theta$~Tohoku~U.\ Sendai, 
$^\kappa$~INFN Pavia, $^\lambda$~U.Bonn, $^\mu$~LAPP Annecy, 
$^\nu$~U.Trieste, $^\rho$~LAL Orsay.)
}}

\author{J. Kalinowski\thanks{Supported by by the KBN Grant 2 P03B 040
    24 (2003-2005).},  
Institute of Theoretical Physics, Warsaw University, Warsaw, Poland}

\maketitle

\begin{abstract}
 This report summarizes the progress in SUSY
 studies performed during the Extended ECFA/DESY Workshop 
 since the TESLA TDR \cite{Aguilar-Saavedra:2001rg}. Based on
 accurate future measurements of masses of SUSY particles and the
 determination of the couplings and mixing properties of sfermions and
 gauginos, we discuss how the low-energy Lagrangian parameters can be
 determined. In a `bottom-'­up' approach, by extrapolating to
 higher energies, we demonstrate how model assumptions on SUSY
 breaking can be tested.  To this end precise knowledge of the SUSY
 spectrum and the soft SUSY breaking parameters is necessary to reveal
 the underlying supersymmetric theory.
\end{abstract}

\section{INTRODUCTION}
An $e^+e^-$ linear collider in the 500 - 1000 GeV energy range (LC) is
widely considered as the next high-energy physics machine
\cite{Snowmass}. 
One of the many arguments for its construction is the possibility of
exploring supersymmetry (SUSY).    
Of the many motivations for the supersymmetric  extension of the
Standard Model, perhaps the most important, next to the connection to
gravity,  is the ability to  
stabilize the electroweak scale. If the electroweak scale is not
fine-tuned, the superpartner masses (at least some of them) 
need to be in the TeV range.  
In such a case the  Large Hadron Collider (LHC) will certainly see SUSY. 
Many  different channels, in particular   
from squark and gluino decays will be explored and many 
interesting quantities measured. In specific scenarios  characterized
by a handful of free parameters 
some of the elements of  supersymmetry 
can be reconstructed \cite{LHCrec}. However, to
prove SUSY one has to scrutinize its characteristic features in as
model-independent a way as possible. We will have to:
\begin{itemize}
\item measure masses of new particles, their 
decay widths, production cross sections,
mixing angles etc.,
\item verify that they are superpartners, {\it i.e.}~measure
their spin and parity, gauge quantum numbers and couplings, 
\item reconstruct the low-energy SUSY breaking parameters 
without assuming a specific scenario,
\item and ultimately unravel the SUSY breaking mechanism and shed
light on physics at the high (GUT?, Planck?) scale.
\end{itemize}
In answering all the above points an   $e^+e^-$ LC   
would be an indispensable tool. Therefore the concurrent running
of the LHC and the LC  is very much welcome \cite{JKsusy03}. First, 
the LC will provide 
independent checks of  the LHC findings. Second, thanks to the LC  unique
features: clean environment, tunable collision energy, 
high luminosity, polarized
incoming beams, and additional $e^-e^-$, $e\gamma$ and
$\gamma\gamma$ modes, it will offer 
precise measurements of masses, couplings, quantum numbers, 
mixing angles, CP phases etc. Last, but not least, it will 
provide additional experimental input to the LHC analyses, like the
mass of the lightest supersymmetry particle (LSP). 
Coherent analyses of data from the LHC {\it and} LC would thus allow for a 
better, model independent reconstruction of low-energy  SUSY
parameters, and  connect  low-scale phenomenology with the high-scale
physics.  The interplay between LHC and LC is investigated in detail in the
LHC/LC Study Group~\cite{lhclc}.

During the Extended ECFA/DESY Workshop\footnote{ 
The SUSY WG group was very
 active: the members have given 14 talks in Cracow, 15 in St.\ Malo,
 11 in Prague and 15 in Amsterdam, and the transparencies can be found
 in \cite{meetings}.} 
the discovery potential of
TESLA \cite{Aguilar-Saavedra:2001rg} - design of the superconducting   
LC -  
for SUSY particles has been further studied. 
In particular, 
it has been demonstrated that the expected  high luminosity 
(${\cal L}\sim    300$ fb$^{-1}$ per year)  
and availability of polarized electron (up tp 80\%) and positron 
(up to 60\%) beams 
makes precision experiments possible. The virtues of polarized beams
are investigated in the POWER Study Group~\cite{power}. 
Here we will summarize  in some detail how accurate
measurements of the masses of SUSY particles and the determination of
the couplings 
and mixing properties of sleptons, charginos, neutralinos and scalar
top quarks can be performed. 

We will start the
discussion with
the Minimal  Supersymmetric Standard Model considered as 
an effective low energy model with a)~minimal particle content, 
b)~$R$-parity conservation, 
c)~most general soft supersymmetry breaking terms.
Since the mechanism of SUSY breaking is
unknown, several Snowmass benchmark scenarios, so-called 'Snowmass
Points and Slopes' (SPS) \cite{sps},  with 
distinct signatures have been studied. Although each benchmark scenario is
characterized by a few parameters specified at high energies (for
example at the GUT scale), most of the phenomenological analyses have been
performed strictly on low-energy supersymmetry.  

A word of caution is in order here. 
The deduction of low-energy parameters from high-scale assumptions
(and vice-versa) inevitably involves theoretical errors coming from
the level of approximation used, neglected higher order terms etc. 
The SPS benchmarks, while motivated in terms of specific
SUSY-breaking scenarios (like the mSUGRA scenario), have explicitly
been defined in terms of the low-energy MSSM parameters. Therefore it is
not necessary in the SPS benchmarks to refer to any particular program 
for calculating the SUSY spectrum from high-energy parameters.
Studies during the Workshop \cite{GMAKP1,AKP2} 
showed large differences between various calculations of the MSSM spectrum. 
Recent analysis \cite{AKP2} of  the most
advanced modern codes for the MSSM spectra: ISAJET 7.64, SOFTSUSY
1.71 \cite{softsusy}, SPHENO 2.0 \cite{spheno} and SUSPECT 2.101
\cite{suspect}, shows that the 
typical relative uncertainty in mSUGRA and mGMSB scenarios in
generic (i.e. not tricky) regions of parameter space
is about 2\,--\,5\%. In some cases, 
in particular in focus point, high $\tan\beta$  and mAMSB scenarios, 
the relative uncertainty is larger, about 5\,--\,10\%
For the focus point  and high $\tan \beta$ scenarios,  
sparticle masses are particularly sensitive to the values 
of the Yukawa couplings (especially the top Yukawa for the focus point, 
and the bottom Yukawa for the high $\tan \beta$ regime). 
Slightly different treatments of top and bottom masses 
can lead to large differences in mass predictions.
In the mAMSB scenario  larger differences between various programs  
are due to a different implementation of  GUT-scale boundary
conditions. Nevertheless, even in these particular 
cases, comparison   
with previous versions of the codes ~\cite{GMAKP1} (where   
SUSYGEN3.00 \cite{susygen}, 
PYTHIA6.2 \cite{pythia} and the mSUGRA Post-LEP
benchmarks \cite{postlep} have also been investigated)  
shows a significant improvement. 
Differences in the
results between the codes (which may 
be interpreted as very conservative upper bounds 
on  current theoretical uncertainties \cite{AKP2} 
as some programs are more advanced than others) should be reduced  by
future higher--order theoretical calculations.

After extensive discussion of 
experimentation and extraction of SUSY parameters in
the MSSM, we will go to   'beyond the MSSM' scenarios by considering
$R$-parity violating couplings and/or extended gaugino sector. 
Finally, in a `bottom-'­up' approach, by extrapolating  to higher
energies the SUSY parameters 
determined at the electroweak scale with certain errors,  
we  demonstrate  how model assumptions on SUSY breaking can be
tested.    
It will be seen  that precise knowledge of the SUSY spectrum and the soft
SUSY breaking  parameters is necessary to
reveal the underlying supersymmetric theory.

\section{SFERMIONS}

Sfermions $\tilde{f}_L$, $\tilde{f}_R$ are spin-zero superpartners of
the SM chiral fermions ${f}_L$, ${f}_R$.   
The sfermion  mass matrix has the form
\begin{eqnarray} 
&&{\cal M}^2_{\tilde{f}}=\left(\begin{array}{cc}
    m^2_{\tilde{f}_L}      &      a^*_{\ti f} m_f  \\
    a_{\ti f} m_f                &      m^2_{\tilde{f}_R} \end{array}\right)\\
&&m^2_{\tilde{f}_i}={ M^2_{\tilde{F_i}}}
+m^2_Z\cos2\beta\,(I^3_{f_i}-Q_f 
\sin^2\theta_W) +m_f^2\nonumber \\  
&&\;\;a_{\ti f} = {A_{\ti f}} -{ \mu^*}(\tan\beta)^{-2I^3_{f}}\nonumber 
\label{sfermass}
\end{eqnarray}
where $M^2_{\tilde{F_L},\tilde{F_R}}$, $A_{\ti f}$ are soft SUSY
breaking parameters 
(which can be 3$\times$3 matrices in the flavor space),
and $\mu$ is the higgs/higgsino  mass term. Both $A_{\ti f}=|A_{\ti f}
|e^{i\varphi_{A_{\ti f}}}$
and $\mu =|\mu|e^{i\varphi_\mu}$ can be
complex. The mixing between $L$ and $R$ states is important 
when the off-diagonal term is comparable to the splitting of diagonal
ones $\Delta_{\ti f}=m^2_{\tilde{f}_L}-
m^2_{\tilde{f}_R} $, {\it i.e.} $|\Delta_{\ti f}|\leq |a_{\ti f} m_f|$. 
For $\tilde{e}$ and $\tilde{\mu}$
the $L-R$ mixing is therefore usually 
neglected. 

Neglecting inter-generation mixing, the 
masses of physical sfermions $\tilde{f}_{1,2}$ 
\begin{eqnarray}
&&\tilde{f}_{1}=\tilde{f}_L e^{i\varphi_{\ti f}} \cos\theta_{\ti f}
+ \tilde{f}_R \sin\theta_{\ti f}  
\nn \\
&&\tilde{f}_{2}=-\tilde{f}_L \sin\theta_{\ti f}+ \tilde{f}_R 
e^{-i\varphi_{\ti f}} \cos\theta_{\ti f} 
\end{eqnarray}
and the mixing angle $\theta_{{\ti f}}$ and the phase $\varphi_{\ti f}$ 
are given by 
\begin{eqnarray}
&&m^2_{\tilde{f}^\pm_{1,2}}
   =(m^2_{\tilde{f}_L}+m^2_{\tilde{f}_R} 
\mp [\Delta_{\ti f}^2+4 |a_{\ti f}m_f|^2 
   ]^{1/2})/2 \nonumber\\
&&\tan \theta_{\ti f}=(m^2_{\tilde{f}_1}-m^2_{\tilde{f}_L})/
|a_{\ti f} m_f|\nn \\
&& \varphi_{\ti f}=arg({A_{\ti f}} -{ \mu^*}(\tan\beta)^{-2I^3_{f}})
\label{sfermix}
\end{eqnarray}
Thus reconstructing the sfermion sector requires  $m^2_{\tilde{f}_L},\,
m^2_{\tilde{f}_R},\, a_{\ti f}$  to be decoded from measurements of
sfermion masses, cross sections, decay widths etc. 
\cite{jejuslep}.

With the anticipated experimental precision, however,  
higher order corrections will have 
to be taken into account. A current summary of theoretical
progress in this direction can be found in Ref.\cite{WMsusy02}.  
Complete one-loop calculations have been performed for $\smu\smu$
and $\se\se$ production~\cite{freitas} and for sfermion masses and
their decays \cite{0207364}. For a relatively light SUSY
spectrum and a high--energy LC ($M_{\rm SUSY} \ll \sqrt{s} \lsim $ 2 -- 3
TeV), the simple one--loop approximation may turn out to be inadequate
and resummation of higher--order effects might be necessary to obtain
good theoretical predictions \cite{BRV}.

\subsection{ Study of  selectrons/smuons} 
At $e^+e^-$ collisions charged sleptons are produced in pairs via the
s-channel $\gamma/Z$ exchange; for the first generation there is
additional t-channel neutralino exchange. Different states and their
quantum numbers can be disentangled by a proper choice of the beam
energy and polarization.
\begin{figure}[htb]
\centering
\includegraphics*[width=40mm,height=35mm]
      {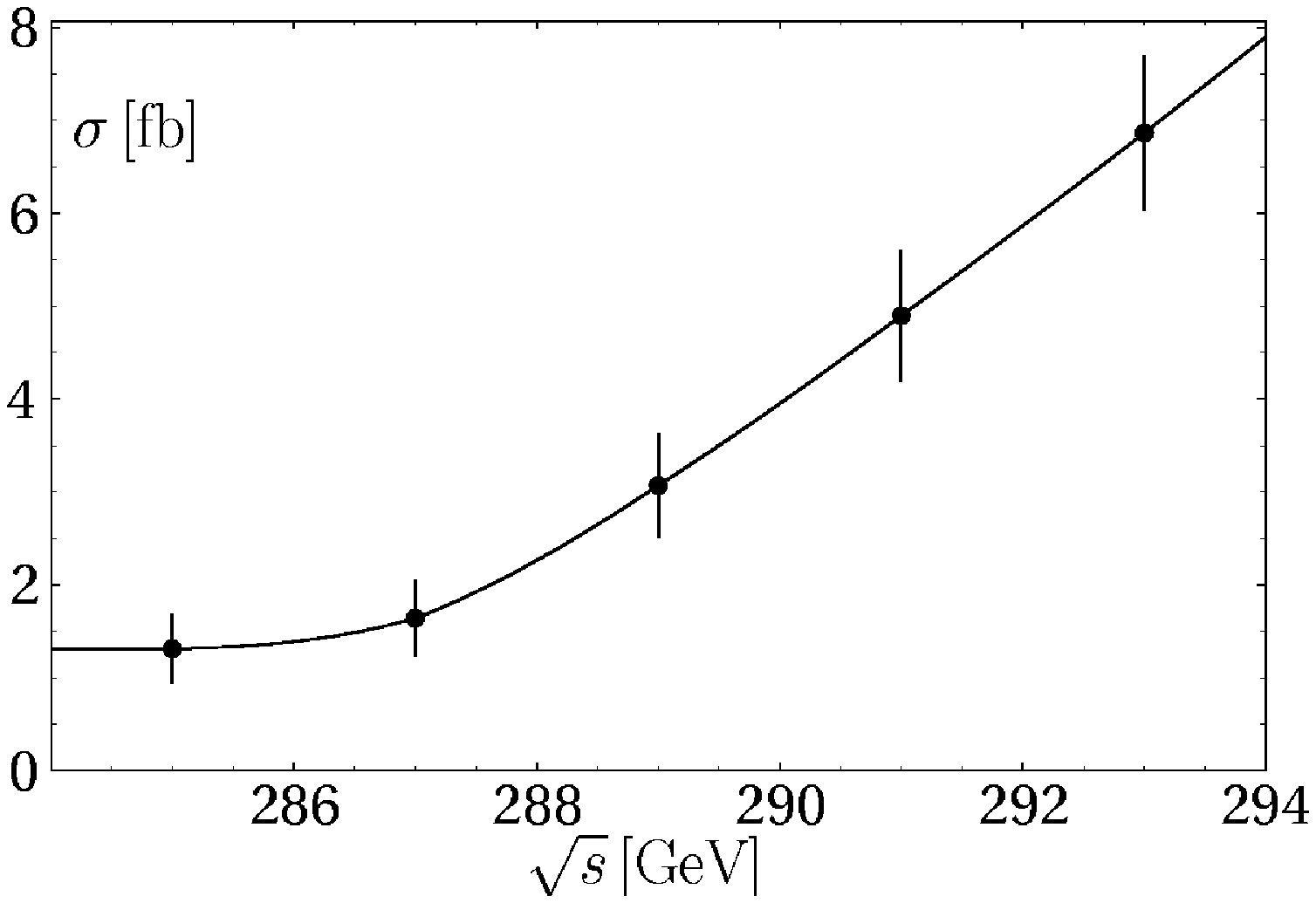}
\includegraphics*[width=40mm,height=35mm]
      {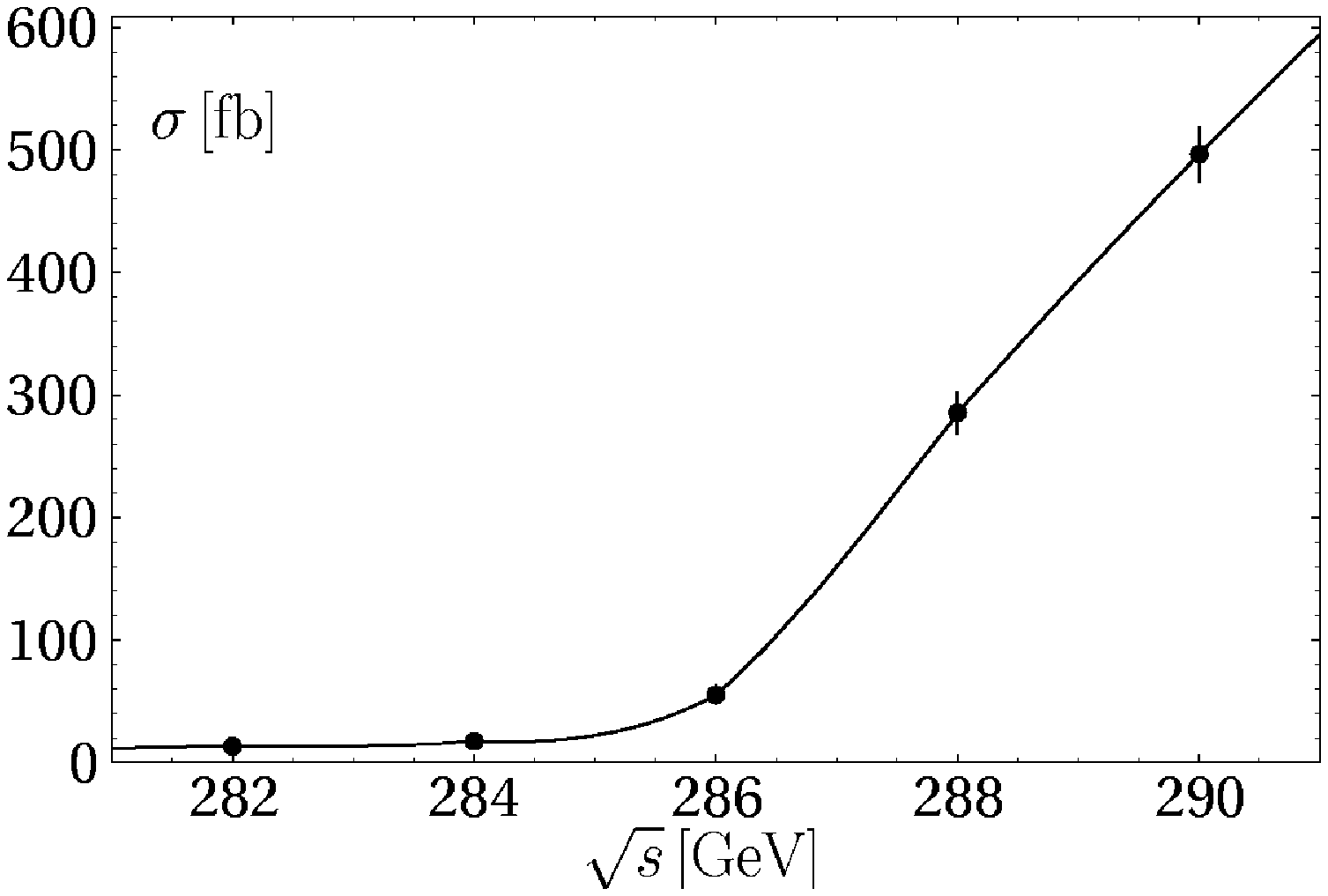}  
\caption{Cross sections at threshold for the reactions	
   $e^+_Le^-_R\to\serp\serm$ (left) and 
  $e^-_Re^-_R\to\serp\serm$ (right) in the SPS\#1a scenario,  
  including background~\cite{freitas}.
  Error bars correspond to a luminosity
  of $10~\fbi$ (left) and  $1~\fbi$ (right) per point.}
\label{scans}
\end{figure}

Slepton masses can be measured in threshold scans or in continuum. 
At threshold:
$\tilde{\mu}^+_L\tilde{\mu}^-_L\, , \, 
\tilde{\mu}^+_R\tilde{\mu}^-_R\, , 
\sel^+ \sel^- $ and $ \ser^+ \ser^-$ pairs are excited in a 
P-wave characterized by a slow rise of the cross section 
$\sigma \sim \beta^3$ with slepton
velocity $\beta$. On the other hand, in 
$e_L^+ e_L^- \,/\, e_R^+ e_R^- \to \ser^+ \sel^- \,/\, \sel^+
\ser^-$ and $e_L^- e_L^- \,/\, e_R^- e_R^- \to \sel^- \sel^- \,/\,
\ser^- \ser^-$ 
sleptons are excited in the S-wave  giving steep
rise of the cross sections $\sigma\sim \beta$. 
Therefore the shape of the cross section near threshold is sensitive
to the masses and quantum numbers. 

\begin{figure}[htb]
\centering
\includegraphics*[width=35mm,height=40mm,angle=-90]
      {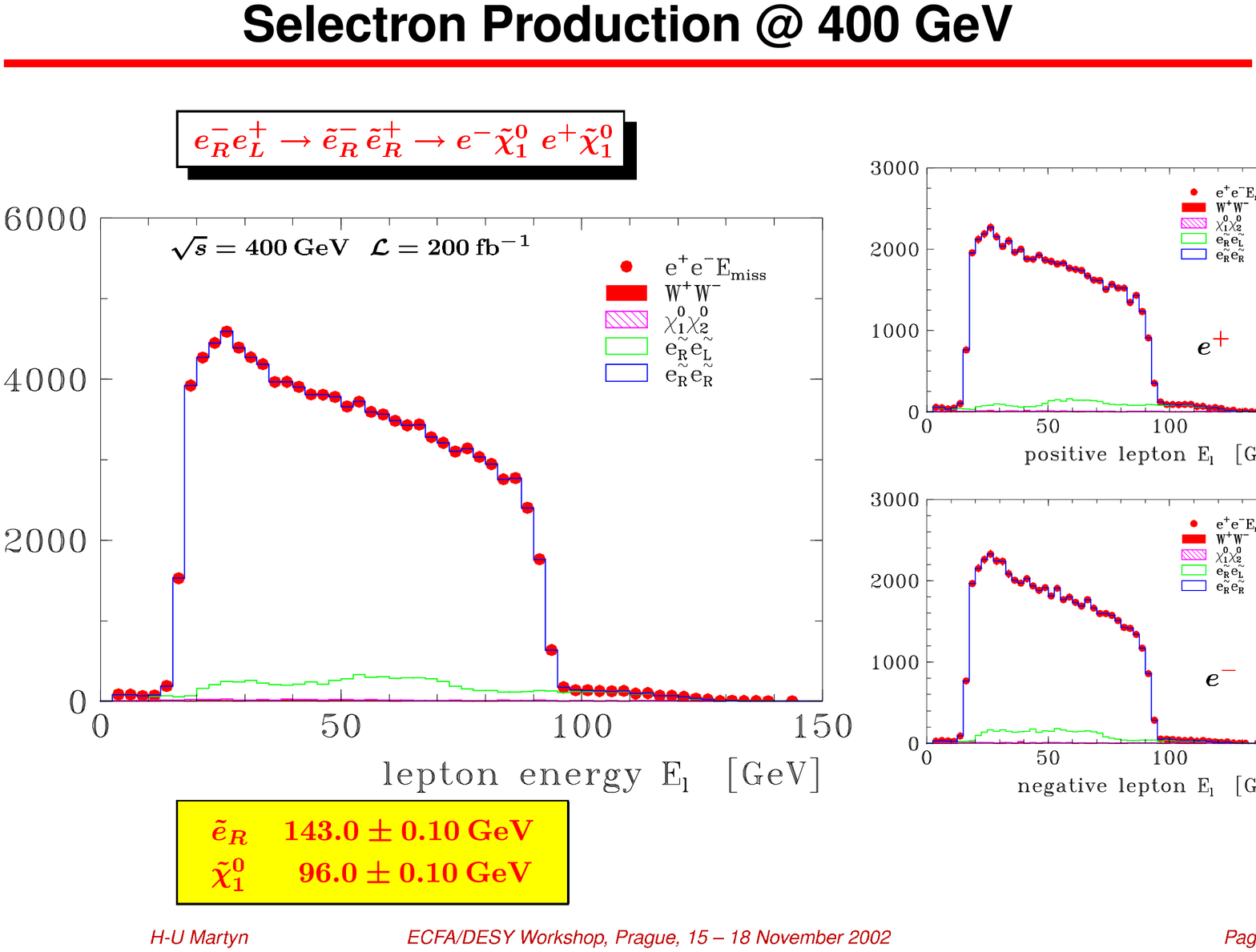}
\includegraphics*[width=35mm,height=40mm,angle=-90]
      {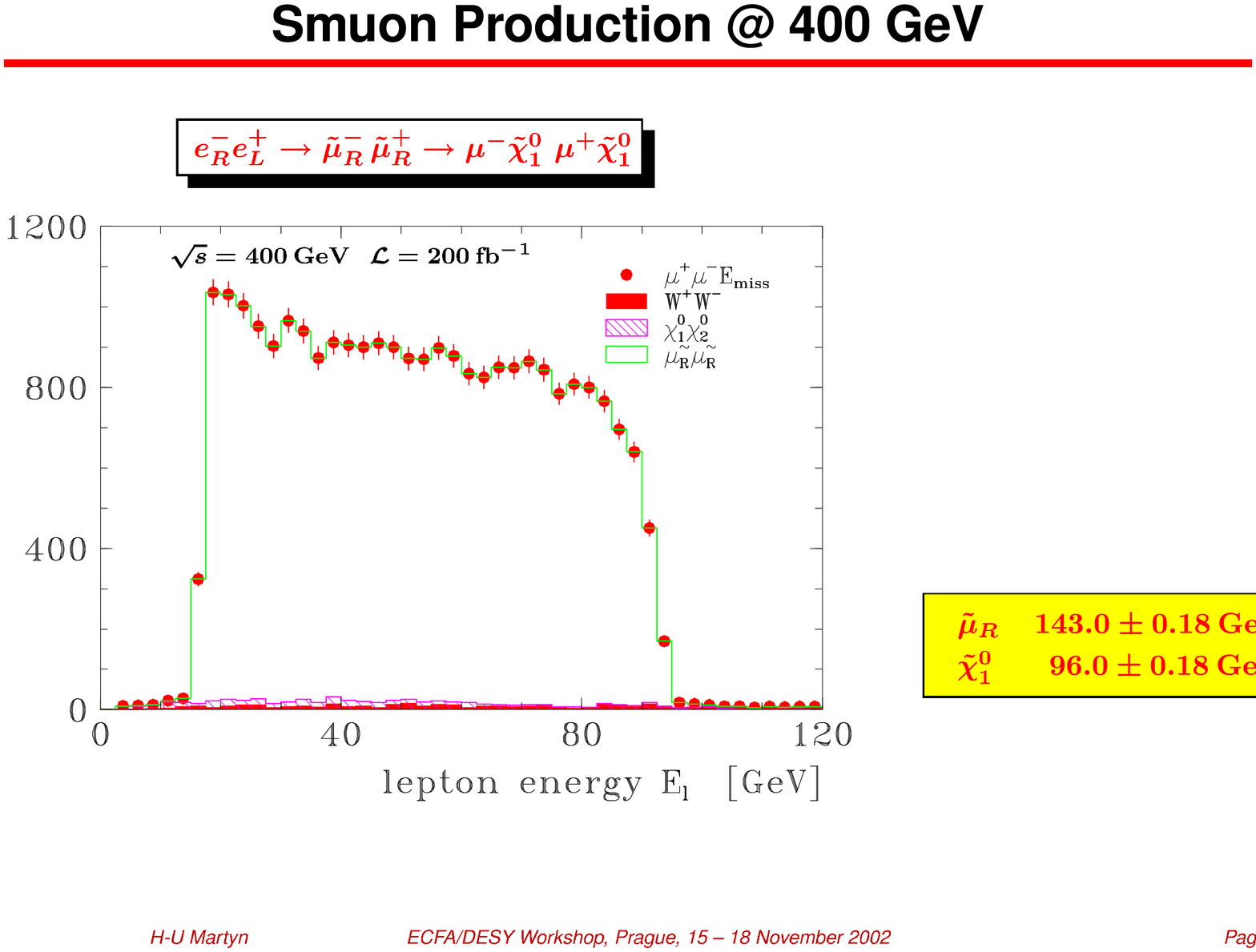}
  \caption{Lepton energy spectra in the processes
    $e^-_R e^+_L \to \serm  \, \serp   
        \to e^- \nt_1 e^+\nt_1 $ (left) 
    and
    $e^-_R e^+_L  \to \smurm  \smurp 
          \to\mu^-\nt_1 \, \mu^+\nt_1 \to         
        \mu^-\nt_1 \mu^+ \nt_1 $ (right) at  
    $\sqrt{s} = 400~\GeV, \cL = 200~\fbi$;
    scenario SPS\#1a \cite{martyn-prague}.  }
  \label{esmu}
\end{figure}

The expected experimental precision requires higher order corrections,
and finite sfermion width effects  to be included.  Examples of
simulations for the SPS\#1a point are shown in \fig{scans}.  Using
polarized $e^+e^-$ beams and $\cL=50~\fbi$ a (highly correlated)
2-parameter fit gives $\delta m_{\ser} = 0.20~\GeV$ and
$\delta\Gamma_{\ser}=0.25~\GeV$; the resolution deteriorates by a
factor of $\sim2$ for $\smur\smur$ production.  For
$e^-_Re^-_R\to\ser\ser$ the gain in resolution is a factor $\sim 4$
with only a tenth of the luminosity, compared to $\ee$ beams.

Above the threshold, slepton masses can be obtained from the endpoint
energies of leptons coming from slepton decays. In the case of
two-body decays, $
   \sell^-  \to  \ell^-\nt_i \ $ and 
  $   \snu_\ell  \to  \ell^-\ch^+_i $ 
the lepton energy
spectrum is flat with endpoints (the minimum $E_-$ and maximum $E_+$
energies)  
\begin{eqnarray}  
  E_{\pm}  & = &
       {\textstyle \frac{1}{4}} \sqrt{s} \, (1 \pm \beta) 
        ( 1 - {m_{\cx}^2}/{m_{\sl}^2} ) 
  \label{eminmax}
\end{eqnarray}
providing an accurate determination of
the masses of the primary slepton and the secondary neutralino/chargino.
   
Simulations of the $e$ and $\mu$ energy spectra of
$\ser\ser$ and $\smur\smur$ (respectively) production, 
including beamstrahlung, QED
radiation, selection criteria and detector resolutions, 
are shown in \fig{esmu} assuming mSUGRA scenario SPS\#1a 
\cite{martyn-prague}. 
With a moderate luminosity of $\cL = 200~\fbi$ at $\sqrt{s}=400$ GeV 
one finds $m_{\ser}=143\pm 0.10$ GeV, $m_{\smur}=143\pm 0.10$ GeV 
and $m_{\nt_1}=96 \pm 0.10$ GeV from selectron, or 
$m_{\nt_1}=96 \pm 0.18$ GeV from smuon production processes. 
Assuming the neutralino mass is known, one can improve slepton mass
determination by a factor 2 from reconstructed kinematically allowed
minimum $m_{min}(\sell)$. 
A slightly better experimental error for the neutralino mass 
$\delta m_{\nt_1}=0.08$ GeV from the smuon production has recently 
been reported
in \cite{nieto-ams}. 
The partner $\smul$ is more difficult to detect because of large
background from $WW$ pairs and SUSY cascades. However, with the high
luminosity of TESLA one may select the rare decay modes
$\smul \to \mu \nt_2$ and  $\nt_2  \to  \ell^+ \ell^-\,\nt_1$,
leading to a unique, background free 
signature $\mu^+\mu^-\,4 \ell^\pm \Eslash$.
The achievable mass resolutions for $m_{\smul}$ and $m_{\nt_2}$ is of the
order of 0.4 GeV \cite{martyn-susy02}.

\begin{figure}[htb]
\centering
\includegraphics*[width=30mm,height=40mm,angle=-90]
{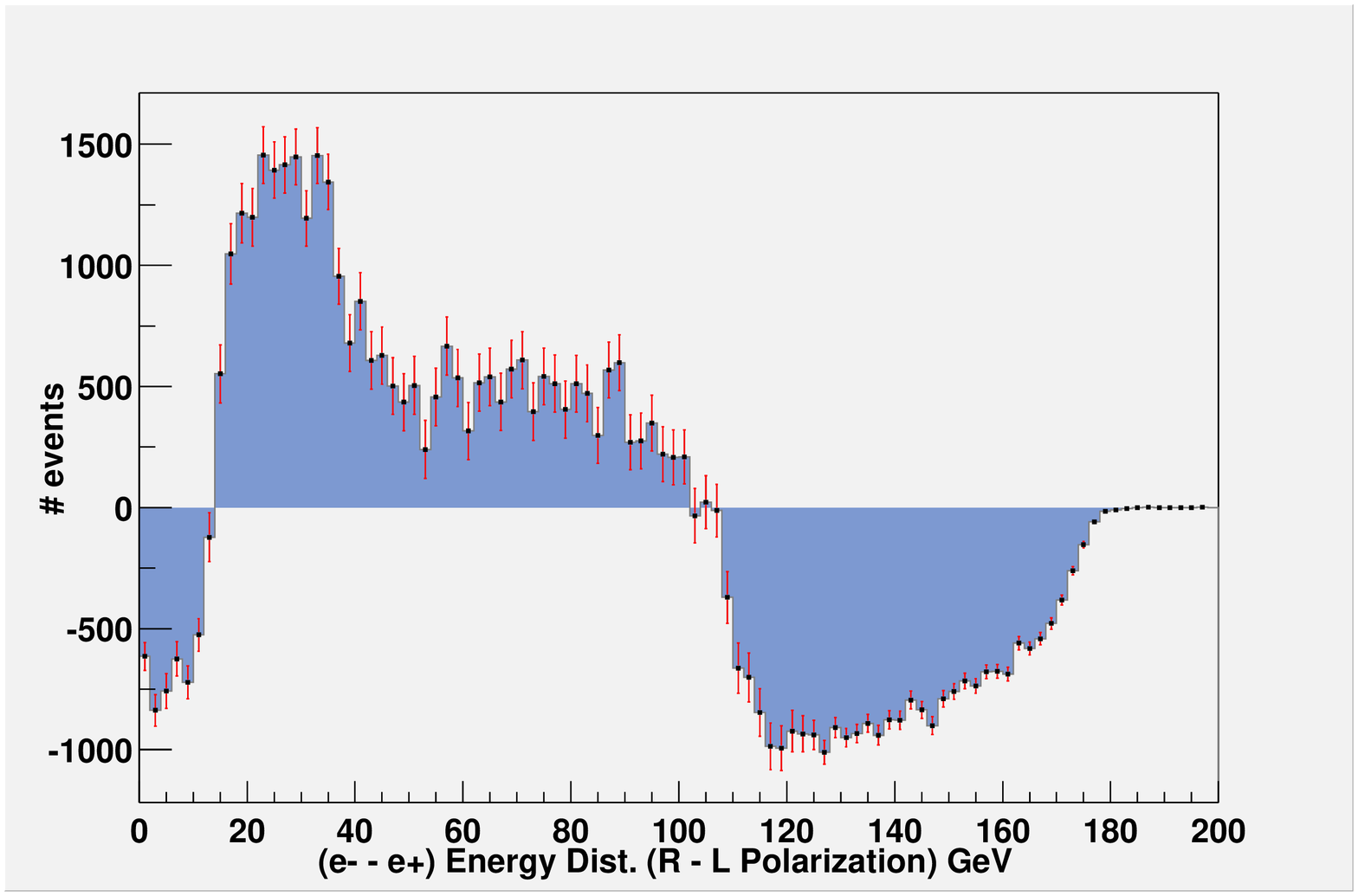}~\begin{minipage}[t]{36mm}
  \caption{Energy spectrum 
    $(E_{e^-} - E_{e^+})_{e^-_R} - (E_{e^-} - E_{e^+})_{e^-_L}$
    for $e^-_{R,\,L} e^+ \to \ser\sel$ in the model
    SPS\#1a at $\sqrt{s}$=500 GeV,   ${\cL}$=2$\cdot$500 $\fbi$
    \cite{Dima:01}.   \label{eserl} }
\end{minipage}
\end{figure}

One should keep in mind that the 
measurement of selectron masses is  
subject to two experimental difficulties: an overlap of flat energy
distributions of leptons from $\ser^-\sel^+, \ser^-\ser^+,
\sel^-\sel^+, \sel^-\sel^+$, and large SM background.   
Nevertheless, it has been demonstrated~\cite{Dima:01} 
that thanks to larger cross sections, both problems can be
solved by a double subtraction of
$e^-$ and $e^+$ energy spectra and opposite electron beam polarizations
$\cP_{e^-} = +0.8$ and $\cP_{e^-} = -0.8$, symbolically
$(E_{e^-} - E_{e^+})_{e^-_R} - (E_{e^-} - E_{e^+})_{e^-_L}$. 
Such a procedure eliminates all charge symmetric
background and clearly exhibits 
endpoints from the $\ser$ and $\sel$ decays, as seen in 
\fig{eserl}. Simulations   at $\sqrt{s}=500~\GeV$
in the SPS\#1a scenario~\cite{Dima:01} show that both selectron masses can
be determined to an
accuracy of $\delta m_{\ser,\,\sel} \sim 0.8~\GeV$.

\subsection{Sneutrino production}
\begin{figure}[tb]
\centering
\includegraphics*[width=38mm,height=83mm,angle=90]{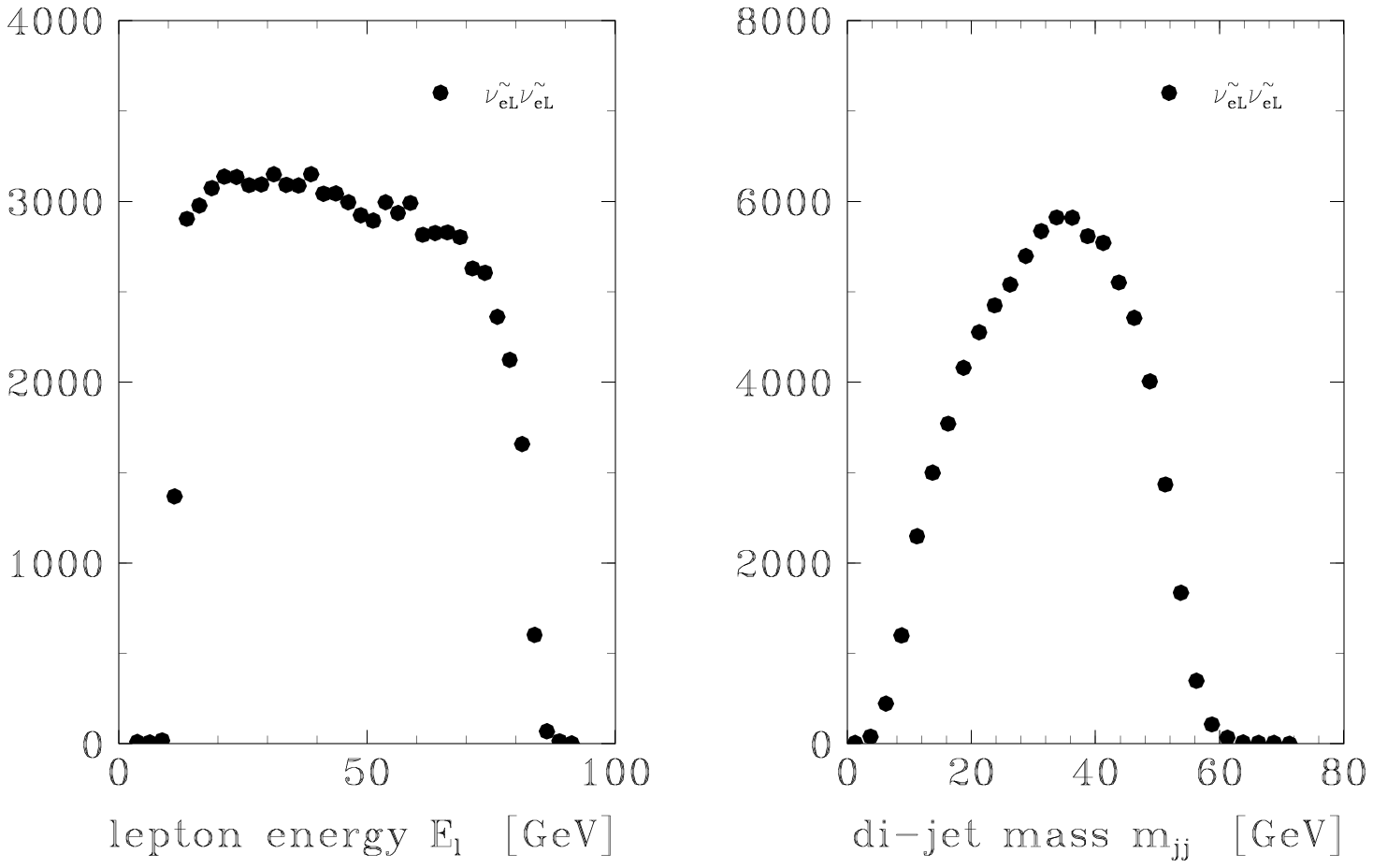}  
\caption{Lepton energy and di-jet mass spectra of 
  $e^-_L e^+_R  \to \sne  \sne \to e^-\cx^+_1\, e^+\cx^-_1$ (left)
  with subsequent decay
  $\cx^\pm_1  \to  q\bar{q}' \,\nt_1 $  (right) \cite{martyn-susy02} } 
\label{esnu}
\end{figure}

At $e^+e^-$ collisions sneutrinos are produced in pairs via the
s-channel $Z$ exchange; for the $\sne$ production there is
additional t-channel chargino exchange. 
Their decay into the corresponding
charged lepton and chargino, and the subsequent chargino decay, 
make the final topology, e.g. 
$\sne\sne \to e^+e^-\ell^\pm 2j\, \Eslash$,
very clean. 
The primary charged lepton energy,  and di-jet energy
and mass spectra, see \fig{esnu}, 
can be used to
determine $m_\snu$ and $m_{\cx_1^\pm}$ to 2~per~mil (or better), and  
to measure the chargino
couplings and the $\cx^\pm_1 - \nt_1$ mass difference; 
a resolution below 50~MeV, given essentially by detector
systematics, appears feasible \cite{martyn-susy02}.
The detection and measurement of tau-sneutrinos $\snt$ is more
problematic, due to neutrino losses in decay modes and decay energy spectra.

\subsection{Study of  staus   }  
In contrast to the first two generations, the $L-R$ mixing for the
third generation sleptons can be non-negligible due to the large $\tau$ 
Yukawa coupling.  
Therefore the $\tilde{\tau}$'s are  
very interesting to study since 
their production and decay is different from $\tilde{e}$
and $\tilde{\mu}$. 

\begin{figure}[tb]
{\small 
$\cos2\theta_{\tilde \tau}$ ~~~~ ~~~~~ ~~~~~ ~~~~~~~~~~ ~~~~ 
$\tan\beta$ ~~~~~~~~~} \\[-1mm]
\centering
\includegraphics*[width=40mm,height=36mm]{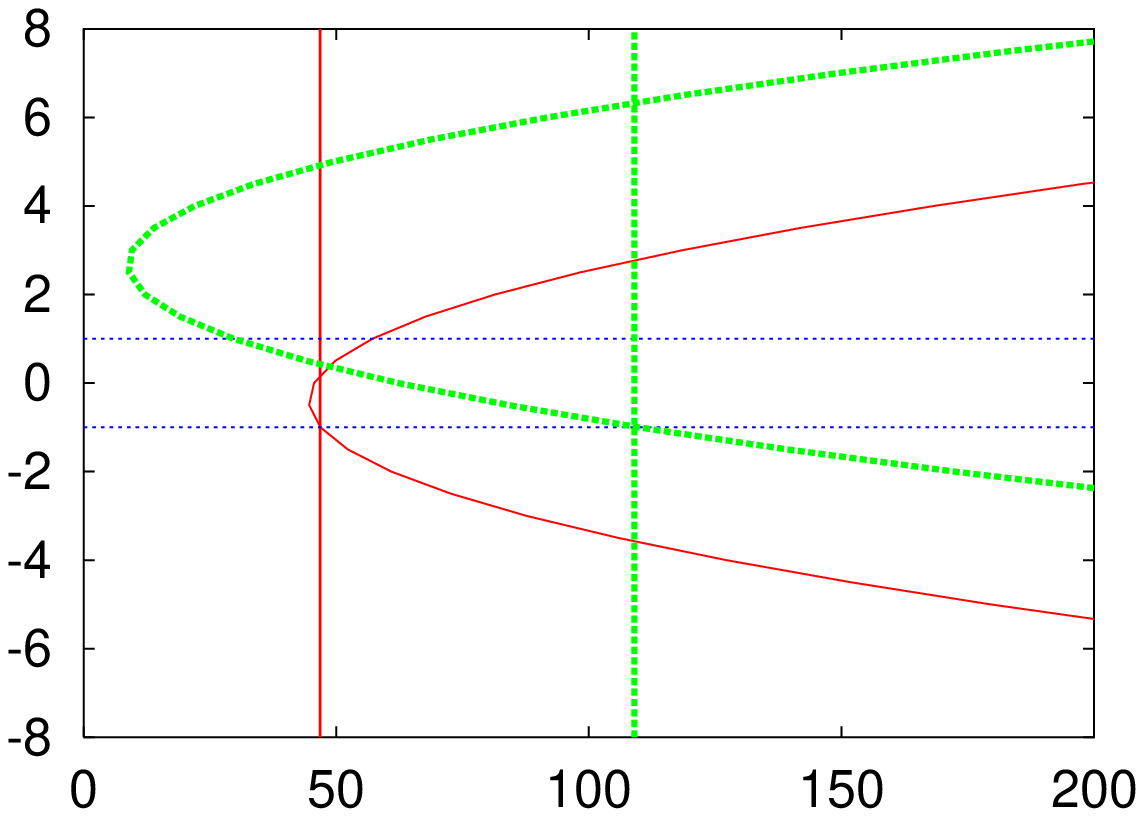}
\includegraphics*[width=40mm,height=36mm]{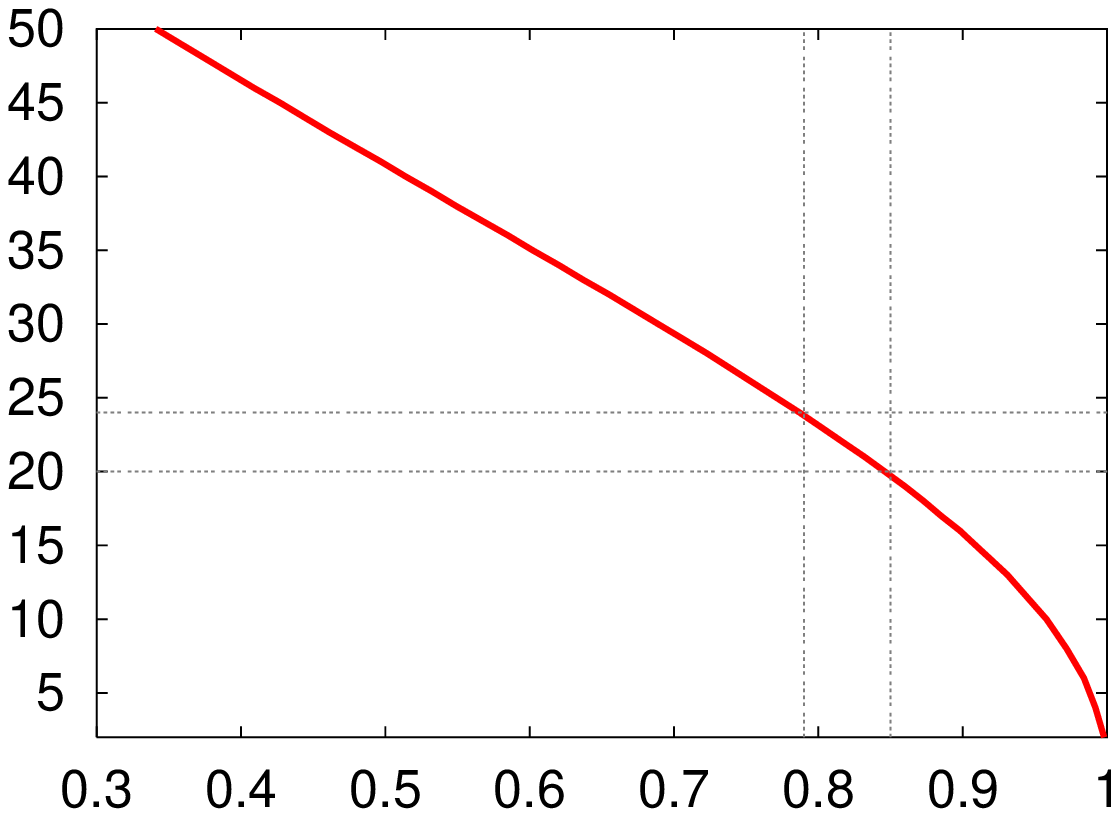}\\[-5mm]
{\small \phantom{} ~~~ ~~
 $\sigma(e^+ e^- \to \stau_1\stau_1)$ [fb] ~~~~~~~~~~~~~~~~~~~~
~~~~~~~~~~~~~~ $P_{\tilde{\tau}_1\to\tau\nt_1}$ }
\caption{Left: $\cos 2\theta_\stau$ versus $\sigma(e^+ e^- \to
\stau_1\stau_1)$ at $\sqrt{s}=500~\GeV$ for polarized
(green, upper curve) and unpolarized
(red, lower curve) beams; the
expected cross sections shown by vertical lines. Unpolarised beams
give 
a two-fold ambiguity in $\cos2\theta_{\tilde{\tau}}$,
while polarized
beams give a unique physical solution.  
Right: $\tan{\beta}$ as a function of $\tau$ polarization.
From simulations $P_{\tau}=0.82\pm0.03$ leading to
$\tan\beta=22\pm 2$ \cite{0303110}.}
\label{taumixpol}
\end{figure}

The $\stau$ masses can be determined with the usual techniques of
decay spectra  or
threshold scans at the per~cent level, while 
the mixing angle $|\cstau|$ can be extracted with high accuracy
from cross section
measurements with different beam polarisations.
In a case study \cite{0303110} for 
$m_{\tilde{\tau}_1}=155$ GeV, $m_{\tilde{\tau}_2}=305$ GeV, 
$\mu=140$ GeV, $\tan\beta=20$, $A_\tau=-254$ GeV it has been found
that at $\sqrt{s}=500$ GeV, ~${\cal L}=250$ fb$^{-1}$, ~${\cal
P}_{e^-}=+0.8$, ~${\cal P}_{e^+}=-0.6$, the  
expected precision is as follows: $m_{\tilde{\tau}_1}=155\pm 0.8$ GeV, 
~ $\cos2\theta_\tau= -0.987\pm0.08$, left panel of \fig{taumixpol}.

The dominant decay mode  
$\tilde{\tau}_1\to \tilde{\chi}^0_1 \tau$ can be exploited 
to determine $\tan\beta$ if $\tan\beta$ turns to be large \cite{taupol}.    
In this case the 
non-negligible $\tau$ Yukawa coupling makes $\tilde{\tau}$ 
couplings sensitive to the neutralino composition in the decay
process. Most
importantly, if the higgsino component of the neutralino is sufficiently
large,  
the polarization of $\tau$'s from the $\tilde{\tau}$   
decay turns out to be a sensitive function of $\tilde{\tau}$
mixing, neutralino mixing {\it and} $\tan\beta$ \cite{0303110}. This is
crucial since for large $\tan\beta$ other SUSY sectors are not very 
sensitive to $\tan\beta$   and therefore cannot provide
a precise determination of this parameter.

The $\tau$ polarization can be measured using the energy distributions
of the decay hadrons, e.g. $\tau\to \pi\nu$ and
$\tau\to \rho\nu \to\pi^\pm\pi^0\nu$.
Simulations show that the $\tau$ polarization can
be measured very accurately, $\delta\cP_\tau
=0.82\pm0.03$, which in turn allows to determine $\tan\beta=20\pm2$,
as shown in the right panel of \fig{taumixpol}. 

\subsection{Squarks   }
For the third generation squarks, $\tilde{t}$ and  $\tilde{b}$, 
~~the  $L-R$ mixing is also 
expected to be important.   As a result of the large 
top quark Yukawa coupling, it is possible that the lightest superpartner
of the quarks  is the
stop $\st_1=\st_L \cos\tst + \st_R \sin\tst$.  
If the mass $m_{\st_1}$ is below 250~GeV,
it may escape detection at the LHC, 
while it
can easily  be discovered at the Linear Collider.
\begin{figure}[htb]
{\small \phantom{} \hspace*{-1cm}
$\cos\theta_{\ti t}$~~~~~ ~~~~~~~ ~~~~~~~~ ~~~~~~~~~
 ~~~~~~~~ $\tan\beta$}\\[-1mm] ~~~~~~~~~~
\centering
\includegraphics*[width=40mm,height=33mm]{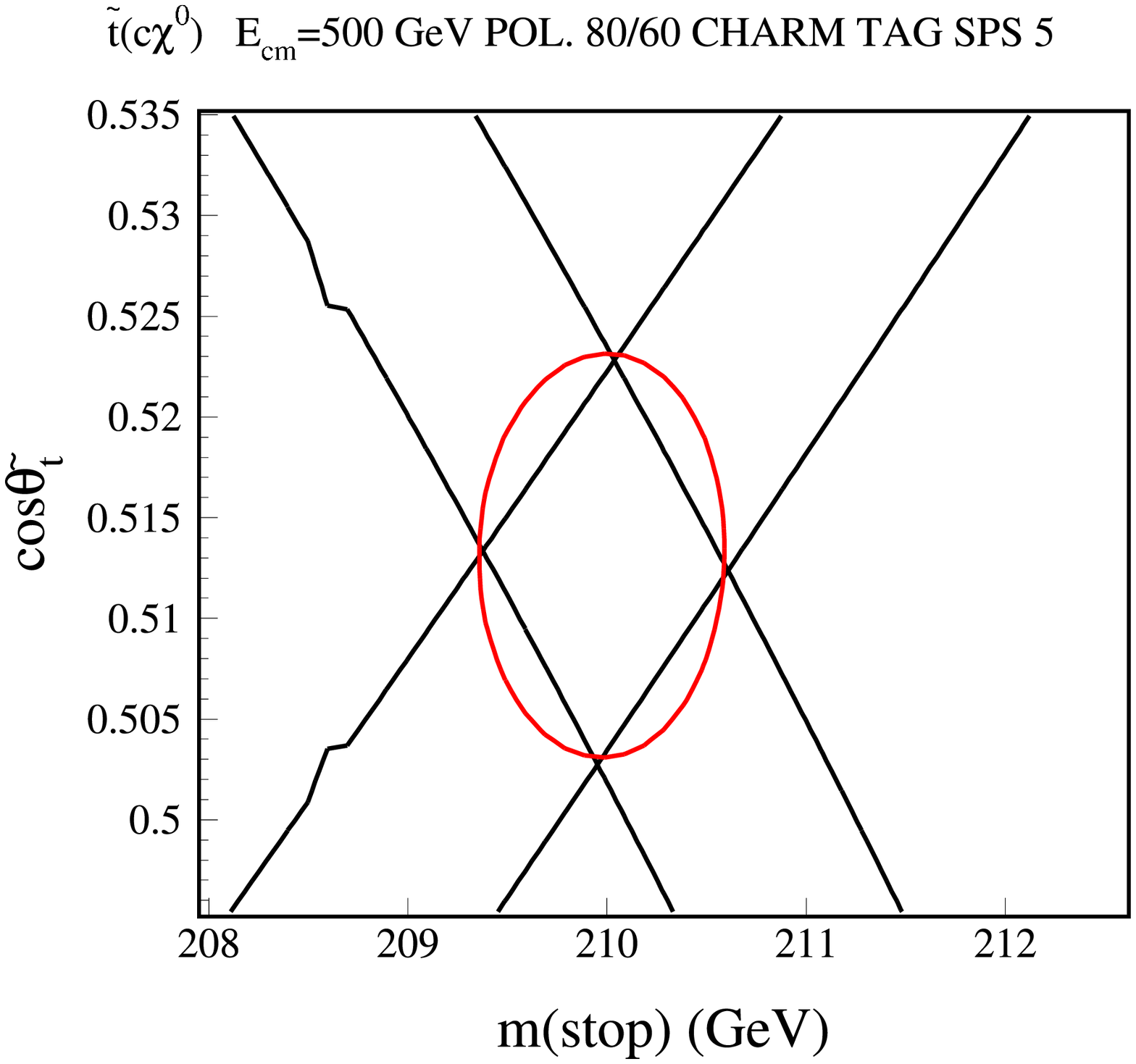}
\includegraphics*[width=40mm,height=35mm]{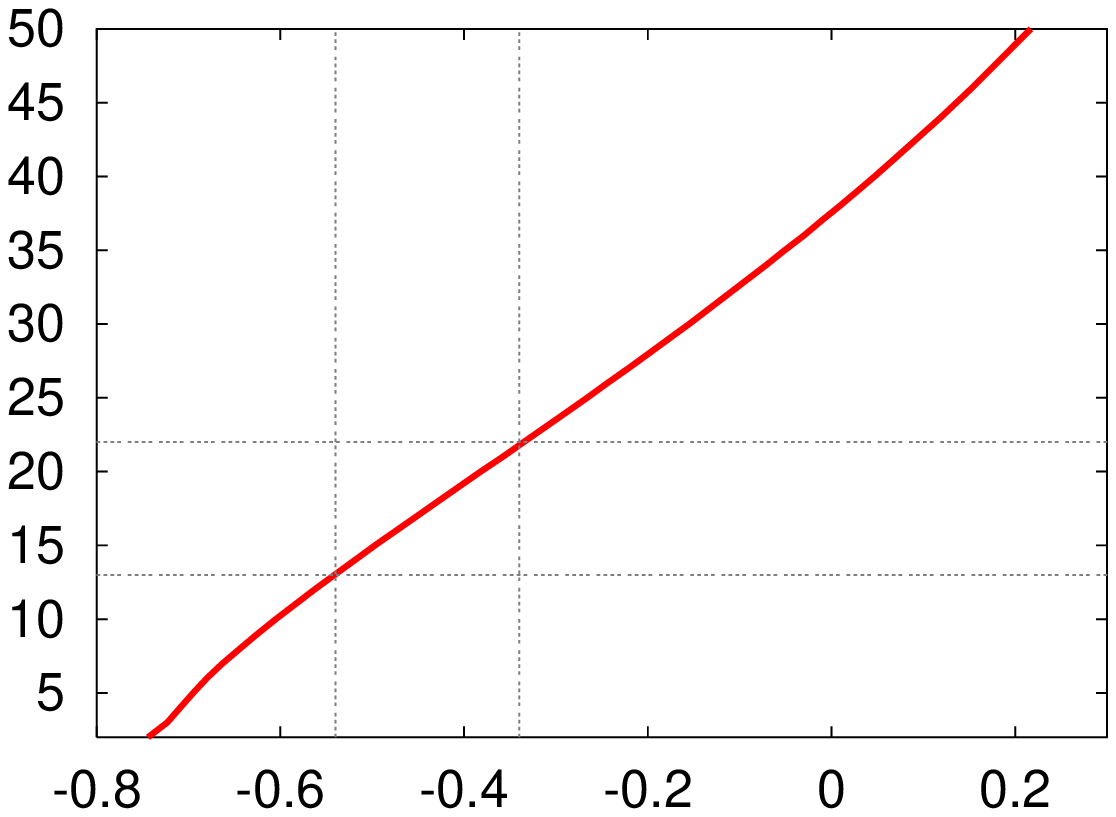}\\[-5mm] {\small
\phantom{} ~~~~~~~~ $m_{\ti t}$ [GeV]~~~~~~~~~~~~~~~ ~~~~~~~~~
~~~~~ ~~~~ ~~~~~~ $P_{\sb_1\to t
\cx_1^\pm}$} \caption{Left: Contours of $\sigma_{R}(\st_1\st_1)$ and
$\sigma_{L}(\st_1\st_1)$ as a function of $m_{\st_1}$ and $\cos\tst$
for $\sqrt{s} = 500~\GeV$, $\cL = 2\cdot 500~\fbi$ \cite{FNS}.
Right: $\tan\beta$ as a function of top polarization.  From
simulations $P_t = -0.44\pm 0.10$ leading to $\tan\beta=17.5\pm4.5$
\cite{0303110}.  } \label{stopmasspol}
\end{figure}

The $\st$ and $\sb$ phenomenology is analogous to that of the
$\stau$ system. The masses and  
mixing angles can be extracted from production cross sections
measured with polarized beams. 
The production cross sections for $e^+e^- \to \st_1 \bar \st_1$ with
different beam polarizations, $\sigma_R=\sigma_{e^-_R e^+_L}$ and
$\sigma_L=\sigma_{e^-_L e^+_R}$, have been studied for $\st_1\to
b\,\ch^\pm_1$ and $\st_1 \to c\,\nt_1$ decay modes including
full-statistics SM background.  New analyses have been performed for
the SPS\#5-type point: a dedicated ``light-stop'' scenario with
$m_{\st_1}=210$~GeV, $m_{\nt_1}=121.2$~GeV \cite{FNS}. For this point
the decay $\st_1\to b\,\ch^\pm_1$ is not open, and the SUSY background
is small. The charm tagging, based on a CCD detector, helps to enhance
the signal from the decay process $\st_1 \to c\,\nt_1$.  Generated
events were passed through the SIMDET detector simulation.  The
results, shown in the left panel of \fig{stopmasspol}, provide high
accuracies on the mass $\Delta m_{\st_1}\sim 0.7$ GeV and mixing angle
$\Delta\cos\tst\sim 0.01$.

\begin{figure}[htb]
\centering
\includegraphics*[width=45mm,height=35mm]
{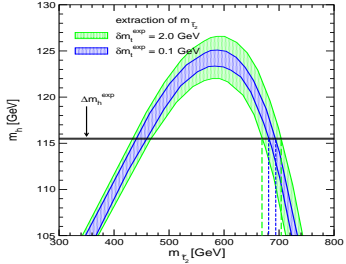}~~\begin{minipage}[b]{35mm} 
\caption{Indirect determination of $m_{\st_2}$ 
from the $m_h$ measurement   for
$\delta m_t$=2 GeV (LHC) and  0.1 GeV (LC) 
 \cite{0306181}. ~~~~~~~~~~~~~~~~~~~~~~~~~~~~~~~~~~~~~~~~~~~~~~~ 
\label{fig:mhMSt2}} \end{minipage}
\end{figure}

If the heavier stop $\st_2$ is too heavy to be produced at the LC, the
precise measurement of the Higgs boson mass $m_h$ together with
measurements from the LHC can be used to obtain indirect limits on
$m_{\st_2}$ \cite{0306181}. 
Assuming 
$m_{\st_1} = 180 \pm 1.25$ GeV, $\cos\tst = 0.57 \pm 0.01$, 
$M_A = 257 \pm 10$ GeV, 
$\mu = 263 \pm 1$ GeV, $m_{\sg} = 496 \pm 10$ GeV, $A_{\ti b} = 
A_{\ti t} \pm 30\%$ 
and $m_{\sb_1}>200$ GeV, \fig{fig:mhMSt2} shows the allowed region in the
$m_{\st_2}$--$m_h$ plane. 
Only a lower bound  $\tan\beta > 10$ has been assumed, which
could for instance be inferred from the gaugino/higgsino sector. 
Intersection of the assumed measured value $m_h = 115.5 \pm 0.05$ GeV
with the allowed $m_{\st_2}$--$m_h$ region gives  an indirect determination of 
$m_{\st_2}$, yielding 670 GeV $\lsim m_{\st_2} \lsim$ 705 GeV
for the LHC precision  
$\delta m_t = 2$ GeV (${\st_2}$ must be above the LC reach).
The LC precision of $\delta m_t = 0.1$ GeV 
reduces the range to 680 GeV $ \lsim
m_{\st_2} \lsim$ 695 GeV, i.e.\ by a factor of more than~2.

Similarly to the $\stau$, the measurement of top quark polarization
in the squark decay can provide information on $\tan\beta$. 
For this purpose  the decay $\tilde{b}_1\to t \tilde{\chi}^{\pm}_1$ 
is far more useful than $\tilde{t}_1\to t \tilde{\chi}^0_k$  
since in the latter the $t$ polarization depends on 
$1/\sin\beta$   and therefore is only weakly sensitive to large $\tan\beta$.

A feasibility study of the reaction
\begin{eqnarray}
  e^+_Le^-_R & \to & \sb_1\bar \sb_1
  \ \to \ t\cx_1^- + \bar{t} \cx_1^+
  \quad  \ 
  \label{sb11}
\end{eqnarray}
has been performed in \cite{0303110}. 
A fit to the angular distribution $\cos\theta^*_s$, 
where $\theta^*_s$ is the angle between the 
$\bar{s}$ quark and the primary $\sb_1$ in the top rest frame in the
decay chain $e^+e^-\to \bar \sb_1 + t\,\cx_1^- \to \bar \sb_1 + b c
\bar{s}\,\cx_1^-$,
yields $P_t = -0.44 \pm 0.10$,  consistent with the input
value of $P_t^{th} = -0.38$. 
From such a measurement one can derive $\tan\beta = 17.5 \pm 4.5$, 
as illustrated in the right panel of \fig{stopmasspol}.
After $\tan\beta$ is fixed, measurements of stop masses and mixing
allow us to determine the trilinear coupling $A_{\ti t}$ at the
ten-percent level \cite{0303110}.


\subsection{Quantum numbers}
An important quantity is the spin of the sfermion which 
can directly be determined from the
angular distribution of sfermion pair production in $e^+e^-$
collisions \cite{Aguilar-Saavedra:2001rg,martyn-susy02}.

\begin{figure}[htb]
\centering
\includegraphics*[width=45mm,height=40mm]
{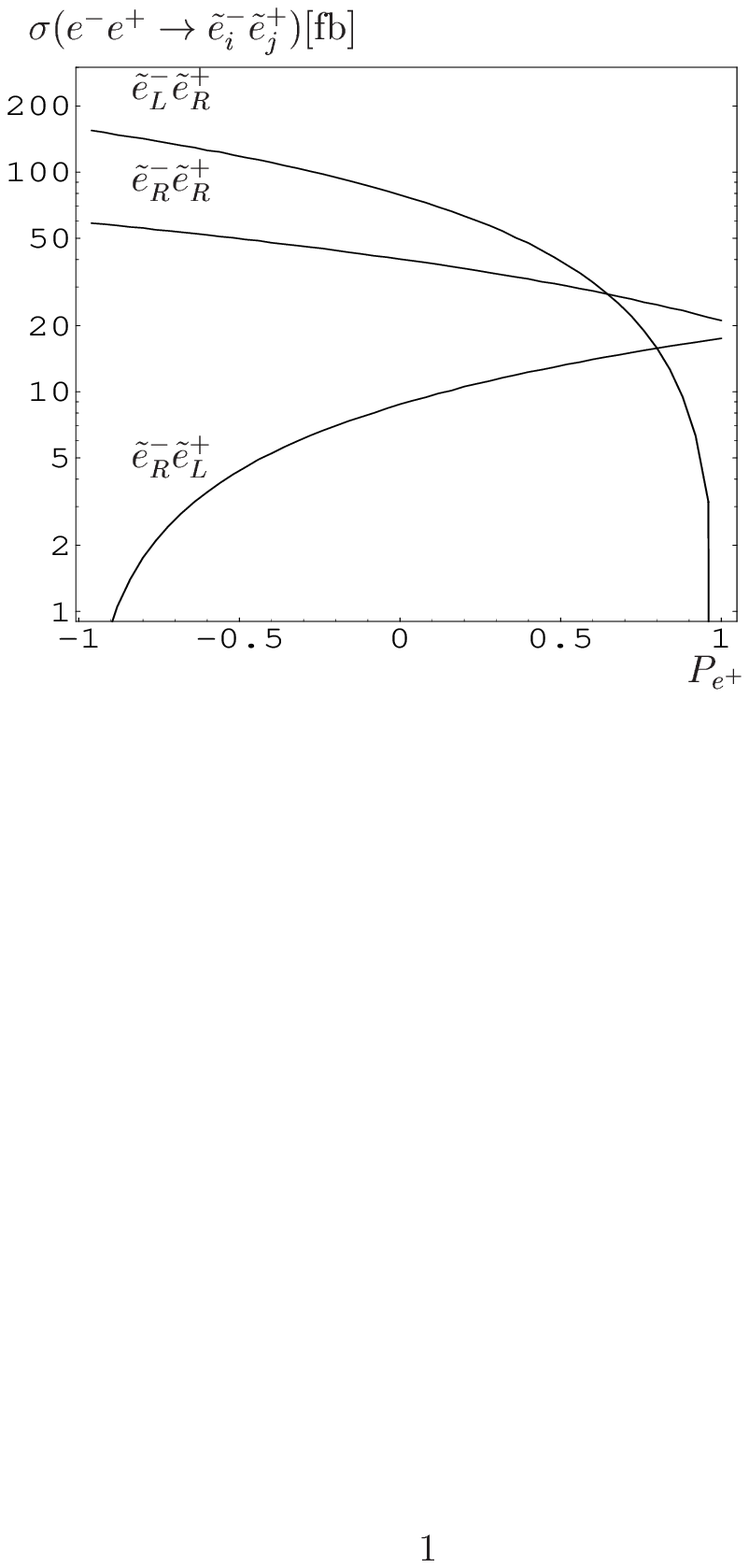}
 \begin{minipage}[b]{35mm}
  \caption{Production cross sections as a function of $P_{e^+}$ for
    $\sqrt{s}$ =350~GeV, $P_{e^-}$=-0.8. 
  ISR and beamstrahlung are included \cite{Bloechi:02}. ~~~~~~~~~ 
~~~~~~~~~~~~~~~~~~~~~~~~~~~~~~~~~~~~~~~~~~~~~~~~~~ \label{elqn} }
\end{minipage}
\end{figure}
Due to small $L-R$ mixing  of the first two generation sfermions, 
the mass eigenstates are 
chiral.  As a result, of particular interest is the
associated production of
\bea
   e^-_R e^+_R \to \ser^-\sel^+ & {\rm and} &
   e^-_L e^+_L \to \sel^-\ser^+
   \label{seproduction}
\eea
via $t$-channel $\nt$ exchange for the sfermion 
quantum number determination. 
For polarized beams the charge of the observed lepton is directly
associated to the $L,\,R$ quantum numbers of the selectrons and the
energy spectrum uniquely determines whether it comes from the $\ser$
or the $\sel$ decay. However, 
in order to separate the t-channel neutralino exchange
from the s-channel photon and Z-boson exchange,
both the electron and positron beams must be
polarized. 
By comparing the
selectron cross-section for different beam polarizations the chiral
quantum numbers of the selectrons can be disentangled, as can be seen
in \fig{elqn}, where other parameters are $m_{\tilde{e}_R}=137.7$ GeV,
  $m_{\tilde{e}_L}=179.3$ GeV, $M_2=156$~GeV, $\mu=316$~GeV and
  $\tan\beta=3$~\cite{Bloechi:02}. 

\subsection{Sfermion Yukawa couplings}
Supersymmetry enforces 
gauge couplings and their supersymmetric Yukawa counterparts to be
exactly equal at  tree level. 
For example,
the Yukawa coupling $\hat{g}_{\tilde{V}f\tilde{f}}$ between the
gaugino partner $\tilde{V}$ of the vector boson $V$, the fermion $f$ and the
sfermion $\tilde{f}$ must be equal to the corresponding gauge coupling
$g_{Vff}$. 

\begin{figure}[htb]
\centering
\includegraphics*[width=40mm,height=30mm]{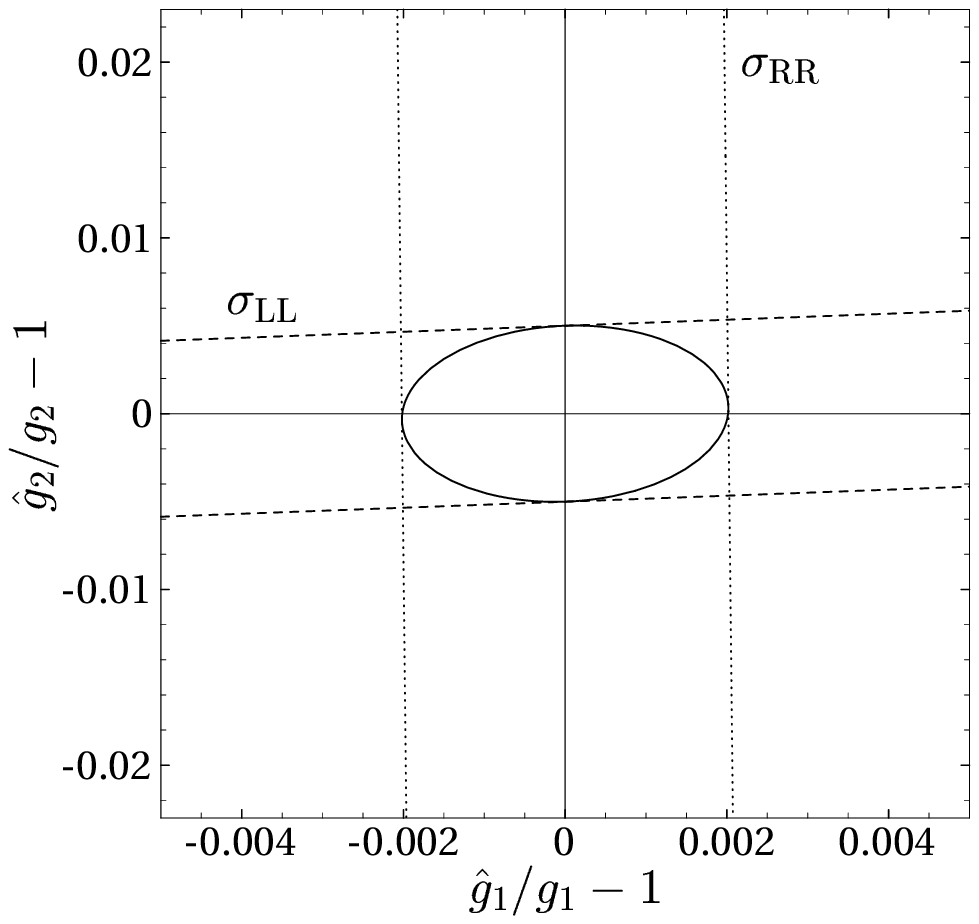} 
\includegraphics*[width=40mm,height=30mm]{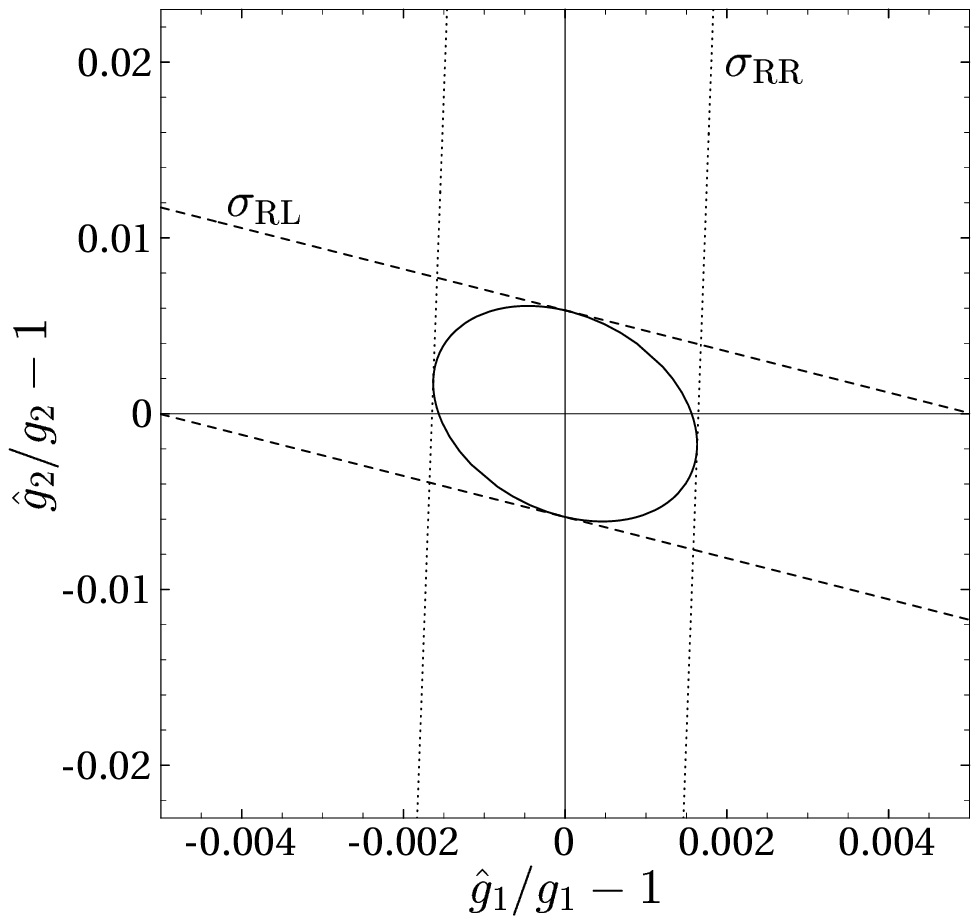}
\caption{The 1$\sigma$ bounds on the supersymmetric Yukawa couplings 
$\hat{g}_1$ and $\hat{g}_2$ in the SPS\#1a scenario 
from  $e^-e^-$ with ${\cal L}=50$
  fb$^{-1}$ (left) and   $e^+e^-$ with ${\cal L}=500$
  fb$^{-1}$ (right), both running at $\sqrt{s}=500$ GeV  
\cite{freitas-phd}.
\label{fig:yukawa}}
\end{figure}

The Yukawa couplings of selectrons 
can best be probed in the production of selectrons via the t-channel
neutralino exchange contributions. 
For this purpose one can exploit 
the $e^-e^-$ collider mode due to reduced background, 
larger production cross-sections, higher beam 
polarizability and no interfering s-channel contributions.
Simulations have shown that these couplings can be determined with high
accuracy \cite{freitas,freitas-phd}. 
For example, errors for the extraction of the
supersymmetric 
Yukawa couplings $\hat{g}_1$ and $\hat{g}_2$ (corresponding to the U(1)
and SU(2) gauge couplings $g_1$ and $g_2$) are expected in the range
$\delta \hat{g}_1/\hat{g}_1 \approx 0.2 \%$ and $\delta \hat{g}_2/\hat{g}_2
\approx 0.8 \%$.  
The values are for the SPS\#1a scenario and  integrated luminosity
of 50 fb$^{-1}$ of the $e^-e^-$ collider running 
at $\sqrt{s} = 500 \GeV$, with no
detector simulation included. Similar precision in the $e^+e^-$ mode
requires integrated luminosity of 500 fb$^{-1}$, see \fig{fig:yukawa}.  

Such a high  experimental precision requires radiative
corrections to be included in the theoretical predictions for the
slepton cross-sections.  
Far above threshold the effects of the non-zero slepton width
are small, of the order $\Gamma_{\tilde{f}}/m_{\tilde{f}}$, and 
the production and decay of the sleptons
can be treated separately. As mentioned, for both sub-processes 
the complete electroweak
one-loop corrections in the MSSM have been computed \cite{freitas,0207364}.
The electroweak corrections were found to be
sizable, of the order of 5--10\%. They include important effects from
supersymmetric particles in the virtual corrections,
in particular non-decoupling logarithmic contributions,
e.g. terms $\propto \log m_{\tilde{f}}/m_{\rm weak}$ 
from fermion-sfermion-loops.

The equality of gauge and Yukawa couplings in the SU(3)$_{\rm C}$ gauge sector
can be tested at a linear collider by investigating the associated production
of quarks $q$ and squarks $\tilde{q}$ with a gluon $g$ or gluino $\tilde{g}$.
While the processes $e^+e^- \to q \bar{q} g$ and $e^+e^- \to \tilde{q}
\bar{\tilde{q}} g$ are sensitive to the strong gauge coupling of quarks and
squarks, respectively, the corresponding Yukawa coupling can be probed in
$e^+e^- \to q \bar{\tilde{q}} \tilde{g}$.
In order to obtain reliable
theoretical predictions for these cross-sections it is necessary to include
next-to-leading order (NLO) supersymmetric QCD corrections. These
corrections are generally expected to be rather large and they are necessary
to reduce the large scale dependence of the leading-order result.
The NLO QCD corrections to the process $e^+e^- \to q \bar{q} g$
within the Standard Model have been  known for a long time.
Recently, the
complete ${\cal O}(\alpha_{\rm s})$ corrections to all three processes in the
MSSM have been calculated \cite{BMW}. 
The NLO contributions enhance the cross-section in the peak region by roughly
20\% with respect to the LO result. Furthermore, the scale dependence is
reduced by a factor of about six when the NLO corrections are included.

\subsection{Mass universality   }  
Most analyses are performed with a simplifying assumption of 
universal mass parameters at some high energy scale $G$: 
$\delta m^2(G)=m^2_{\tilde{l}_R}(G)-m^2_{\tilde{l}_L}(G)$=0. 
This assumption can be tested at the LC. 
For example,
in \cite{Baer:01} a quantity 
\begin{eqnarray}
\Delta^2
= & m^2_{\tilde{e}_R}-m^2_{\tilde{e}_L}+\frac{m^2_{\tilde{\chi}^\pm_1}}
{2\alpha^2_2}
[\frac{3}{11}(\alpha^2_1-\alpha^2_1(G))\nn \\
&-3(\alpha_2^2-\alpha^2_2(G))],
\end{eqnarray}  
defined at the electroweak scale, 
is proposed as a probe of  non-universality of slepton masses 
if only both selectrons
and the light chargino are accessible at a linear collider ($\alpha_1$
and $\alpha_2$ are the U(1)  and SU(2) couplings). 
It turns out that $\Delta^2$    
is strongly correlated with   
the slepton mass  splitting, $\Delta^2 \sim 0.76\; \delta m^2(G)$.
Assuming SUSY masses in the 150  GeV range to be measured with
an experimental error of 1\%, it has been found~\cite{Baer:01} that 
the non-universality can be detected for $|\delta m^2(G)| \geq  2500$
GeV$^2$;  
knowing the gaugino mass $M_2$ to 1\% increases the sensitivity down to
$\delta m^2(G) 
=1400$ GeV$^2$.

\subsection{Sfermions with complex CP phases}
The soft SUSY breaking parameters: the gaugino masses and trilinear
scalar couplings, and   the Higgsino mass
parameter $\mu$, can in general be  complex and the presence of  non-trivial
phases violates CP. This generalization is quite natural and is
motivated by the analogy between fermions and sfermions: in the SM the
CKM phase is quite large and the smallness of CP-violating observables
results from the structure of the theory. Furthermore, large leptonic
CP-violating  phases together with leptogenesis 
may explain the baryonic asymmetry of the Universe.

\begin{figure}[hbt]
\centering
\includegraphics*[width=40mm,height=35mm]{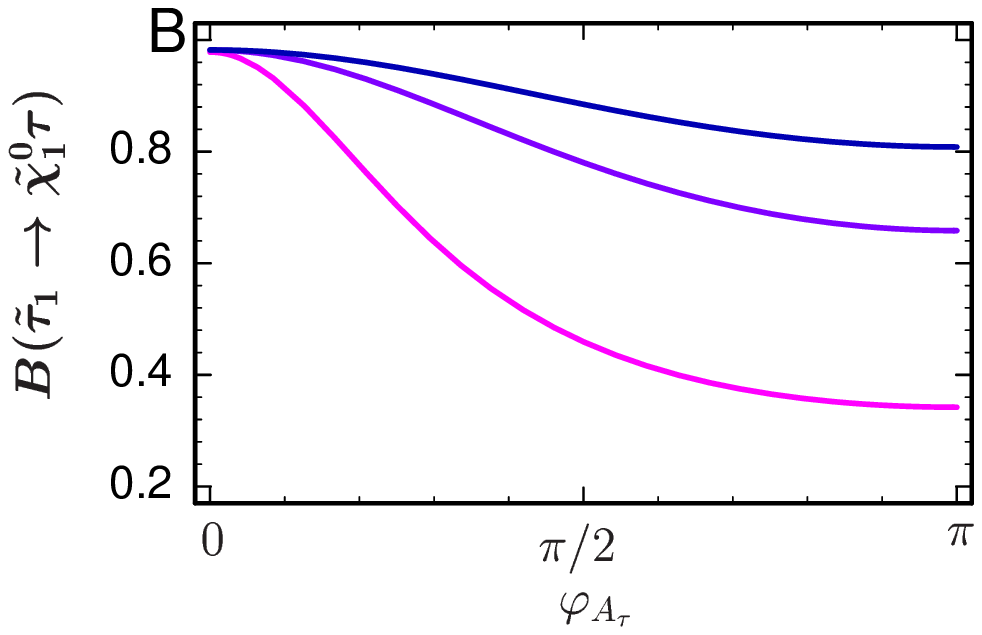}
\begin{minipage}[b]{38mm}
\caption{Branching ratios of $\stau_1\to \tilde{\chi}^0_1\tau $ for 
$m_{\snu}=233$, 
$238$, $243$ GeV (from bottom to top) \cite{0207186}. \label{CPstau}
~~~~~~~~ ~~~~~~~~~~ ~~~~~~~~ ~~~~~~~~ ~~~~~~~~ ~~~~~~~~}
\end{minipage}
\end{figure}

In mSUGRA-type models the phase $\varphi_{\mu}$ of $\mu$ is
restricted by the experimental data on electron, neutron and mercury
electric dipole moments (EDMs) to a range $|\varphi_{\mu}| \lsim 0.1$
-- 0.2 if  a universal scalar mass parameter $M_0 \lsim 400$~GeV is
assumed.
However, the restriction due to the electron
EDM can be circumvented if complex lepton flavour violating terms
are present in the slepton sector \cite{sleptonedm}.
The phases of the parameters $A_{{\ti t},{\ti b}}$ enter 
the EDM calculations only at two-loop level, resulting
in much weaker constraints  \cite{edmAf}. 

\begin{figure}[hbt]
\centering
\includegraphics*[width=40mm,height=35mm]{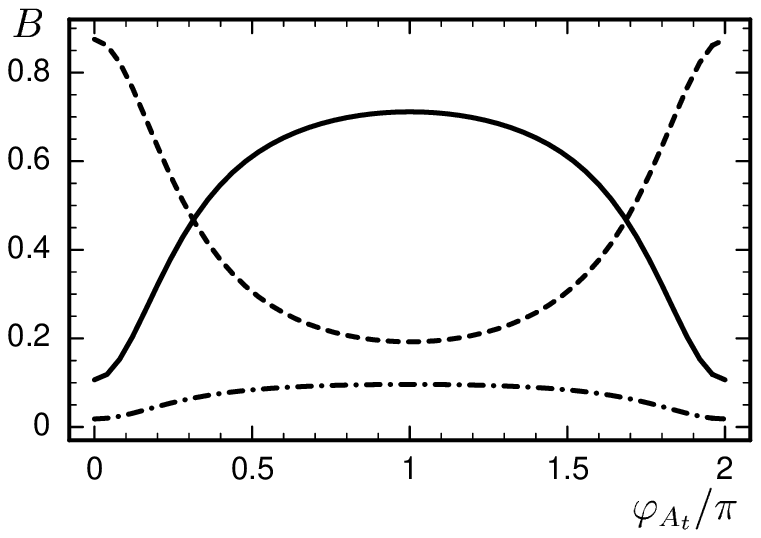}
\includegraphics*[width=40mm,height=35mm]{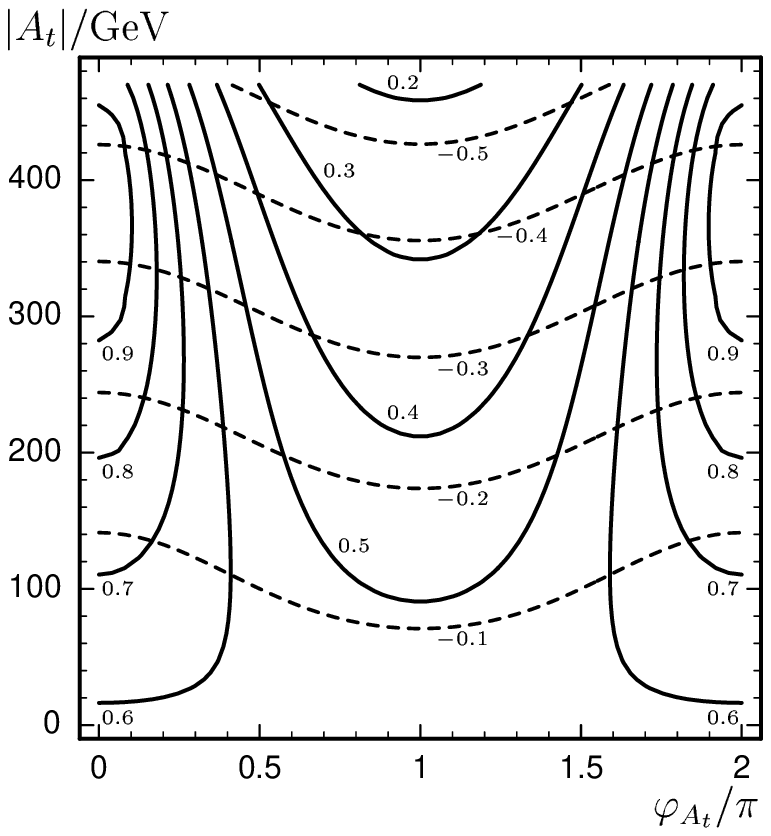}
\caption{
Left: Branching ratios of $\st_1\to \tilde{\chi}^+_1 b$ (solid),
$\tilde{t}_1 \to \tilde{\chi}^0_1 t$ (dashed),
$\tilde{t}_1 \to \tilde{\chi}^+_2 b$ (dashdotted). 
Right: Contours of 
$B(\tilde{t}_1 \to \tilde{\chi}^0_1 t)$ in the SPS\#1a inspired
scenario.
The dashed lines denote the contours of $\cos\theta_{\tilde{t}}$  
\cite{0306281}. }
\label{CPstop}
\end{figure}

In the pure sfermionic sector the phases of $A_{\ti f}$ and $\mu$, 
\eq{sfermix}, 
enter the masses  $m_{\sf_{1,2}}^2$ and mixing angle 
$\theta_{\ti f}$ only through a
term { $m_f^2|A_{\ti f} \mu|(\tan\beta)^{-2I^3_{f}}
\cos(\varphi_{A_{\ti f}} + \varphi_{\mu})$}. Therefore the ${\ti
  t}_{1,2}$ masses are more    sensitive to phases than  
masses of ${\stau_{1,2}}$  and
$\sb_{1,2} $ because of the mass hierarchy of the corresponding fermions.
The phase dependence of $\theta_{\ti f}$  is strongest 
if { $|A_{\ti f}|
\simeq |\mu| (\tan\beta)^{-2I^3_{f}}$} and $|m_{\sf_L}^2 - m_{\sf_R}^2|
\lsim |a_{\ti f}
m_f|$ \cite{0207186}.  Since the  $Z\sf_i\sf_i$ couplings are real, and
for$\sf_1\bar{\sf}_2$ production  only $Z$ exchange 
contributes, the $\sf_i\bar{\sf}_j$
production cross
sections do not explicitly depend on the phases -- dependence enters
only through the shift of sfermion masses and mixing angle. 
However, the various $\tilde f$ 
decay branching ratios depend in a
characteristic way on the complex
phases. This is illustrated in \fig{CPstau}, where branching ratios
for $\stau_1$ are shown for   
$m_{\stau_1}=240$ GeV, $\mu = 300$ GeV, $|A_{\ti \tau}| = 1000$ GeV, 
$\tan\beta = 3$, and
$M_2 = 200$ GeV \cite{0207186}. The branching ratios 
for the light $\st_1$ in the SPS\#1a inspired
scenario  are shown in \fig{CPstop}, including the contour
plot for the mixing angle $\cos\theta_{\ti t}$   
\cite{0306281}. 
A simultaneous measurement of 
$B(\tilde{t}_1 \to \tilde{\chi}^0_1 t)$ and $\cos\theta_{\tilde{t}}$ might be 
helpful to disentangle the phase of $A_{\ti t}$ from its absolute value.
As an example a measurement of 
$B(\tilde{t}_1 \to \tilde{\chi}^0_1 t)=0.6\pm 0.1$ and
$|\cos\theta_{\tilde{t}}|=0.3\pm 0.02$ would allow to determine
$|A_{\ti t}|\approx 320$~GeV with an error $\Delta(|A_{\ti t}|)\approx 
20$~GeV and
$\varphi_{A_{\ti t}}$ with a twofold ambiguity $\varphi_{A_{\ti t}} 
\approx 0.35\pi$
or $\varphi_{A_{\ti t}} \approx 1.65\pi$ with an error
$\Delta(\varphi_{A_{\ti t}}) \approx 0.1\pi$, see \fig{CPstop} (right).
\begin{figure}[htb]
\centering
\includegraphics*[width=42mm,height=35mm]{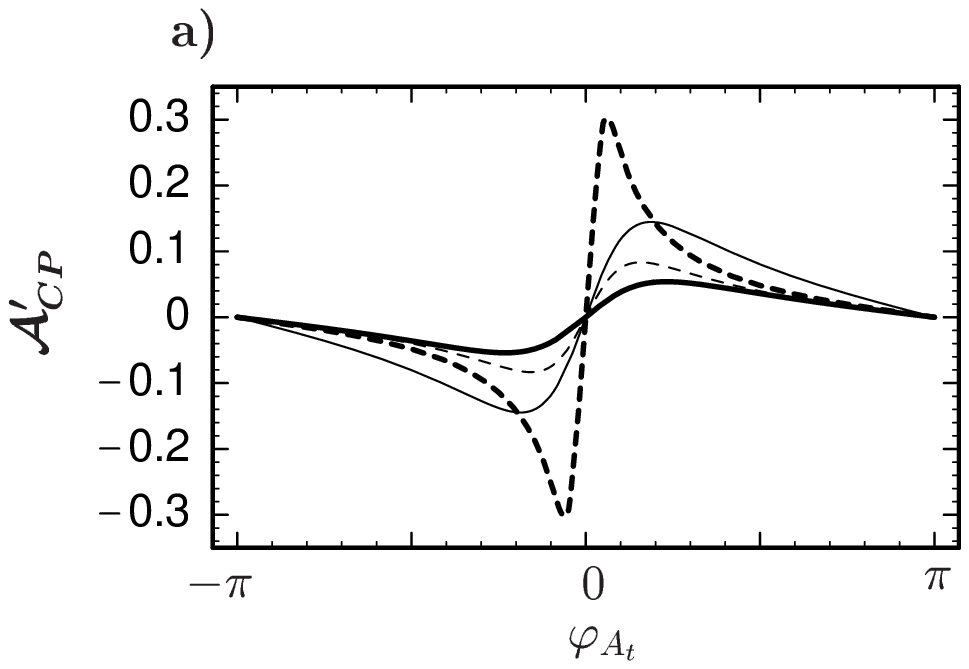}
\begin{minipage}[b]{38mm}
\caption{The $CP$ sensitive asymmetry
as a function of $\varphi_{A_{\ti t}}$;  
$\tan\beta$=3 (thick), $\tan\beta$=10 (thin), $\mu$=400 GeV (solid),
$\mu$=700 GeV (dashed) \cite{0202198}. \label{fig-CPtasym} }
\end{minipage}
\end{figure}

In principle, the imaginary parts of
the complex parameters involved could most directly and unambiguously
be determined by measuring suitable $CP$ violating observables.
For example, the polarization
of the $\tau^+$ normal to the $\ti t_1$ decay plane in the decay 
$\tilde t_1\to b \snu  \tau^+$ is sensitive to $CP$
violation.
The asymmetry of the $\tau$ polarization perpendicular to the decay
plane can go up to $30\%$ for some SUSY parameter points where  
the decay
$\tilde t_1\to b \snu  \tau^+$ has a sufficient branching ratio allowing for
the measurement of this asymmetry, see \fig{fig-CPtasym} 
where other parameters are 
taken as $m_{\st_1}= 240$ GeV, $m_{\st_2}=800$ GeV,
$m_{\snu}=200$ GeV, $M_2=350$ GeV, $|A_{\ti t}|=1000$ GeV \cite{0202198}.

CP violation in the stau sector can generate electric and weak dipole
moments of the taus. The CP-violating tau dipole form factors can be
detected up to the level of $(3 - 5) \cdot 10^{-19} e
{\rm cm}$ \cite{Anan:02} at a linear collider with high luminosity and
polarization of both $e^+$ and $e^-$ beams.  Although such a precision
would improve the current experimental bounds by three orders of
magnitude, it  still remains by an order of magnitude above the
expectations from supersymmetric models with CP-violation.

\subsection{Lepton flavour violation}
\def\mueg{\mu^- \to e^- \gamma}
\def\taueg{\tau^- \to e^- \gamma}
\def\taumug{\tau^- \to \mu^- \gamma}
There are stringent constraints on lepton flavour violation (LFV) 
in the charged
lepton sector, the strongest being 
$BR(\mueg) < 1.2 \times 10^{-11}$ \cite{Hagiwara}. 
However, neutrino oscillation experiments have established the existence
of LFV in the neutrino sector with $\tan^2\theta_{Atm}\simeq 1$, 
$\tan^2\theta_{\odot} = 0.24-0.89$ and $\sin^2(2\theta_{13}) \lsim 0.1$  
\cite{nuosc}.
\begin{figure}[htb]
{\small $\sigma(e^\pm \tau^\mp
\Eslash)$ [fb] ~~~~~~~~~~~~~~~~~~~~~ $\Delta m_{23}$ [GeV]}\\
\centering
\includegraphics*[height=3.5cm,width=4cm]{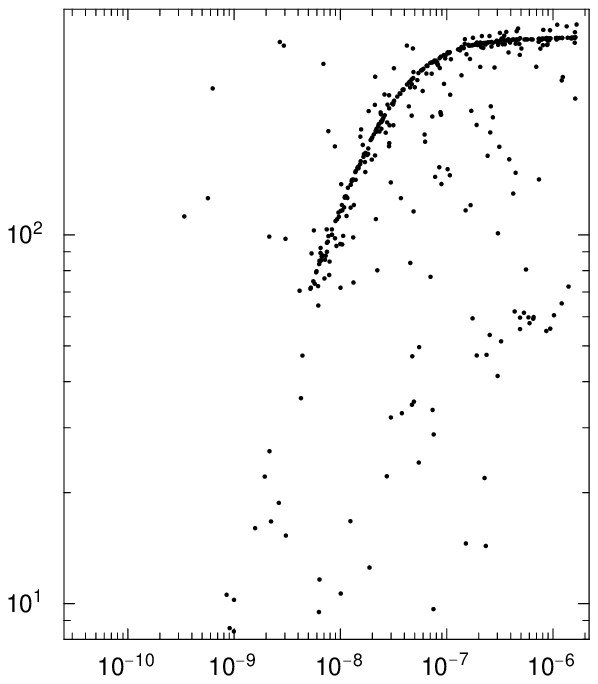}
\includegraphics*[height=3.5cm,width=4cm]{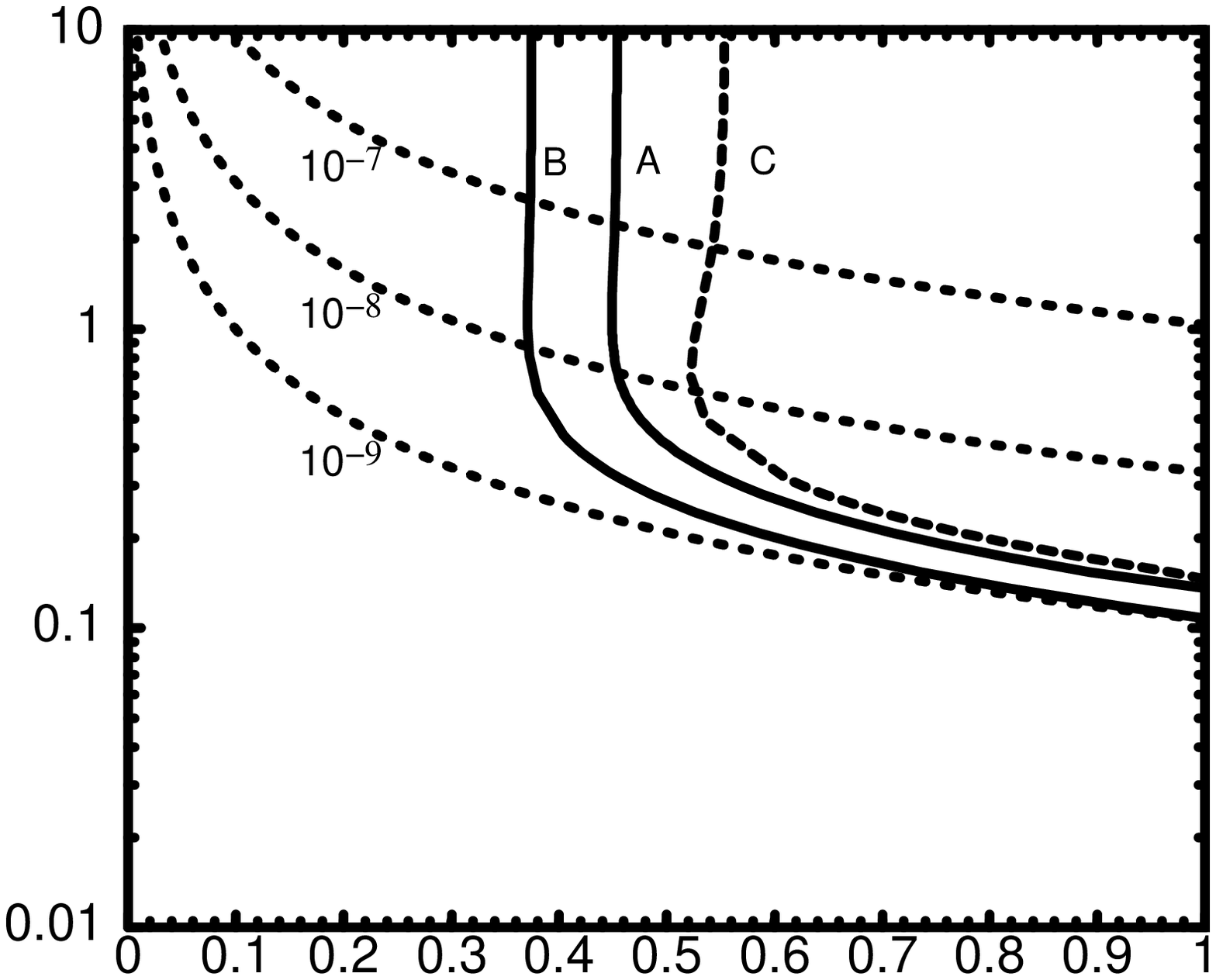}\\[-2mm]
{\small  BR$(\tau \to e \gamma)$ ~~~~~~ ~~~~~ ~~~~~~~~~ ~~~~~~
~~~ ~~~~~~ $\sin2\theta_{23}$} 
\caption{Left: Cross section for the signal $e^\pm \tau^\mp
\Eslash$ as a function of BR$(\tau \to e \gamma)$ for $\sqrt{s} =
500$~GeV \cite{0210326}.  
Right: 3$\sigma$ significance contours for $\sqrt{s}=$500
GeV and $\int{\cal L}$= 500 fb$^{-1}$ (A), =1000 fb$^{-1}$ (B).  Line C:
$\tilde\nu \tilde\nu^*$ contribution with luminosity 500 fb$^{-1}$.
Dotted lines: BR($\tau\to\mu\gamma$)=10$^{-7}$, 10$^{-8}$, 10$^{-9}$ 
\cite{jk}.}
\label{fig:LFV}
\end{figure} 

In the MSSM the R--parity symmetry 
forces total lepton number conservation but still allows the violation of
individual lepton number, e.g.~due to loop effects in $\mueg$ \cite{bm}.
Moreover, a large $\nu_\mu$-$\nu_\tau$ mixing can lead to a 
large  $\tilde \nu_\mu$-$\tilde \nu_\tau$ mixing
via renormalization group equations. Therefore one can
expect  clear LFV
signals in slepton and sneutrino production
and in the decays of neutralinos and charginos into sleptons and sneutrinos
at future  colliders \cite{lfv-lc}. 

For the reference point SPS\#1a a scan over the flavour non-diagonal
($i\neq j$) entries of slepton mass matrix \eq{sfermass} shows
\cite{0210326} that values for $|M^2_{R,ij}|$ up to $8 \cdot
10^3$~GeV$^2$, $|M^2_{L,ij}|$ up to $6 \cdot 10^3$~GeV$^2$ and
$|A_{ij} v_d|$ up to 650~GeV$^2$ are compatible with the current
experimental constraints. In most cases, one of the mass squared
parameters is at least an order of magnitude larger than the
others. However, there is a sizable part in parameter space where at
least two of the off-diagonal entries have the same order of
magnitude.

Possible LFV signals at an 
$e^+ e^-$ collider include  $e \mu \,  \Eslash$, 
$e \tau \, \Eslash$, $\mu \tau \,  \Eslash$ in the final state 
plus a  
possibility of additional jets.  
In \fig{fig:LFV} the cross section of $e^+ e^- \to
e^\pm \tau^\mp \Eslash$   
at $\sqrt{s}=500$ GeV versus BR($\tau\to e\gamma$) 
is shown for points consistent with the experimental LFV data    
which are  randomly generated 
in the range $ 10^{-8} \le |A_{ij}|
\le 50$~GeV, $ 10^{-8} \le M^2_{ij} \le 10^4$~GeV$^2$. 
The accumulation of points  along a band is due to
a large $\tilde e_R$-$\tilde \tau_R$
mixing which is less constrained by $\taueg$ than the corresponding
left-left or left-right mixing.

Note that the collider LFV signals can be very competitive to those
from rare charged lepton decay, like $\tau\to \mu \gamma$. This is
illustrated in \fig{fig:LFV}, where for simplicity the LFV has been
restricted to the 2-3 generation subspace of sneutrinos with the
mixing angle $\theta_{23}$ and $\Delta m_{23}=|m_{\snu_2}-m_{\snu_3}|$
as free parameters. 
\cite{jk}.

\subsection{Sgoldstinos}
In the GMSB SUSY, not only  the mass splittings $\Delta m^2$, but also the 
supersymmetry-breaking scale $\sqrt{F}$ is close to the weak scale: 
$G_F^{-1/2} \sim \Delta m^2 \lsim \sqrt{F}$. Then the gravitino $\sG$
becomes very light,
with $m_{\sG}={F}/{\sqrt{3}M'_P} ={F}/{(10~{\rm TeV})^2}\times 0.03$
eV.    The appropriate effective low-energy theory must then contain, 
besides the goldstino, also its supersymmetric partners, called 
sgoldstinos \cite{prz}. The spin-0 complex component of the chiral
goldstino superfield has two degrees of freedom, giving rise to two
sgoldstino states: a CP-even state $S$ and a CP-odd state $P$. 
In the simplest case it is assumed that there is no 
sgoldstino-Higgs mixing, and that squarks, sleptons, gluinos, 
charginos, neutralinos and Higgs bosons are sufficiently heavy 
not to play a r\^ole in sgoldstino production and decay. Thus  
the $S$ and $P$ are mass eigenstates and, being  R--even, 
they can be produced
singly together with the SM particles. 

During the Workshop new results on massive sgoldstino production at $\ee$
and $\pp$ colliders have been presented \cite{checchia}. 
The most interesting channels for the production of such
scalars ($\phi$ will be used to indicate a generic state) are
the process  $\ee\to \phi\gamma$, and the fusion $\pp\to\phi$,    
followed by the  $\phi$ decay to photons or gluons.

\begin{figure}[tb]
\centering
\includegraphics*[height=4cm,width=4cm]{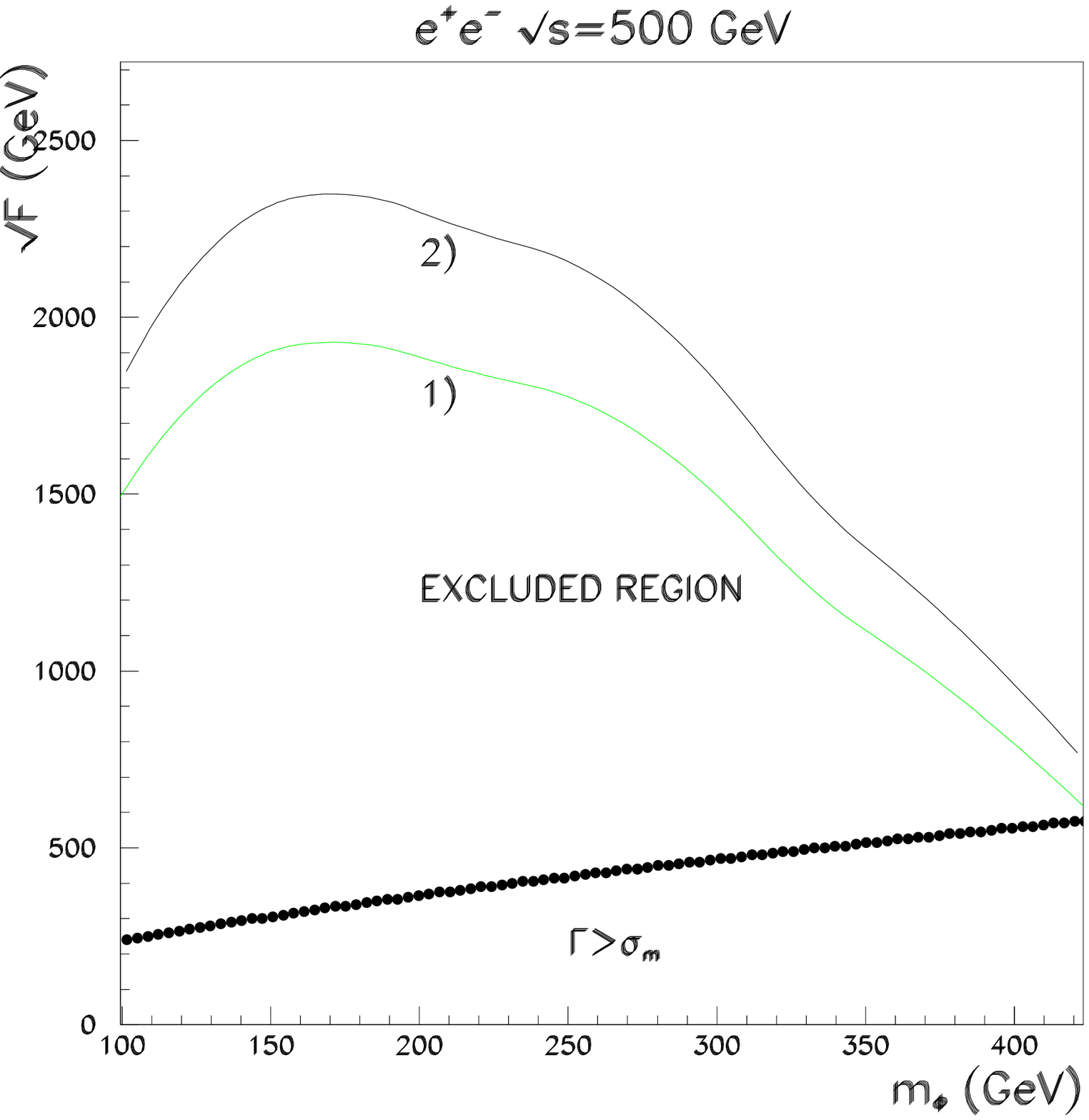} 
\includegraphics*[height=4cm,width=4cm]{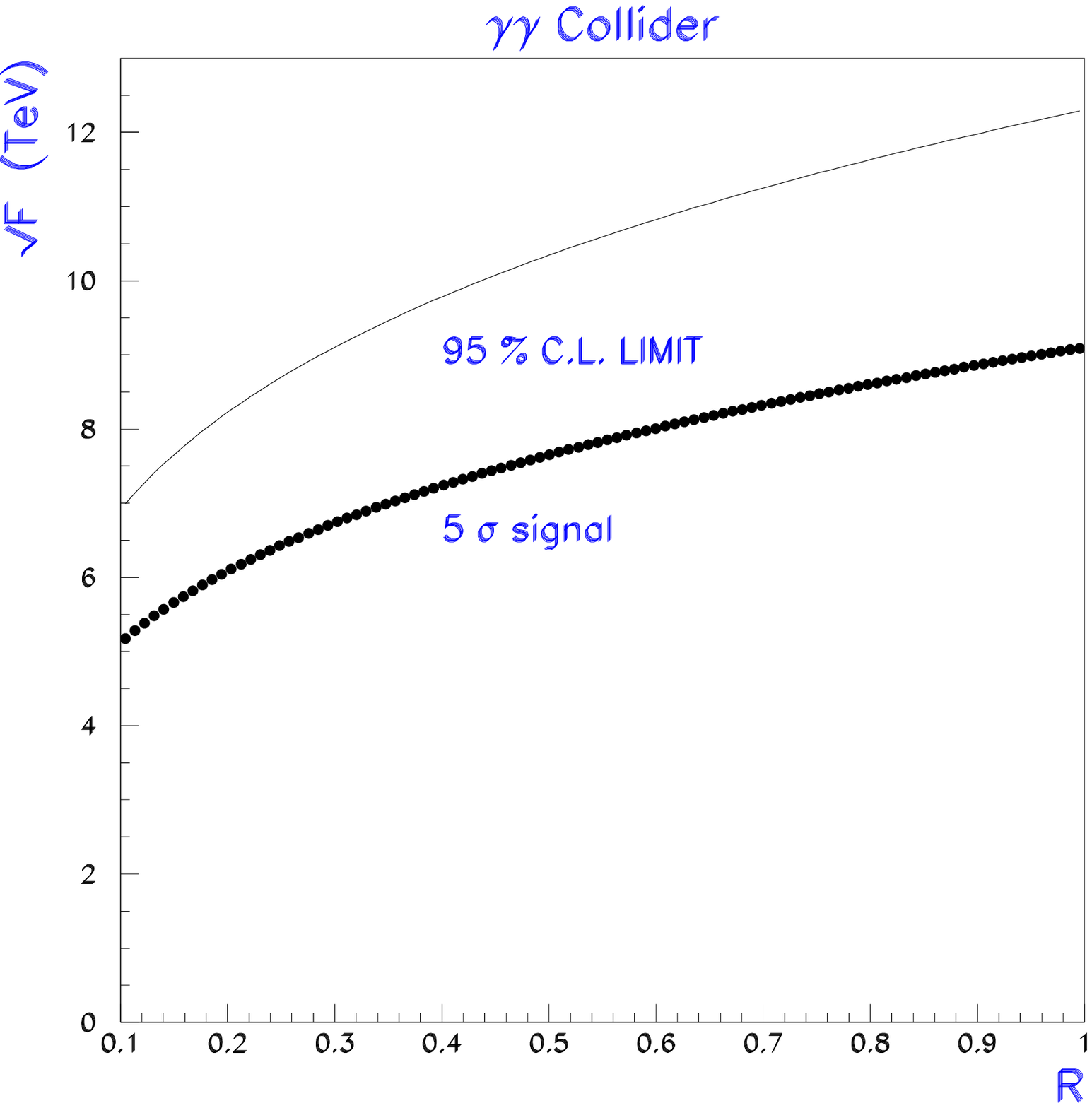}
\caption{Left: Exclusion region at 95\% CL at a 500 GeV $\ee$ collider.
Right: Exclusion region at 95\% CL and 5$\sigma$ discovery at a $\pp$
collider \cite{checchia}.}
\label{fig:sgold}
\end{figure}

The $\ee\to\phi\gamma \to  g g\gamma$ process gives rise to events with one
monochromatic 
photon and two jets. However, the brems- and beamstrahlung  
induces a photon energy smearing comparable to or larger than
the experimental resolution. 
On the other hand, the signal can be searched for directly in the jet-jet  
invariant mass distribution. Results of the simulation are presented  in
 \fig{fig:sgold} where the 
exclusion region at the 95\% CL is shown in the $m_\phi$--$\sqrt{F}$
plane 
for two parameter sets: 1) $M_1$ = 200 GeV, $M_2$ = 300 GeV, $M_3$ =
400 GeV, 2)
$M_1=M_2=M_3$ = 350 GeV. 

For the $\pp$ collider, despite the smaller decay branching ratio,      
only the two-photon final state has been considered since it 
has a very little SM background. Taking as a reference point 
the value $(\sigma B)_0$ obtained
for $M_{\gamma \gamma} =350$ GeV
and a 10$\%$ branching ratio to two photons, the 95 \% CL 
exclusion limit on
and the $5~\sigma$ discovery line 
for $\sqrt{F}$ is shown in \fig{fig:sgold} in terms of the ratio 
$R=\sigma \times BR(\phi \rightarrow \gamma \gamma)/(\sigma B)_0$.
Thus the sensitivity at a photon collider 
obtained from the same electron-positron beam energy
is expected to be much higher for 
$m_{\phi}\sim 300-400$ GeV.

\section{GAUGINOS AND HIGGSINOS}
Supersymmetric partners of electroweak gauge and Higgs bosons mix  
due to the gauge symmetry breaking. The mass-eigenstates (with
positive mass eigenvalues)  are 
charginos ($\tilde{\chi}^\pm_i$, $i$=1,2,  
mixtures of the wino and charged higgsino) and  neutralinos
($\tilde{\chi}^0_i$, $i$=1,2,3,4,  mixtures of  
$\tilde{B}$, $\tilde{W}^3$, $\tilde H_1^0$ and 
$\tilde H_2^0$). At tree level the chargino sector depends on $M_2$,
$\mu$ and $\tan\beta$; the neutralino sector depends in addition on
$M_1$. The gaugino and higgsino mass parameters can be complex;
without loss of generality $M_2$ can be assumed real and positive, and
the non-trivial CP-violating phases may be attributed to
$\mu=|\mu|e^{i\varphi_\mu}$ and $M_1=|M_1|e^{i\varphi_1}$. The
chargino mass matrix is diagonalized by  two 
unitary matrices acting on left- and
right-chiral weak eigenstates (parameterized by two  
mixing angles $\phi_{L,R}$
and three CP   phases $\beta_{L,R}$ and 
$\gamma$) \cite{cdsz,cdgksz}. 
The neutralino mass matrix is diagonalized by a
4$\times$4   unitary rotation $N$ parameterized in terms of  
6 angles and 9 phases (three Majorana $\alpha_i$ and six Dirac
$\beta_{ij}$ phases) \cite{six,ckmz}
\begin{equation}
N= {\rm diag}\{1,\, 
                   {\rm e}^{i\alpha_1},
                   {\rm e}^{i\alpha_2},
                   {\rm e}^{i\alpha_3}\} {\rm R}_{34}\, {\rm
                   R}_{24}\,{\rm R}_{14}\,{\rm R}_{23}\,{\rm
                   R}_{13}\, {\rm R}_{12}  
\label{eq:Mdef} 
\end{equation}
where ${\rm R}_{jk}$ are rotations in 
the [$jk$] plane characterized by a mixing angle $\theta_{jk}$ and a
(Dirac) phase $\beta_{jk}$. 

Charginos and neutralinos are produced in pairs
\begin{eqnarray}
    e^+e^- & \to & \cp_i \cm_j, \quad  \nt_{i} \nt_{j}
\end{eqnarray}
via $s$-channel $\gamma/Z$ and $t$-channel $\sne$ exchange 
for $\cpm$, and via $s$-channel $Z$ and $t$- and $u$-channel $\se$ 
exchange for $\nt$ production. 
Beam polarizations are very important to study the $\cx$ properties
and couplings.  The
polarized differential cross section for the $\cx_i
\cx_j$ production can be written as \cite{ckmz}
\begin{eqnarray}
&& \frac{{\rm d}\sigma^{\{ij\}}}{{\rm d}\cos\theta \,{\rm d}\phi}
  =\frac{\alpha^2 \lambda^{1/2}}{16\, s}  [
     (1-P_l\bar{P}_l)\Sigma_u+(P_l-\bar{P}_l)\Sigma_l
     \nonumber\\
&& 
  +P_t\bar{P}_t\cos(2\phi-\eta)\Sigma_t
  +P_t\bar{P}_t\sin(2\phi-\eta)\Sigma_n]\ \label{eq:diffx}
\end{eqnarray}
where $\lambda=[1-(\mu_i+\mu_j)^2][1-(\mu_i-\mu_j)^2]$ 
is the two--body phase space function with
$\mu_i=m_{\tilde\chi_i^0}/\sqrt{s}$,  
$P$=$(P_t,0,P_l)$ [$\bar{P}$=$(\bar{P}_t
\cos\eta,\bar{P}_t\sin\eta, -\bar{P}_l)$] is the electron [positron]
polarization vector; the electron--momentum direction defines the
$z$--axis and the electron transverse polarization--vector the
$x$--axis.  The coefficients $\Sigma_u$, $\Sigma_l$, $\Sigma_t$ and
$\Sigma_n$ depend only on the polar angle $\theta$ and their explicit
form is given in \cite{cdgksz} for charginos, and in \cite{ckmz} for 
neutralinos. The $\Sigma_n$, present only for
non-diagonal neutralino production, is particularly interesting
because it is non-vanishing only in the CP-violating case. 

Given the  high experimental precision in mass and cross section
measurements expected at the LC, the radiative corrections will have to be
applied to the above expressions. Recently full one-loop
corrections to  chargino and neutralino sector 
have been calculated \cite{0207364,hollikgaug,vienna,0205257}.
The numerical analysis based on a complete one loop calculation has
shown that the corrections to the chargino and neutralino masses can
go up to 10\% and the change in the gaugino and higgsino components
can be in the range of 30\%, and therefore will have to be taken into
account.   

\subsection{Charginos}

\begin{figure}[htb]
\centering
\includegraphics*[width=35mm,height=45mm,angle=90]{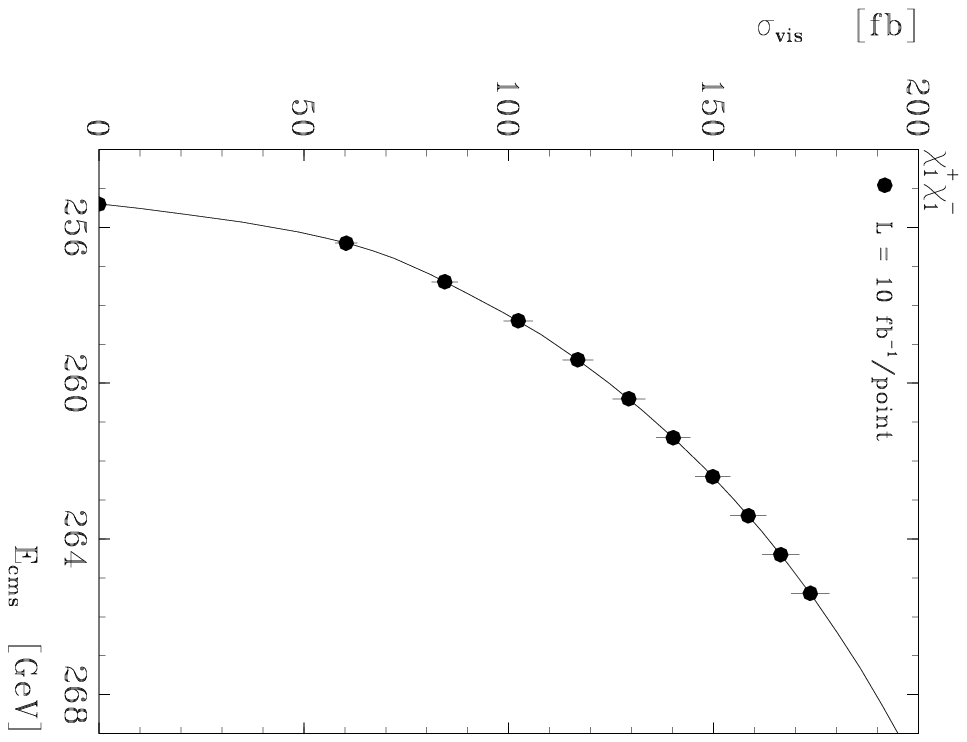}
\begin{minipage}[b]{35mm}
\caption{Cross section for 
    $e^+_R e^-_L \to \cx^+_1 \cx^-_1 \to \ell^\pm \nu_\ell\cx^0_1\, 
    q\bar q' \cx^0_1$ at threshold (in the RR~1
scenario~\cite{Aguilar-Saavedra:2001rg,martyn-susy02}, errors for 
    $10~\fbi$ per point).   \label{c11} }
\end{minipage} 
\end{figure}

\begin{figure}[thb]
\centering
\includegraphics*[width=45mm,height=35mm]{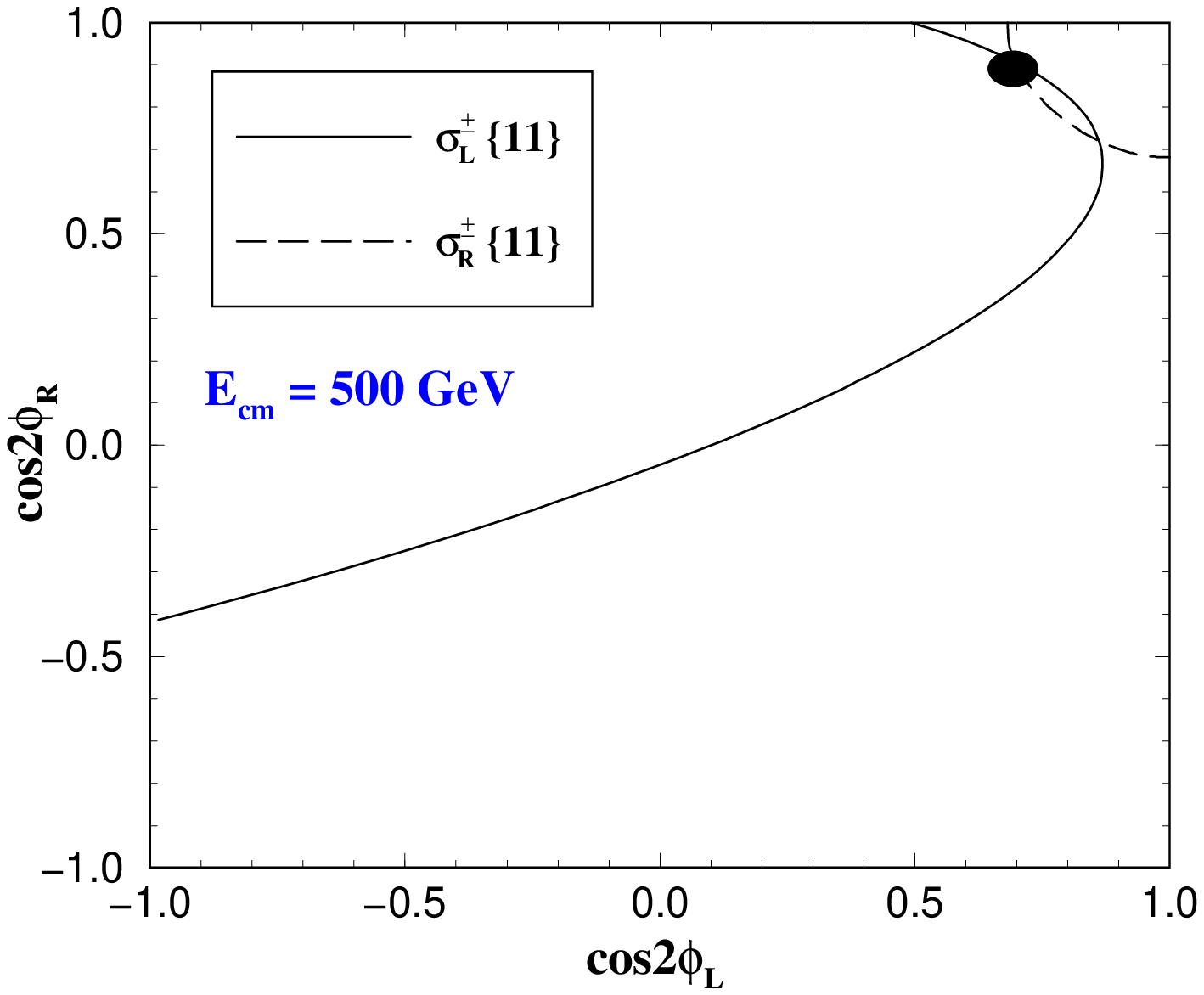}
\begin{minipage}[b]{35mm}
\caption{Contours of $\sigma(\cx^+_1\cx^-_1)$ with
polarized beams in the plane 
$[\cos2\phi_L,\cos2\phi_R]$  \cite{add}. ~~~~~~~ ~~~~~~~~~ ~~~~~~~~~
~~~~~~~~~ ~~~~~~~ ~~~~~~~~ \label{fig:char} }
\end{minipage}
\end{figure}

Experimentally the chargino 
masses can be measured very precisely at  threshold since 
the production cross
section for spin 1/2 Dirac fermions rises as $\beta$ leading to steep
excitation curves. Results of a simulation for the reaction
$e^+_R e^-_L \to \cx^+_1 \cx^-_1 \to \ell^\pm \nu_\ell\cx^0_1\, 
q\bar q' \cx^0_1$, \fig{c11}, show that the mass resolution is
excellent of ${\cal O}(50~\MeV)$, degrading to the per~mil level for the
higher $\cx^\pm_2$ state. Above threshold, from 
the di-jet energy distribution one
expects a mass resolution of $\delta m_{\cx_1^\pm}=0.2~\GeV$,
while the di-jet mass distributions constrains the 
$\cx^\pm_1 - \nt_1$ mass splitting within about $100~\MeV$. 
Since the chargino
production cross sections are simple binomials of $\cos2\phi_{L,R}$,
              see \fig{fig:char}, the 
mixing angles can be determined in a model independent way 
using  polarized electron beams   \cite{add}. 

Once masses and mixing angles are measured, the fundamental SUSY 
parameters of the chargino sector  
can be extracted to lowest order in analytic form \cite{add,KM}
\begin{eqnarray}
&&M_2=M_W[\Sigma - \Delta[\cos2\phi_R+\cos2\phi_L]]^{1/2}\label{eq:m2}\\
&&\left|\mu\right|=M_W[\Sigma + \Delta[\cos2\phi_R+\cos2\phi_L]]^{1/2}
\label{eq:mu}\\
&&\cos\Phi_\mu= [ \Delta^2
                   -(M^2_2-\mu^2)^2-4m^2_W(M^2_2+\mu^2)\qquad {}\nonumber \\
 & & \qquad  \qquad     -4m^4_W\cos^2 2\beta]/8 m_W^2M_2|\mu|\sin2\beta 
\label{eq:cos}\\
&&\tan\beta=\left[\frac{1+\Delta (\cos 2\phi_R-\cos 2\phi_L)}
           {1-\Delta (\cos 2\phi_R-\cos 2\phi_L)}\right]^{1/2} 
\label{eq:tan}
\end{eqnarray}
where $\Delta =
 (m^2_{\tilde{\chi}^\pm_2}-m^2_{\tilde{\chi}^\pm_1})/4M^2_W$
and 
$\Sigma =  (m^2_{\tilde{\chi}^\pm_2}+m^2_{\tilde{\chi}^\pm_1})/2M^2_W -1$. 
However, if $\tilde{\chi}^\pm_2$ happens to be beyond the kinematic
reach at an early stage of the LC, 
it depends on the 
CP properties of the higgsino sector whether they can uniquely 
be determined in the light chargino system alone: 

(i) If $\mu$ is real, $\cos\Phi_\mu =\pm 1$ determines 
$m_{\tilde{\chi}^\pm_2}$ up to at most 
a two--fold ambiguity \cite{cdgksz}; 
this ambiguity can be resolved if other observables
can be measured, e.g.  the mixed--pair $\tilde{\chi}^0_1
\tilde{\chi}^0_2$ production cross sections. 

(ii) In a CP non--invariant theory with complex $\mu$, the parameters
in eqs.(\ref{eq:m2}--\ref{eq:tan}) depend on the unknown heavy
chargino mass $m_{\tilde{\chi}^\pm_2}$.  Two solutions in the 
$\{M_2, \mu, \tan\beta\}$ space are parameterized
by $m_{\tilde{\chi}^\pm_2}$ and classified by the two possible signs
of $\sin\Phi_\mu$.  The unique solution can be found with additional  
information from the two light neutralino states $\tilde{\chi}^0_1$ and
$\tilde{\chi}^0_2$, as we will see in the next section.

\begin{figure}[htb]
\centering
\includegraphics*[width=80mm,height=35mm]{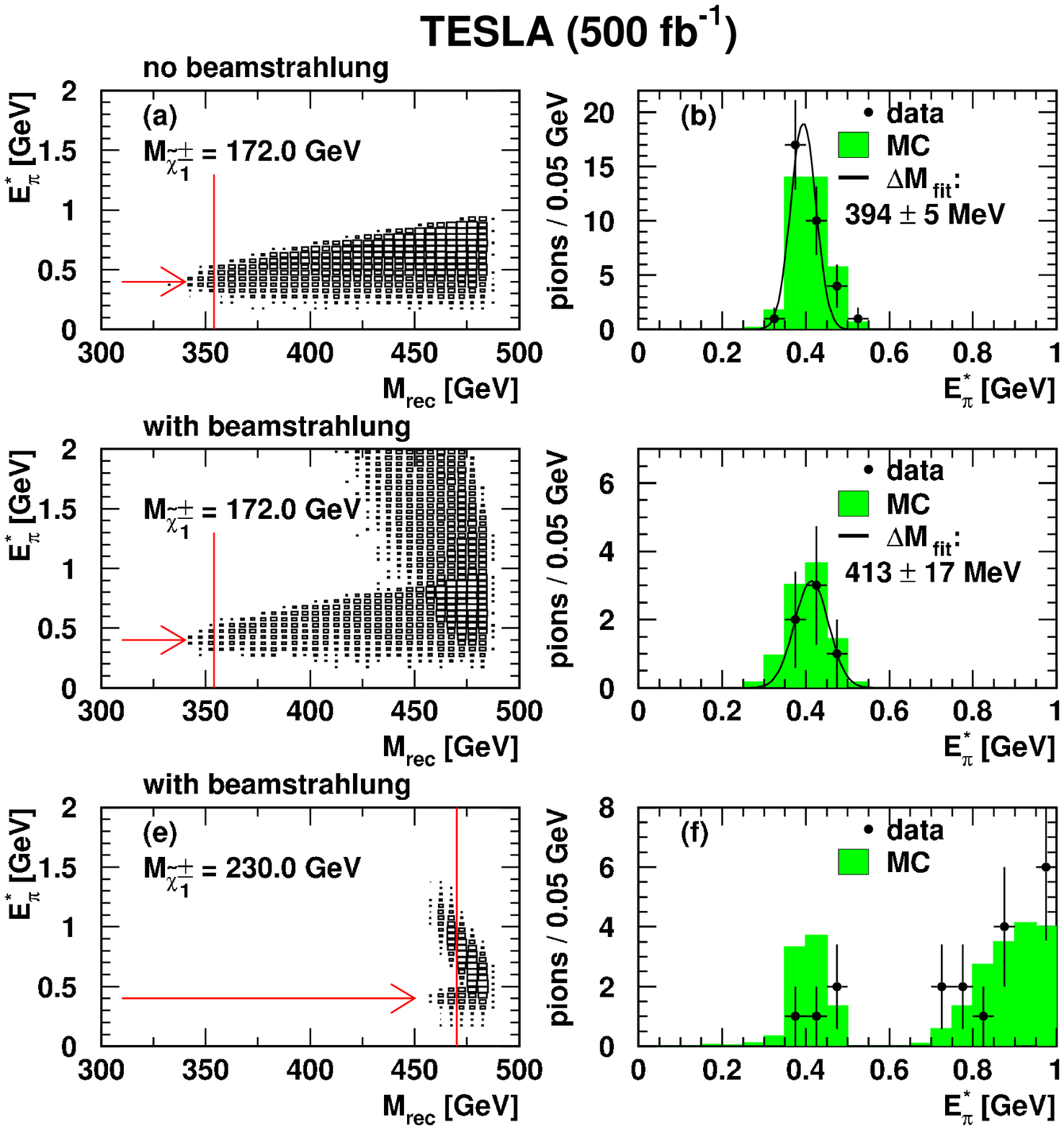} 
\caption{The ISR photon  recoil mass and the pion energy scatter plot
  (left), and the pion energy spectrum across the red line
(right) for  $e^+_R e^-_L \to \cx^+_1 \cx^-_1\gamma \to
\pi^+\pi^-\gamma \Eslash$ \cite{hensel}. }\label{isrcharg}
\end{figure}

The above methods fail for the light chargino 
if it happens to be nearly mass-degenerate with the lightest
neutralino, as predicted in a typical AMSB scenario. 
In this case $\cpm_1\to\nt_1+$ {\it soft pion}, and very
little activity is seen in the final state. However, one can exploit
the ISR photons in $e^+e^-\to \cp_1\cm_1\gamma$ to measure both 
$m_{\cpm_1}$ and the mass splitting $\cpm_1 - \nt_1$ \cite{hensel}. 
The ISR photon recoil
mass spectrum  starts to rise at $2m_{\cpm_1}$ allowing to determine
the chargino mass at a percent level, \fig{isrcharg}.  
Moreover,  
the pion energy spectrum for
events with charginos produced nearly at rest peaks around $\cpm_1 -
\nt_1$ and again precision of order 2 percent is expected.  

Besides the $e^+e^-$ option,  chargino pair 
production
\begin{eqnarray}
\gamma\gamma\rightarrow\tilde{\chi}^+_i\tilde{\chi}^-_i (i=1,2)
\label{ggtochar}
\end{eqnarray}
 in 
the $\gamma\gamma$ mode of a Linear Collider  
has been studied \cite{0209108}.
In this case the production is a pure QED process (at tree level) and
therefore it allows  
the chargino decay mechanism to be studied separately in contrast to 
the $e^+e^-$ mode where both
production and decay are sensitive to the SUSY parameters.  

\begin{figure}[htb]
\centering
\includegraphics*[width=40mm,height=30mm]{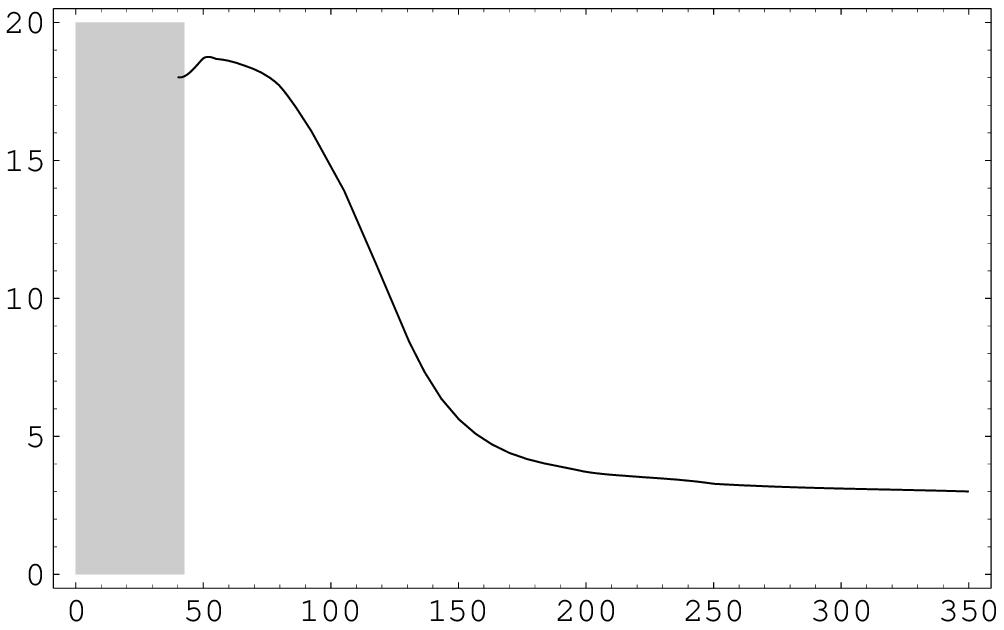} 
\includegraphics*[width=40mm,height=30mm]{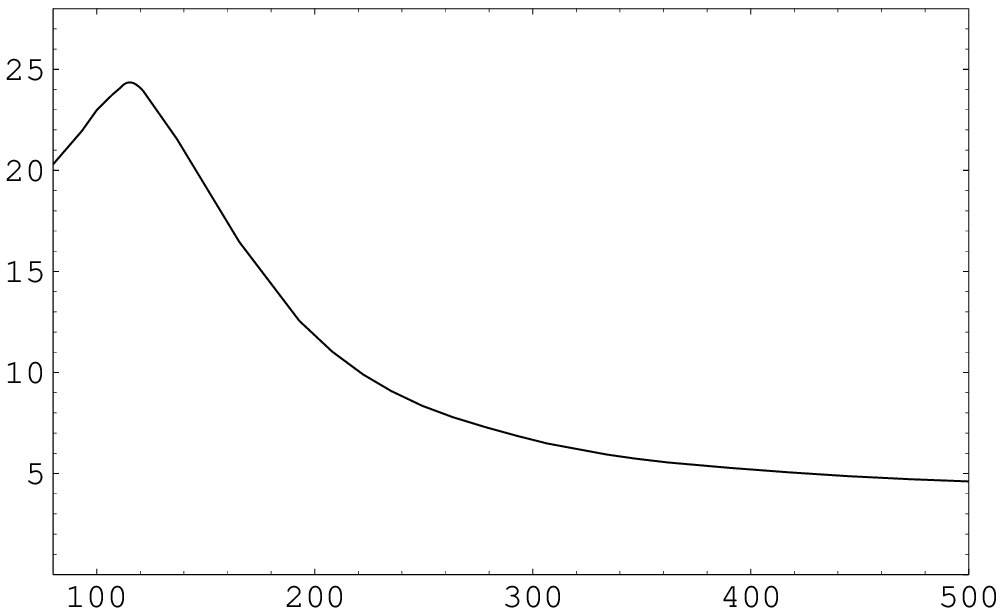}\\[-4mm]
{\small \phantom{} ~~~~~~~~ ~~~ ~~~~~~~~~~ $M_1$ [GeV] ~~~~~~~~~~~~~~~~~~~~ 
~~~~~~ $m_{\snu_e}$ [GeV]}
\caption{The $e^+$ forward-backward asymmetry (in \%) 
in the ee-CMS of the
decay positron from 
  $\gamma\gamma\to\tilde{\chi}^+_1\tilde{\chi}^-_1$,
  $\tilde{\chi}^+_1\to\tilde{\chi}^0_1e^+\nu_e$ as a function of the
  parameter $M_1$ (left) and the
  sneutrino mass $m_{\snu_e}$ (right) at $\sqrt{s_{ee}}=500$ GeV for
  $M_2=152$
  GeV, $\mu=316$ GeV, $\tan\beta=3$. The shadowed region 
  corresponds to the bound $m_{\tilde{\chi}_1^0}>38$ GeV
\cite{0209108}.   } 
\label{m1andsnu}
\end{figure}

Provided the chargino mass has been measured and the energy spectrum
and polarization of the high energy photons are well under control, the
production cross section and the polarization of the charginos in
reaction \eq{ggtochar} are
uniquely predicted. By manipulating the
polarization of the laser photons and the converted electron beam 
various characteristics  of the chargino decay can be 
measured and exploited to study the gaugino 
system.  
As an example, in
\cite{0209108} 
the forward-backward asymmetry (measured with respect to the $e^+e^-$ beam
direction) 
\begin{equation}\label{asym}
A_{\mathrm{FB}}=\frac{\sigma_e(\cos\theta_{e^+} >
  0)-\sigma_e(\cos\theta_{e^+} < 
  0)}{\sigma_e(\cos\theta_{e^+} > 0)+\sigma_e(\cos\theta_{e^+} < 0)}
\end{equation}
of the positron from the decay
$\tilde{\chi}_1^+\to\tilde{\chi}_1^0e^+\nu_e$, shown in
\fig{m1andsnu},   
has been studied to determine $M_1$ and $m_{\snu_e}$.

\subsection{Neutralinos}
Similarly to the chargino case,  
the di-lepton energy and mass distributions in the reaction
$\ee\to\nt_2 \nt_2\to  4\ell^\pm \Eslash$ can be used to
determine $\nt_1$ and $\nt_2$ masses.
Previous analyses of the di-lepton mass and di-lepton energy  
spectra performed in the $\tan\beta=3$ case showed that uncertainties in
the primary and secondary $\nt_2$ and $\nt_1$ masses of about
2~per~mil can be expected \cite{Aguilar-Saavedra:2001rg,martyn-susy02}. 
Higher resolution of order 100~MeV for $m_{\nt_2} $ can be obtained 
from a threshold scan of $\ee\to\nt_2\nt_2$; heavier states  $\nt_3$ and
$\nt_4$, 
if accessible, can still be resolved with a resolution of a
few hundred MeV.
For the higher values of $\tan\beta\gsim 10$ the dominant decay mode
of $\nt_2$ is to $\tau^+\tau^-\nt_1$. With $\tau$'s decaying in the
final state  the 
experimental  selection of the signal from the SM and SUSY background  
becomes more
difficult. Preliminary analyses nevertheless   show \cite{tevlinball}
that an accuracy
of 1-2 GeV  for the mass determination seems possible from the process
$e^+e^-\to \nt_1\nt_2$.

\begin{figure}[htb]
\centering
\includegraphics*[height=39mm,width=40mm]{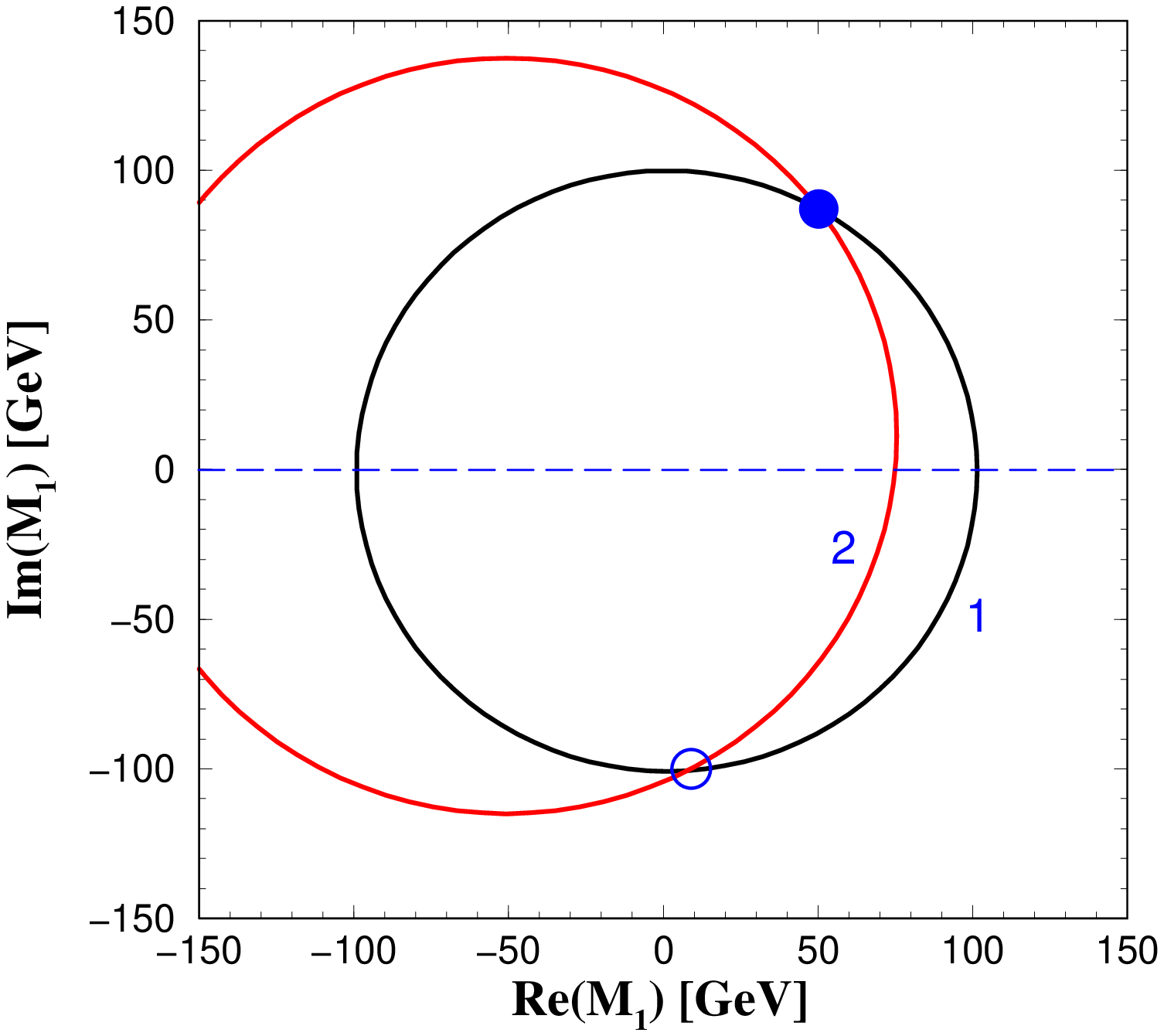}
\includegraphics*[height=40mm,width=40mm]{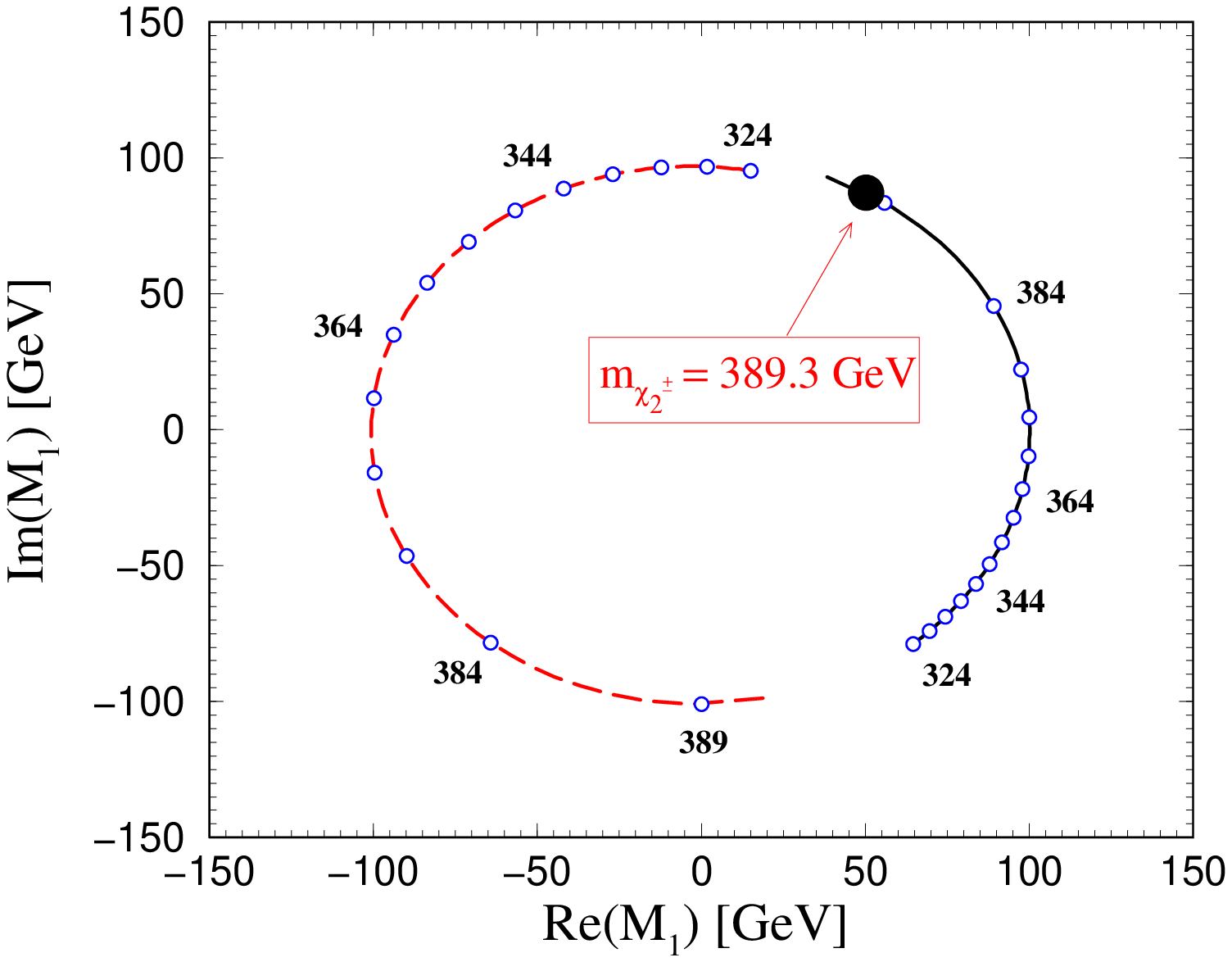} 
\caption{Two crossing points determined by two light neutralinos
(left) and their migration with $m_{\tilde{\chi}^\pm_2}$ (right) 
\cite{add}.}
\label{fig:circle}
\end{figure}

To resolve the light chargino case
in the CP-violating scenario (ii) discussed in the previous section, 
we note that each 
neutralino mass $m_{\nt_i}$ satisfies the characteristic equation
\begin{eqnarray}
(\real{M_1})^2+(\imag{M_1})^2+ u_i\, \real{M_1}+ v_i\, \imag{M_1}
 = w_i 
\label{eq:Mphase}
\end{eqnarray}
where $u_i,\,v_i,\,w_i$ are functions of $m_{\nt_i},\, M_2,\,\mu,\,
\tan\beta$; since physical masses are CP-even, $v_i$ is necessarily
proportional to $\sin\varphi_\mu$. 
Therefore  each neutralino mass  defines a circle in the $\{\real M_1,
\imag M_1\}$ plane, assuming other parameters fixed. 
With  two light neutralino masses two crossing points in the ($\real{M_1}$,  
 $\imag{M_1}$) plane are found, \fig{fig:circle} (left). 
Since from the chargino sector  
$\{M_2,\mu\; \tan\beta\}$ are 
parameterized by the unknown $m_{\tilde{\chi}^\pm_2}$, 
the crossing points will migrate with $m_{\tilde{\chi}^\pm_2}$, 
\fig{fig:circle} (right). 
Using the measured cross section  for
$\tilde{\chi}^0_1\tilde{\chi}^0_2$, a { unique}
solution for { $M_1$} is obtained and the 
heavy chargino mass predicted. If the LC would run concurrently with the LHC, 
the LHC experiments might be able to verify the predicted value of   
$m_{\tilde{\chi}^\pm_2}$.

\subsection{Neutralinos with CP-violating phases}
Particularly interesting is the threshold behavior since due to the Majorana
nature of neutralinos \cite{ckmz},   
a clear indication of non--zero CP violating phases can be provided by
studying the excitation curve for  non--diagonal
neutralino pair production near thresholds. 

Like in the quark sector, it is useful \cite{ckmz,AASB} to represent the 
unitarity constraints 
\begin{eqnarray}
 \label{eq:M}
&&M_{ij}= N_{i1}N^*_{j1}+N_{i2} N^*_{j2} + N_{i3} N^*_{j3}
               +N_{i4}N^*_{j4}\qquad {} \\
&&D_{ij}= N_{1i}N^*_{1j}+N_{2i} N^*_{2j} + N_{3i} N^*_{3j}
               +N_{4i}N^*_{4j} 
 \label{eq:D}
\end{eqnarray}
on the neutralino mixing matrix $N$, \eq{eq:Mdef},  
in terms of unitarity quadrangles. For $i$$\neq$$ j$ we get
$M_{ij}$=$D_{ij}$=0 and  the above equations
define two types of quadrangles in the complex plane. The
$M$-type quadrangles are formed by the sides $N_{ik}N^*_{jk}$
connecting two rows $i$ and $j$, \eq{eq:M}, and the $D$-type by
$N_{ki}N^*_{kj}$
connecting two columns $i$ and $j$, \eq{eq:D}, of the mixing matrix.
By a proper ordering of sides the quadrangles are assumed to be convex
with areas 
\begin{eqnarray}
&{\rm area}[M_{ij}]={\textstyle \frac{1}{4}}
         (|J_{ij}^{12}|+|J_{ij}^{23}|+|J_{ij}^{34}|+|J_{ij}^{41}|)&
\label{eq:aM}\\
&{\rm area}[D_{ij}]={\textstyle \frac{1}{4}}
         (|J_{12}^{ij}|+|J_{23}^{ij}|+|J_{34}^{ij}|+|J_{41}^{ij}|)&
\label{eq:aD}
\end{eqnarray}
where $J_{ij}^{kl}$ are the Jarlskog--type CP--odd ``plaquettes''
\cite{CJ} 
\begin{equation}
J_{ij}^{kl}=\imag N_{ik}N_{jl}N_{jk}^*N_{il}^*
\label{eq:plaq}
\end{equation}
Note that plaquettes, and therefore the areas of unitarity
quadrangles, are not sensitive to the Majorana phases $\alpha_i$.
Unlike in the quark or lepton sector, the orientation of all
quadrangles is physically meaningful, and determined by the CP-phases
of the neutralino mass matrix.   

For a CP-conserving case with real $M_1,M_2$ and $\mu$, the neutralino
mixing matrix $N$ has all Dirac phases $\beta_{ij}= 0$
mod $\pi$ and Majorana phases $\alpha_i=0$ mod $\pi/2$.  Majorana
phases $\alpha_i=\pm\pi/2$ describe only different CP parities of the
neutralino states.  
In terms of quadrangles, CP is conserved if and only if all 
quadrangles have null area (collapse to lines or points) {\it and} are
oriented along either the real or the imaginary axis.

\begin{figure}[htb]
\centering
\includegraphics*[height=40mm,width=40mm]{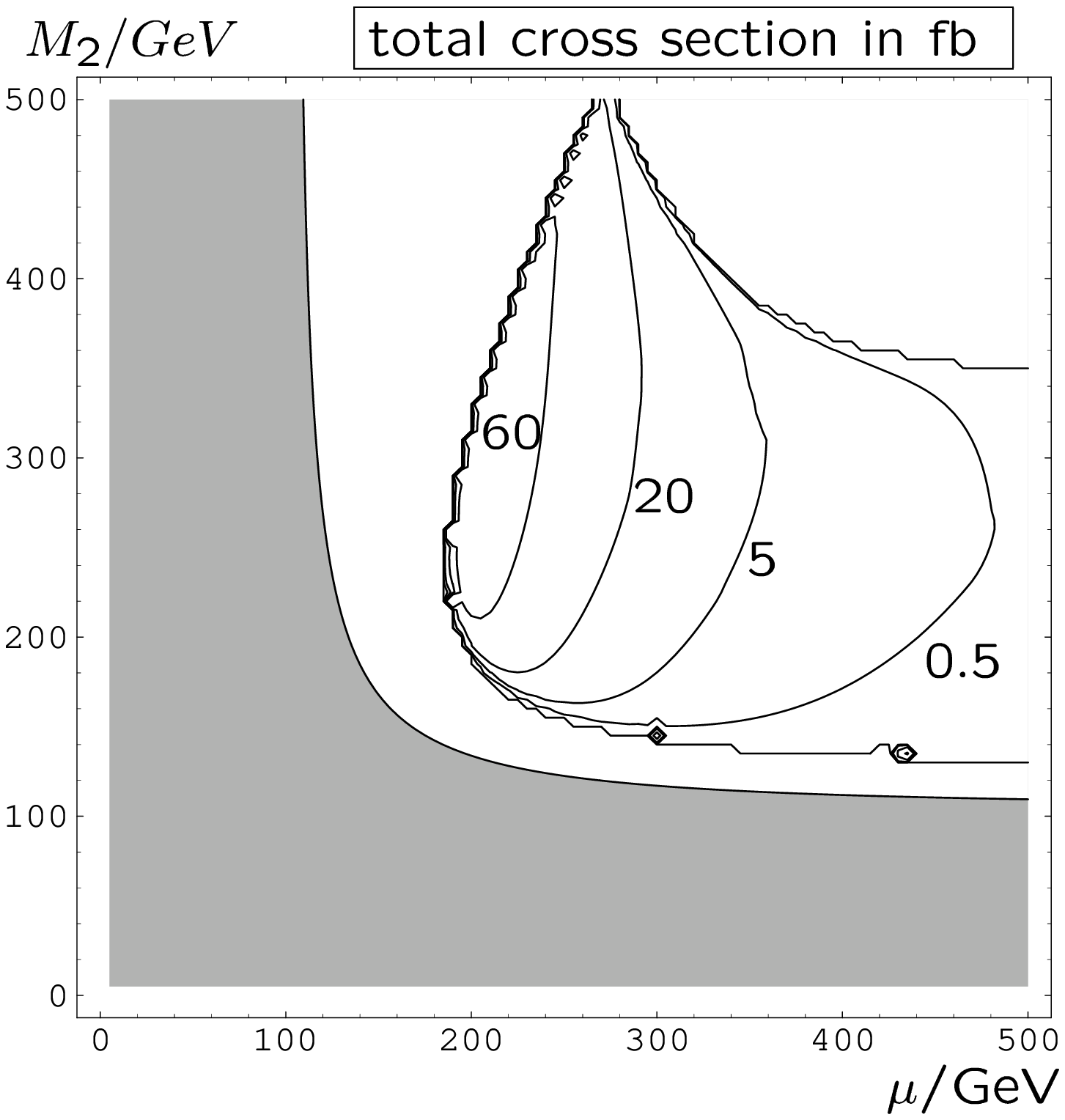}
\includegraphics*[height=40mm,width=40mm]{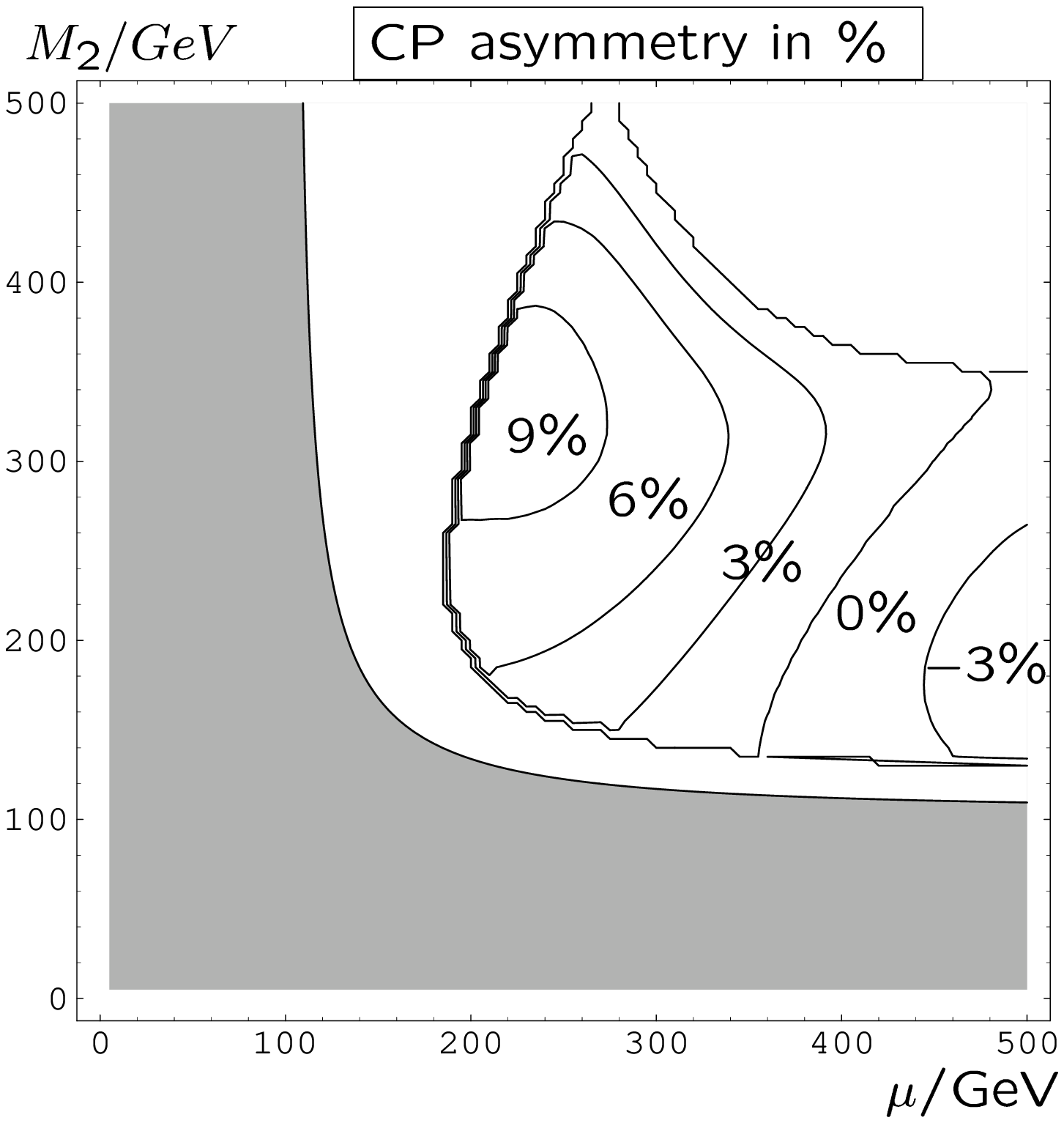} 
\caption{Cross section (left) and CP-odd asymmetry for 
$e^+e^-\to \none \ntwo \to \none \none l_1 l_2$ at $\sqrt{s}$=500 GeV
with $m_0$=100 GeV, $\tan\beta$=10 and gaugino mass universality. In
shaded area $m_{\cpm_1}<$104 GeV \cite{kittel}.}
\label{fig:kittel}
\end{figure}

The non--zero values of CP-odd quantities, like 
$\Sigma_n$ or the polarization of the produced neutralino normal to the
production plane,   would unambiguously indicate CP-violation
in the neutralino sector. 
In \cite{kittel} the CP-odd asymmetry defined as 
\begin{eqnarray}
{\cal A}=\frac{\sigma(T>0)-\sigma(T<0)}{\sigma(T>0)+\sigma(T<0)}
\label{asymkittel}
\end{eqnarray}
where $T=\vec{p}(e^-)\times\vec{p}(l_1)\cdot\vec{p}(l_2)$ for the
process
$e^+e^-\to \none \ntwo \to \none \none l_1 l_2$ with two visible
leptons in the final state has been considered. In \fig{fig:kittel}
the expected cross section (left) and the asymmetry (right) are shown
as functions of $M_2$ and $\mu$ assuming $\varphi_1=\pi/2$.

\begin{figure}[tb]
\centering
\includegraphics*[height=8cm,width=3.5cm,angle=-90]{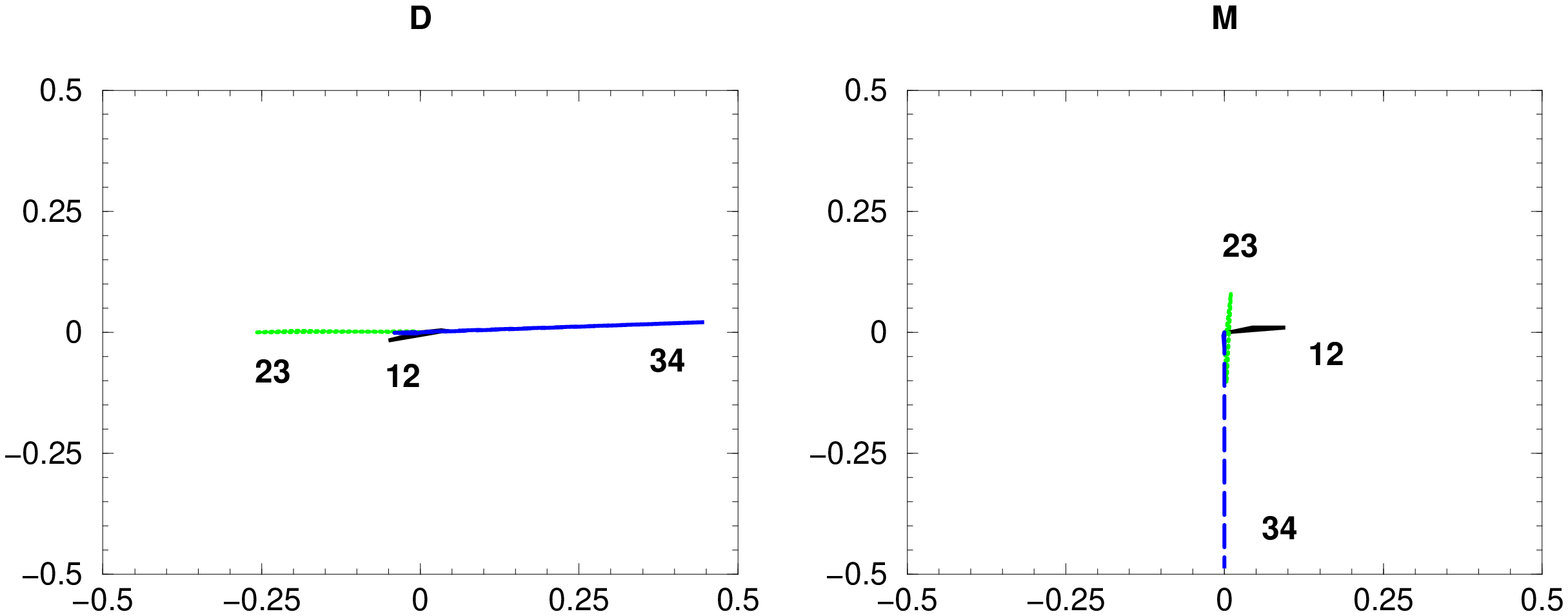} 
\caption{ The $D$--type (left panel) and $M$--type (right panel) 
  quadrangles in the complex plane,
  illustrated for $\tan\beta=10$, $|M_1|=100.5$ GeV, $\varphi_1=\pi/5$,
  $M_2=190.8$ GeV, $|\mu|=365.1$ GeV and $\varphi_{\mu}=0$; 
  $ij$ as indicated in the figure \cite{0306272}.}
\label{fig:quad}
\end{figure}

One can also try to
identify the presence of 
CP-phases by studying their impact on CP-even quantities, like
neutralino masses, branching ratios etc. Since these quantities are
non--zero in the CP-conserving case, the detection of CP-odd phases will
require a careful 
quantitative analysis of a number of physical observables
\cite{gaissmaier}, in particular for 
numerically small CP-odd phases. 
For example, \fig{fig:quad} displays the unitarity
quadrangles for the SPS\#1a point assuming a small non-vanishing phase  
$\varphi_1=\pi/5$ (consistent with
all experimental constraints)   \cite{0306272}. 
The quadrangles are almost degenerate to lines parallel
to either the real or the 
imaginary axis, and revealing a small  phase of $M_1$
will be quite difficult. 
However, studying the threshold behavior of the production cross
sections can be of great help \cite{ckmz,0306272}.

\begin{figure}[tb]
\centering
\includegraphics*[height=3.4cm,width=4cm]{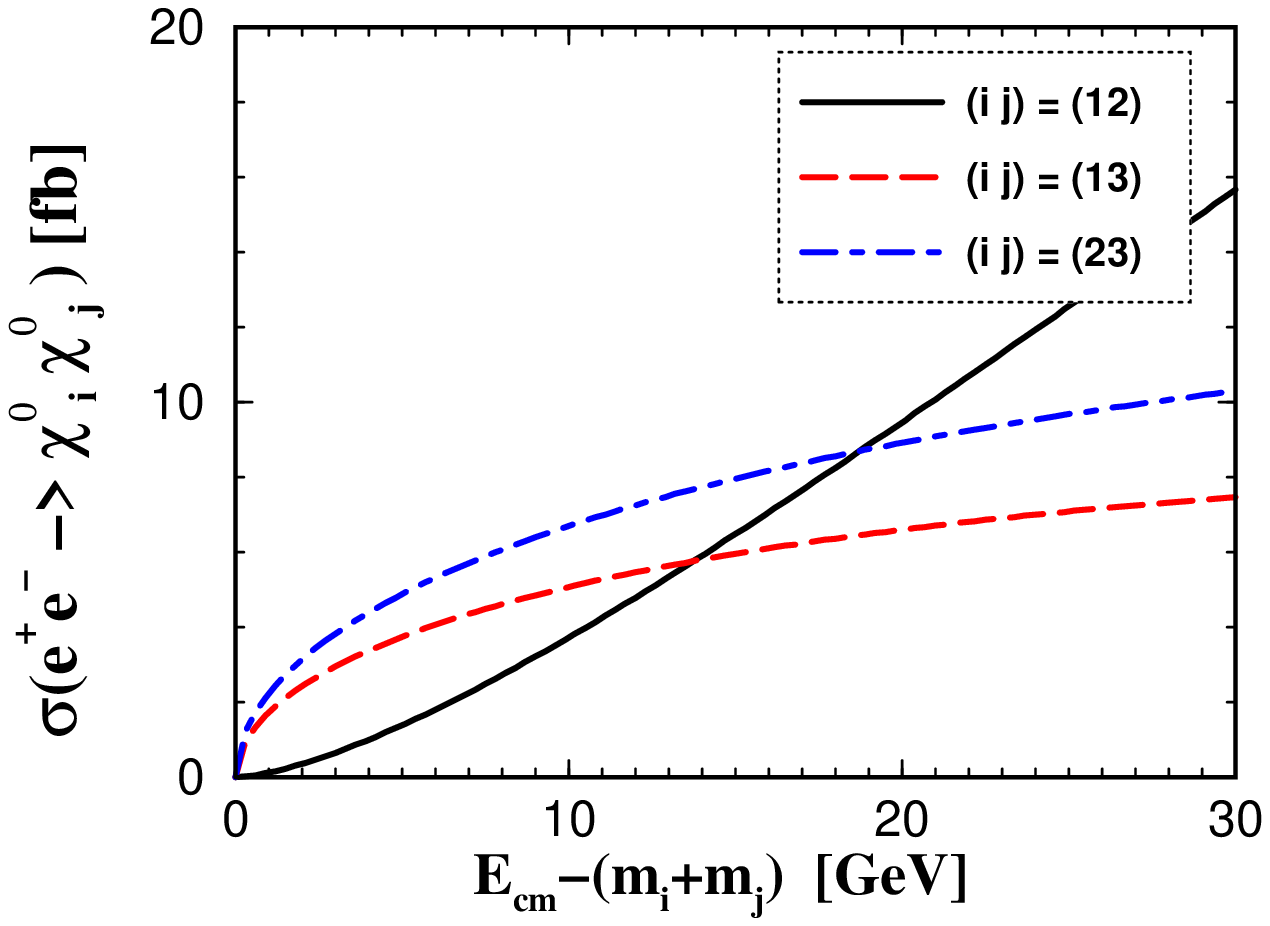} 
\includegraphics*[height=3.4cm,width=4cm]{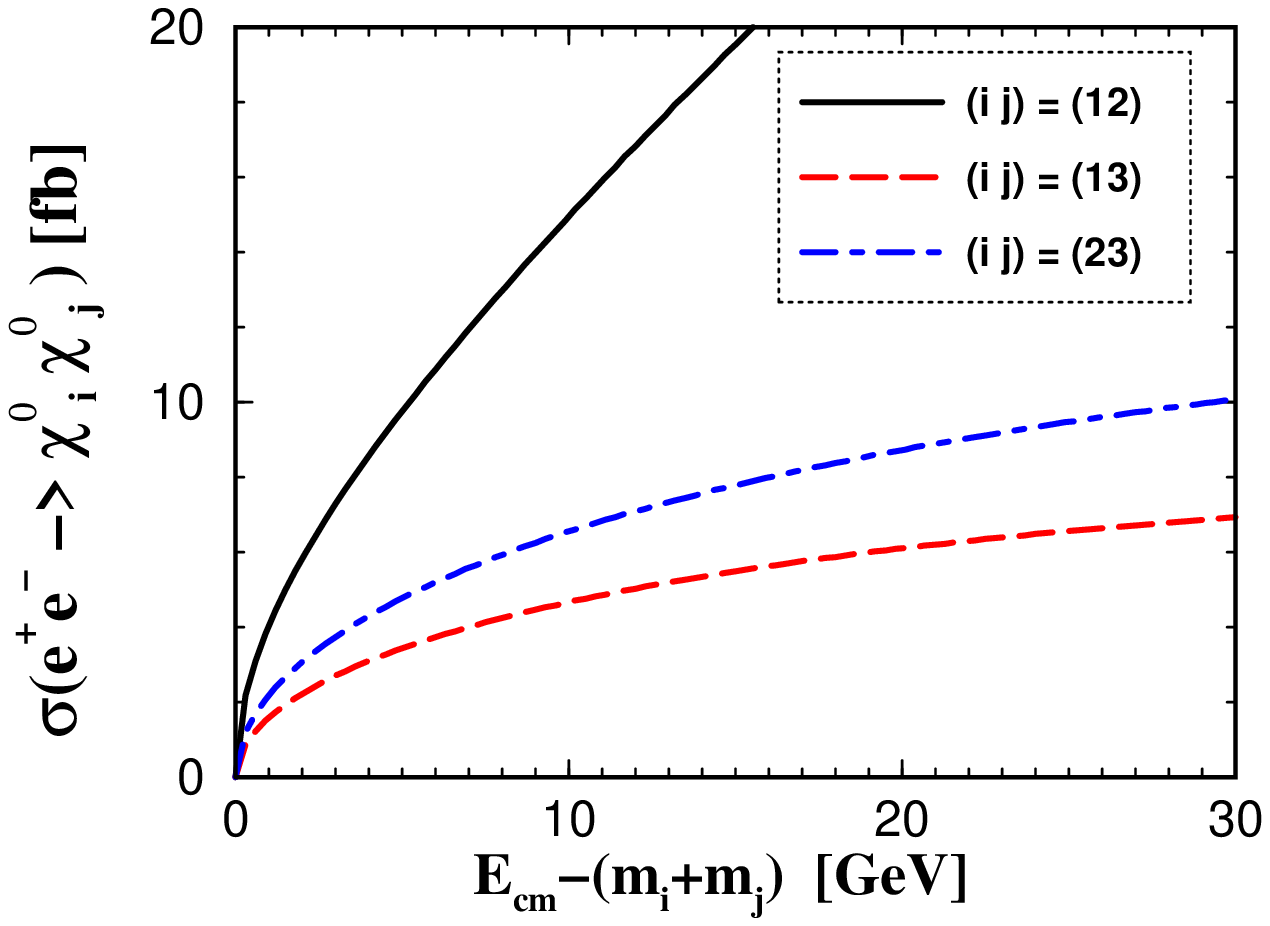} 
\caption{ The threshold behavior of the neutralino  production
  cross--sections $\sigma^{\{ij\}}$ for the CP--conserving (left panel)
and the CP--violating (right panel) cases. Other parameters as in
\fig{fig:quad} \cite{0306272}.} 
\label{fig:th}
\end{figure}

If CP is conserved, the CP parity of a pair of Majorana fermions
$\tilde{\chi}^0_i\tilde{\chi}^0_j$ produced 
in the static limit in $e^+e^-$ collisions 
by a spin-1 
current with positive intrinsic CP must satisfy 
the relation
\begin{eqnarray}
\eta^i\eta^j (-1)^L=1
\label{cpparity}
\end{eqnarray}
where $\eta^i=\pm i$ is the intrinsic 
CP parity of $\tilde{\chi}^0_i$ and $L$ is the
angular momentum \cite{11a}.  Therefore neutralinos with the same CP
parities (for example $i=j$) can only be excited in P-wave. The S-wave
excitation, with the characteristic steep rise $\sim \beta$ of the
cross section near threshold, can occur only for $i\neq j$ with
opposite CP--parities of the produced neutralinos \cite{R2}.  This
immediately implies that if the $\{ij\}$ and $\{ik\}$ pairs are
excited in the S--wave, the pair $\{jk\}$ must be excited in the
P--wave characterized by the slow rise $\beta^3$ of the cross section,
\fig{fig:th}, left panel.

If CP is violated, however, the angular momentum of the produced
neutralino pair is no longer restricted by the \eq{cpparity} and
all non--diagonal pairs can be excited in the S--wave.  This is
illustrated in \fig{fig:th}, where the threshold behavior of the
neutralino pairs $\{12\}$, $\{13\}$ and $\{23\}$ for the CP-conserving
(left panel) case is contrasted to the CP-violating case (right
panel).  Even for a small CP--phase $\varphi_1=\pi/5$, virtually
invisible in the shape and orientation of unitarity quadrangles in
\fig{fig:quad},   
the change in the energy
dependence near threshold can be quite dramatic.  Thus, observing the
$\{ij\}$, $\{ik\}$ and $\{jk\}$ pairs to be excited {\it all} in
S--wave states would therefore signal CP--violation.
%

\subsection{Gluinos}

Strongly interacting gluinos will copiously be produced at the LHC.
Only for rather light gluinos, $m_{\sg} \sim$ 200 -- 300 GeV, can 
a 1 TeV LC improve on the LHC gluino mass measurement.
  
\begin{figure}[htb]
\centering
\includegraphics*[height=3.4cm,width=4cm]{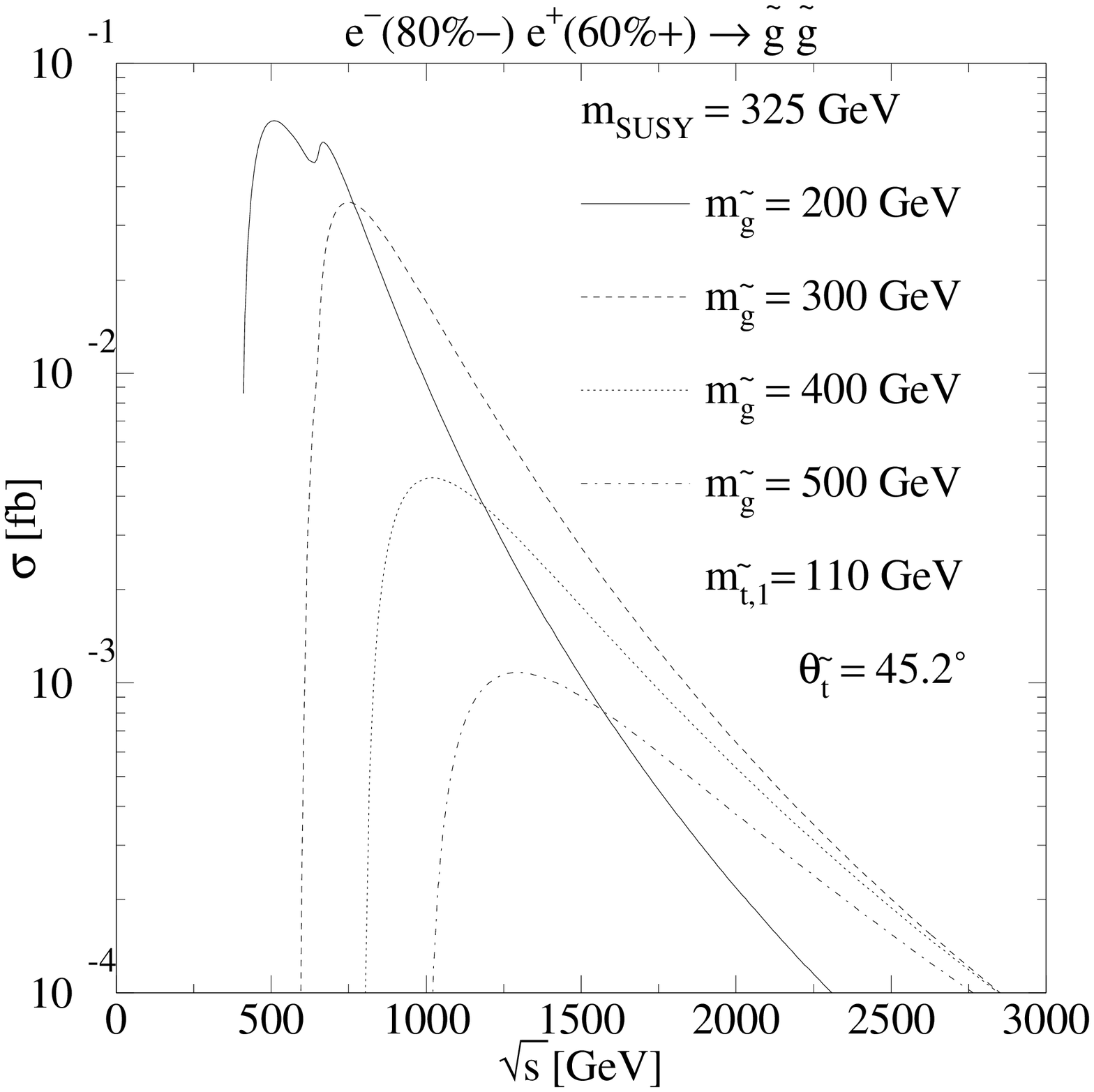} 
\includegraphics*[height=3.4cm,width=4cm]{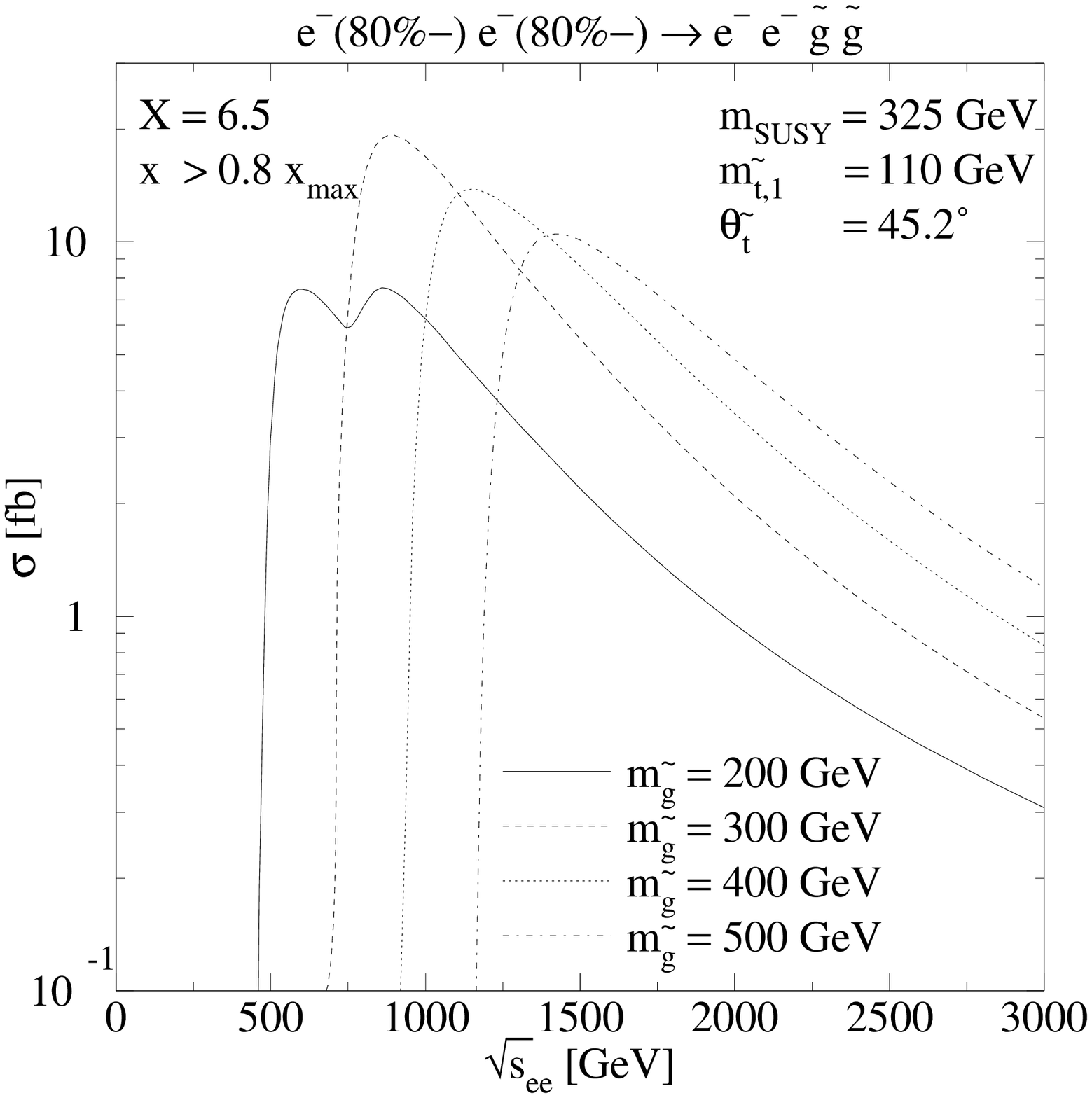} 
\caption{Gluino production cross section in $e^+e^-$ 
  annihilation (left), and in polarized
 direct photon collisions generated in   $e^-e^-$ (right). \cite{gluino}.}
\label{fig:glu}
\end{figure}

In $e^+e^-$ annihilation the exclusive production of gluino pairs
proceeds only at the loop level: $s$-channel photons and $Z^0$ bosons 
couple to the gluinos via triangular quark and squark loops. 
Moreover, near threshold the pairs of identical Majorana gluinos are
excited in a P-wave with a slow rise of the cross section.  
As a
result, the production cross sections are rather small even for
relatively light gluinos, see 
left panel of \fig{fig:glu}. For $m_{\sg}\gsim 500$
GeV, no events at LC with luminosities of 1 ab$^{-1}$ per year are
expected irrespectively of their collision energy.

In the $\gamma\gamma$ option, the chances to observe gluinos are better. 
First, the gluino pairs can be excited in an S-wave with a faster rise of the
cross section. Second, for $m_{\sq} \gg m_{\sg}$ the production can be 
enhanced by resolved photons. As seen in the right panel of
\fig{fig:glu}, the production  
cross sections in the polarized $e^-e^-$  option can reach several fb
in a wider range of gluino masses.

\section{R--parity violating SUSY}

In the MSSM the multiplicative quantum number
R--parity is conserved. Under this symmetry all standard model
particles have $R_p = +1$ and their superpartners $R_p = -1$. As a
result, the
lightest SUSY particle (LSP) is stable, SUSY particles are only
produced in pairs with the distinct signature of missing energy in
an experiment. However,  R--parity
conservation has no strong theoretical justification since the
superpotential admits explicit R--parity violating ($\rpv$) terms 
\begin{eqnarray}
&&W_{\rpv}=      \epsilon_i L_iH_u +
{\textstyle \frac{1}{2}}
\lambda_{ijk} L_i L_j \bar{D}_k \nonumber \\
&& ~~~~~~~~~ + \lambda'_{ijk} L_i Q_j
\bar{D}_k
+ {\textstyle \frac{1}{2}}\lambda''_{ijk} \bar{U}_i \bar{D}_j \bar{D}_k
\label{supWrpv}
\end{eqnarray}
where $H_u,L,Q$ are the Higgs and left--handed lepton and squark
superfields, 
and $\bar{E},\bar{D},\bar{U}$  are the corresponding right--handed
fields. 
R--parity violation changes the SUSY phenomenology drastically. The LSP 
decays, so the characteristic signature of missing energy in the
$\rpv$ conserving  MSSM is replaced by multi--lepton and/or multi--jet
final states.

The couplings $\epsilon$, $\lambda$ and $\lambda'$ violate
lepton number, while $\lambda''$ violate baryon number. If both 
types of couplings were present, they would induce  fast
proton decay. This can be avoided by assuming at most one type of
couplings  to be non-vanishing.

\subsection{Bilinear R--parity violation}
Models with explicit bilinear breaking of R--parity
(BRpV) assume only $\epsilon_i\neq 0$ in \eq{supWrpv} and  
the corresponding terms in the
soft SUSY breaking part of the Lagrangian  
$  {\cal L}_{soft} \ni  B_i \epsilon_i {\tilde
    L}_i H_u$  \cite{hprv}. 
As a result, 
the sneutrinos develop non-zero vacuum expectation $v_i=\langle
{\tilde \nu}_i \rangle$ 
in addition to the VEVs $v_u$ 
and $v_d$ of the MSSM Higgs fields $H_u^0$ and $H_d^0$.  
The bilinear parameters $\epsilon_i$ and $v_i$ induce mixing between
particles that differ only by R--parity: 
charged leptons mix with charginos, neutrinos with
neutralinos, and Higgs bosons with sleptons.  
Mixing between the neutrinos and the
neutralinos generates at tree level a non-zero
mass $m_{\nu_3}\sim M_2|\vec{\Lambda}|^2/{\rm Det}(M_{\nt})$ (where 
$\Lambda_i=\epsilon_iv_d+\mu v_i$) for one of the three neutrinos 
and the mixing angle   
$\tan^2\theta_{ atm}\sim (\Lambda_2/\Lambda_3)^2$;
the remaining two
masses and mixing angles are generated at 1-loop. For example, 
 the solar mixing
angle scales as $\tan^2\theta_{
  sol}\sim (\epsilon_1/\epsilon_2)^2$.  
Thus the model   can 
provide a simple and calculable
framework for neutrino masses and mixing angles in agreement with the
experimental data, and at the same time leads to clear predictions for
the collider physics \cite{BRpVnu-LC}.
\begin{figure}[tb]
\centering
\includegraphics*[height=3.4cm,width=4cm]{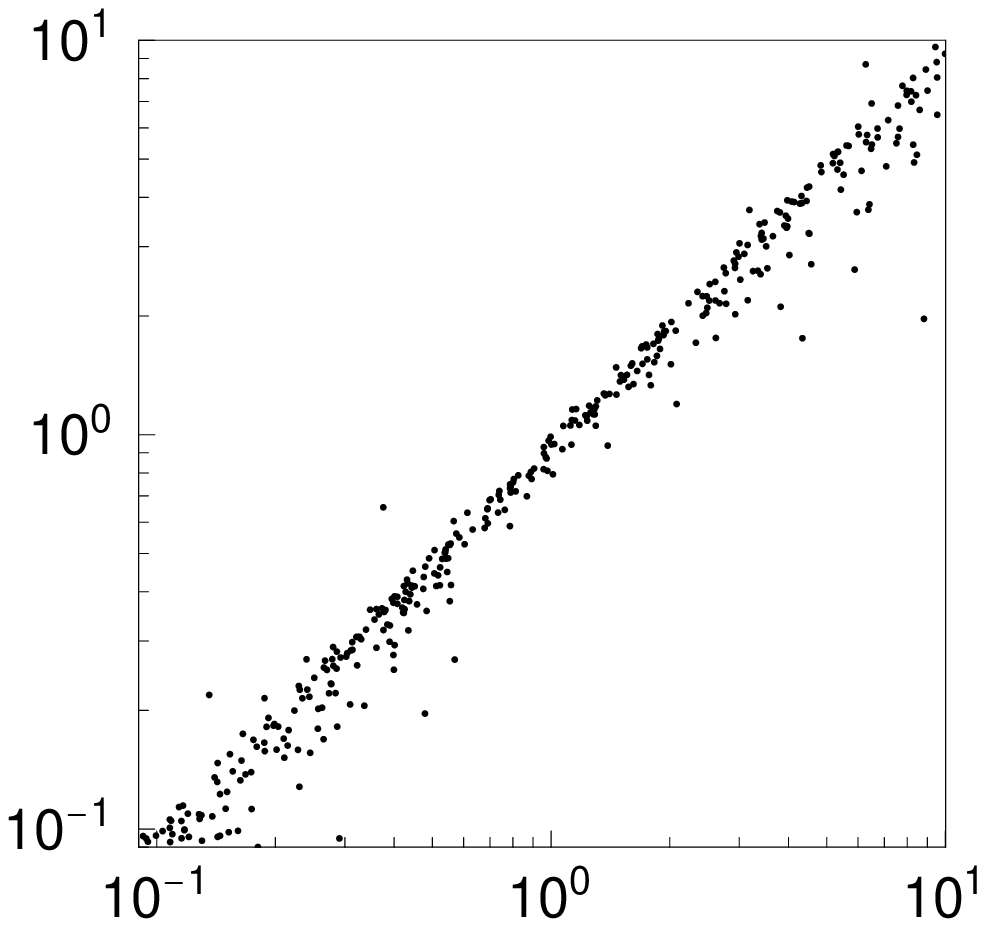} 
\includegraphics*[height=3.5cm,width=4cm]{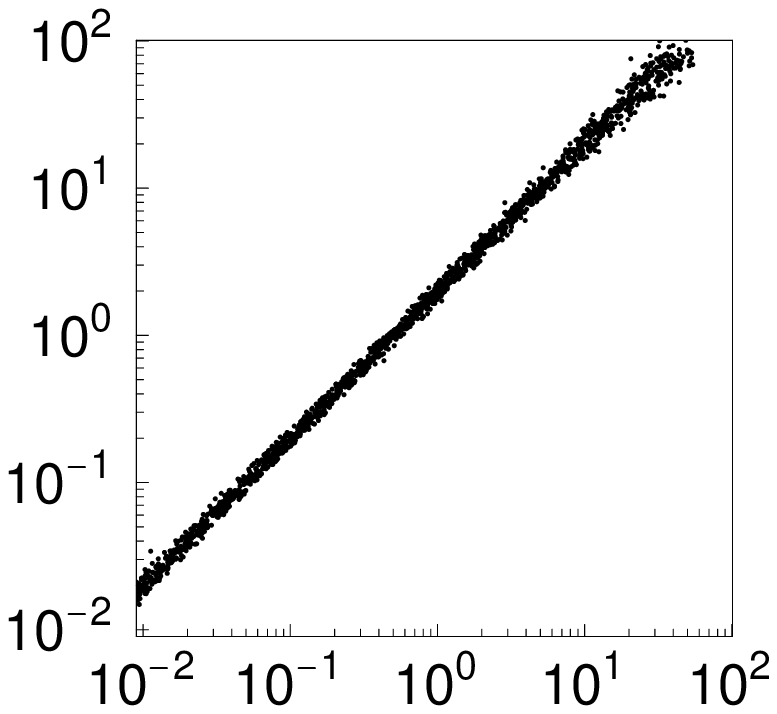}\\[-1mm]
~~~~~~~~~~~~~~~ $\tan^2\theta_{atm}$ ~~~~~~~~~~~~~~~ 
~~~~~~~~ $\tan^2\theta_{sol}$
\caption{Left: BR($\cp_1$$\to$$\mu^+ q\bar{q}$)/BR($\cp_1 
$$\to$$\tau^+q\bar{q}$) 
as a function of $\tan^2\theta_{ atm}$. 
Right: BR($\sb_1$$\to$$
e^+t$)/BR($\sb_1$$\to$$ \mu^+t$)  
as a function of $\tan^2\theta_{ sol}$ \cite{0307364}. }
\label{fig:BRpV}
\end{figure}

For small $\rpv$ couplings, production and decays of SUSY particles is
as in the MSSM except that the LSP decays. Since the astrophysical
constraints on the LSP no longer apply, a priori {\it any} SUSY
particle could be the LSP. 
In a recent study \cite{0307364} 
a sample of the SUSY
parameter space with $\rpv$ couplings consistent with neutrino masses
shows that irrespectively of the LSP nature, there is always at least
one correlation between ratios of LSP decay branching ratios and one
of the neutrino mixing angles. Two examples of chargino and squark
being the LSP are shown in \fig{fig:BRpV}.

\subsection{Bilinear versus Trilinear $\rpv$}
In the case of charged slepton LSP, the collider physics might
distinguish whether  bilinear or trilinear couplings are  dominant
sources of $\rpv$ and the neutrino mass matrix \cite{bi-tri}. 
Possible final states
of the LSP are either $l_j\nu_k$ or $q{\bar q}'$. 
If the LSP is dominantly right-chiral, the former by far dominate over the
hadronic decay mode.
In the case of TRpV, the 
two-body decay width  for ${\tilde l}_i \rightarrow l_j +
\Sigma_k \nu_k$ scales as  $\Gamma \sim \Sigma_k
 \sin^2\theta_{\ti l_i}\lambda^2_{kji}$
provided $\lambda'\lsim \lambda$, 
while for the BRpV one has $\Gamma \sim Y_i \sin^2\theta_{\ti l_i}
\epsilon^2_j$  
for $i\neq j$ ($Y_i $ is the corresponding Yukawa coupling),
and  BR$( {\tilde e}_1 \rightarrow e \Sigma_k \nu_k) \sim 1$.
Immediately one finds then  
\begin{eqnarray}
{\rm BR}( {\tilde e}_1 \rightarrow e \Sigma_k \nu_k)=
\left\{ \begin{array}{l} \sim 1~~~~ ~~~~{\rm for~ BRpV}\\ 
\lsim 0.5 ~~~~{\rm for~ TRpV}
\end{array} \right.
\end{eqnarray}
Therefore, the LC measurements of the ${\tilde l}_i$ decay modes can
distinguish between bilinear or trilinear terms as dominant
contributions to the  neutrino masses \cite{bi-tri}.

\begin{figure}[htb]
\centering
\includegraphics*[width=45mm,height=35mm]
{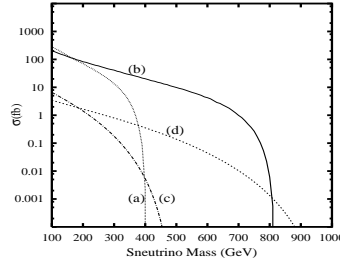}~~\begin{minipage}[b]{35mm} 
\caption{$\sigma(\snu\tau^\pm\mu^\mp)$ at $\gamma\gamma$ (a,b) and
  $e^+e^-$ (c,d) collider with $\sqrt{s}$=100 GeV (a,c) and 1 TeV
  (b,d) \cite{ghosh}. \label{fig:ghosh} ~~~~~ ~~~~~~~ ~~~~~~~}
\end{minipage}
\end{figure}

For trilinear couplings of the order of current experimental upper
bounds, in particular for the third generation (s)fermions,  
additional production as well as decay channels may produce strikingly 
new signatures. For example, sneutrinos could be produced as an
s-channel resonance in $e^+e^-$ annihilation. During this workshop
single sneutrino production in association with  
fermion pairs at polarised photon colliders has been analysed
\cite{ghosh}.  The associate mode
may  also appear with fermions of different  flavour \cite{sourov2}, so
that the signal is  basically
SM background free.
Moreover, the advantage of exploiting $\gamma\gamma$
collisions in place of $e^+e^-$ ones 
in producing single sneutrinos with a fermion pair of differnt flavour 
resides in the fact that the cross sections for the former
are generally larger than those for the latter. As an
example, \fig{fig:ghosh} shows the unpolarised production rates
for both the $\gamma\gamma$ and $e^+e^-$ induced $\snu\tau^\pm\mu^\mp$
modes at $\sqrt{s_{e^+e^-}}=500$ GeV
and 1 TeV. For illustration, the couplings are set 
$\lambda=\lambda'=1$. 

\section{Extended SUSY}
The NMSSM, the  minimal extension of the MSSM, 
introduces  a singlet superfield 
field $S$ in the superpotential
\begin{equation}
W \supset \lambda H_u H_d S - \frac{1}{3} \kappa S^3 \,.
\end{equation}
In this model, an effective $\mu=\lambda x$ term is generated when the
scalar component of the singlet $S$ acquires a vacuum expectation
value $x=\langle S \rangle $.  The fermion component of the
singlet superfield (singlino) will mix with neutral gauginos and
higgsinos after electroweak gauge symmetry breaking, changing the
neutralino mass matrix to the 5$\times$5 form 
which depends on $M_1$, $M_2$, $\tan\beta$, $x$ and the 
trilinear couplings $\lambda$ and $\kappa$.

\begin{figure}[tb]
\centering
\includegraphics*[height=3.8cm,width=3.5cm]{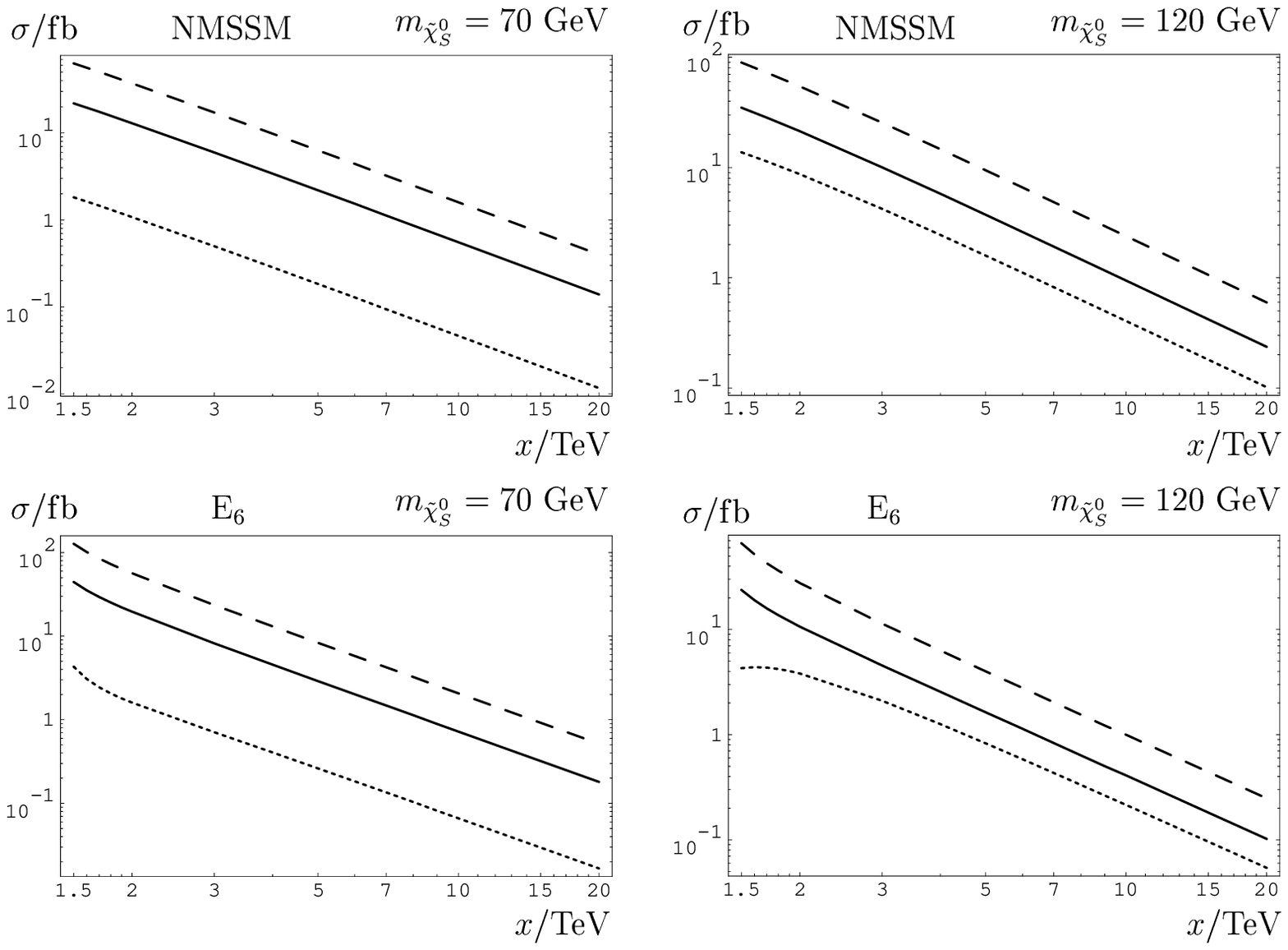} 
\includegraphics*[height=4cm,width=4.5cm]{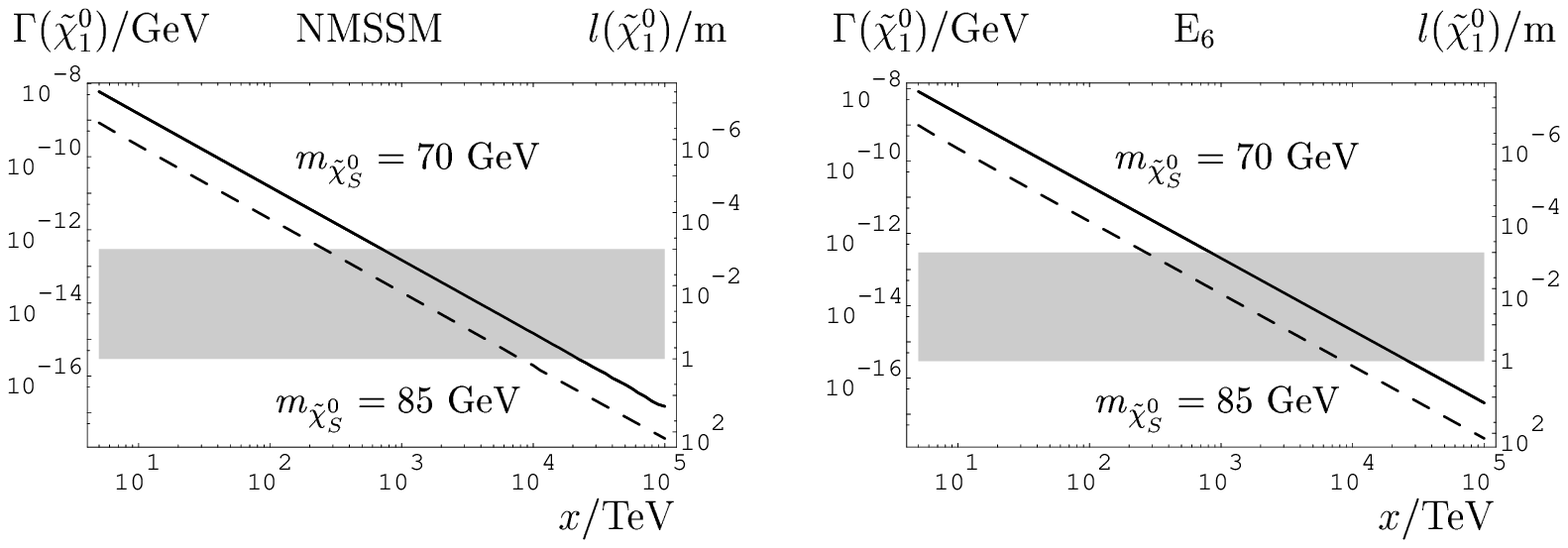}
\caption{Left: Unpolarized $\sigma(\tilde{\chi}^0_S\tilde{\chi}^0_1$)
  at $\sqrt{s}$=500 GeV (solid) and for polarized beams $\cP_-$=0.8,
  $\cP_+$=-0.6   (dashed), $\cP_-$=-0.8, $\cP_+$=0.6 (dotted).      
Right: Total decay widths of the lightest
MSSM-like $\tilde{\chi}^0_1$ decaying into a
singlino-dominated $\tilde{\chi}^0_S$.
The shaded area shows where displaced
vertices exist \cite{Franke}. }
\label{fig:nmssm}
\end{figure}

In some regions of the parameter space  the singlino may
be the lightest supersymmetric particle, weakly mixing with other
states. In the extended SPS\#1a scenario with  large $x \gg |M_2|$,
analysed in \cite{Franke},  
the lightest neutralino $\tilde{\chi}^0_S$ with mass
$\approx 2 \kappa x$ becomes  singlino-dominated 
while the other four neutralinos $\tilde{\chi}^0_{1,\ldots,4}$
have the MSSM characteristics. The exotic $\tilde{\chi}^0_S$ state can be
searched for in the associated production of 
$\tilde{\chi}^0_S$ together with the lightest MSSM-like neutralino
$\tilde{\chi}^0_1$ in $e^+ e^-$ annihilation. The unpolarized cross 
section, shown in
\fig{fig:nmssm} for $m_{\tilde{\chi}^0_S}=70$ GeV, 
is larger than 1~f\/b up to $x< 7.4$~TeV which
corresponds to a singlino content of 99.7\,\%. Polarized beams can
enhance the cross section by a factor 2--3, and provide discriminating
power between different scenarios \cite{hff}. If the couplings of a
singlino-dominated LSP to the NLSP 
are strongly suppressed at large values of $x$, 
displaced vertices in the NMSSM may be generated, \fig{fig:nmssm},    
which would clearly signal the extension of the minimal model.  
For a similar analysis in the E$_6$ inspired model we refer to
\cite{Franke}. 

\begin{figure}[htb]
\centering
\includegraphics*[width=45mm,height=40mm]
{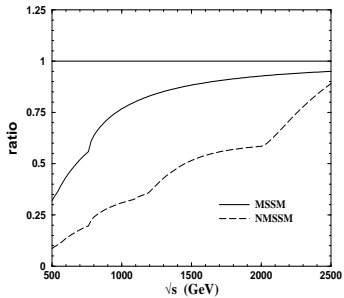}~~\begin{minipage}[b]{35mm} 
\caption{The sum of  
  neutralino--pair production cross sections: all pairs in the MSSM
  (solid), and   
  of the first four neutralino states in the NMSSM (dashed);
both  normalized to the
  asymptotic value \cite{ckmz}. \label{fig:sr}}
\end{minipage}
\end{figure}

However, if the
spectrum of the  four lighter neutralinos in the extended model is similar
to the spectrum in the MSSM, but the mixing is substantial, 
discriminating the models by analysing the mass spectrum becomes very
difficult.   Studying in this
case the summed-up cross sections of the four light neutralinos may
then be a crucial method to reveal the structure of the neutralino
system \cite{ckmz}.
More specifically, in  extended SUSY models with $n$ SU(2)
doublet and $m$ SU(2) singlet chiral superfields, the sum rule reads 
\begin{eqnarray}
&&
\lim_{s\rightarrow \infty}\,s\,
{\textstyle \sum_{i\leq j}}\;      
    \sigma \{ij\} =  {\textstyle \frac{\pi\alpha^2}{48\,
      c^4_W s^4_W} }
\nonumber \\
 && ~~~~        \times  
\left[\, n\,(8 s^4_W-4 s^2_W+ 1)\, + 48 s^4_W + 3\right]
\label{eq:srext}
\end{eqnarray}
The right--hand side of \eq{eq:srext} is independent of the 
number $m$ of 
singlets and it reduces to the sum rule in the
MSSM for $n=2$.
In \fig{fig:sr} the exact sum rules, normalized to the asymptotic
value,   are compared for an NMSSM 
scenario  giving rise to one very heavy neutralino with 
$m_{\tilde{\chi}^0_5}\sim 1000$ GeV, and to four lighter
neutralinos with masses equal within 2 -- 5 GeV to the neutralino
masses in the MSSM.  Due to the incompleteness of these states
below the thresholds for producing the heavy neutralino
$\tilde{\chi}^0_5$,  the NMSSM value
differs significantly from the corresponding sum rule of the MSSM.
Therefore, even if the extended neutralino states are very heavy,
the study of sum rules can shed light on the underlying structure of the
supersymmetric model.

\section{Reconstructing fundamental SUSY parameters} 
Low energy SUSY  particle physics is characterized by energy scales of
order $\lsim$ 1
TeV. However, the roots for all the phenomena we will 
observe experimentally in
this range  may go 
to energies near the Planck or the GUT scale. 
Fortunately, supersymmetry  
provides us with a
stable bridge  between these two vastly different
energy regions \cite{Witten:nf}. 
To this purpose
renormalization group equations (RGE) are exploited, by which parameters
from low to high scales are evolved based on nothing but measured
quantities in laboratory experiments. This procedure has very
successfully been pursued for the three electroweak and strong gauge
couplings, and has been expanded to a large ensemble of supersymmetry
parameters~\cite{bpz} -- the soft SUSY breaking parameters: 
gaugino and scalar
masses, as well as trilinear couplings. 
This bottom-up approach makes use of the low-energy measurements to
the maximum extent possible and it reveals the quality with which the
fundamental theory at the high scale can be reconstructed in a transparent
way.

A set of representative examples in this context has been studied 
\cite{bpz-new}: 
minimal supergravity and
a left--right symmetric extension; gauge mediated supersymmetry breaking;
and superstring effective field theories. The anomaly mediated as well
as the gaugino mediated SUSY breaking 
are technically equivalent to the mSUGRA case and therefore were not 
treated explicitly.

\subsection{Gravity mediated SUSY breaking}
The minimal supergravity scenario mSUGRA is characterized by the universal:
gaugino mass $M_{1/2}$, 
 scalar mass  $M_0$,
 trilinear coupling  $A_0$, sign of $\mu$ (the modulus $|\mu|$ determined
by radiative symmetry breaking)  and $\tan \beta$.
The parameters $M_{1/2}$, $M_0$ and $A_0$
are defined at the GUT scale $M_U$ where 
gauge couplings unify $\alpha_i=\alpha_U$. The RGE are then used to 
determine the low energy SUSY lagrangian parameters.

The point chosen for the analysis is close to 
the Snowmass Point SPS\#1a 
\cite{sps}, except for the scalar mass parameter
$M_0$ which was taken slightly larger for merely
illustrative purpose: $M_{1/2} = 250$~GeV, $M_0 = 200$~GeV, $A_0 = -100$~GeV,
$\tan \beta = 10$ and $sign(\mu) = +$.  

{\small
\begin{table}
\caption[]{Representative gaugino/scalar mass parameters and couplings
as determined
at the electroweak
scale and evolved to the GUT scale in the mSUGRA scenario 
based on LHC
and LC simulations; masses are in GeV.  The errors are 1$\sigma$
\cite{bpz-new}.}
\label{sugradata}
\begin{center}
\begin{tabular}{c|c|c}
 &  Exp.~Input &  GUT Value \\ \hline   \hline
 $M_1$ & 102.31 $\pm$  0.25 &  $250.00 \pm  0.33$ \\
 $M_2 $ &  192.24 $\pm$  0.48      &  $250.00 \pm  0.52$ \\
 $M_3 $ & 586  $\pm$  12   &  $250.0    \pm   5.3$  \\ \hline
$\mu$         & 358.23  $\pm$ 0.28     &  $355.6 \pm  1.2    $  \\
\hline
 $M^2_{L_1} $ & $( 6.768  \pm  0.005)\cdot 10^4$
                &  $(3.99  \pm  0.41) \cdot 10^4$  \\
 $M^2_{E_1} $ & $(4.835  \pm  0.007) \cdot 10^4$
  &  $(4.02  \pm  0.82)  \cdot 10^4 $ \\
 $M^2_{Q_1} $ &  $(3.27 \pm  0.08)\cdot 10^5$
               &  $(3.9  \pm  1.5) \cdot 10^4$ \\
 $M^2_{U_1} $ &  $(3.05 \pm  0.11)\cdot 10^5$
               &  $(3.9  \pm  1.9) \cdot 10^4$ \\
 $M^2_{D_1} $ &  $(3.05 \pm  0.11)\cdot 10^5$
               &  $(4.0  \pm  1.9)  \cdot 10^4$
\\ \hline 
 $M^2_{H_1} $ &  $(6.21 \pm  0.08)\cdot 10^4$  &
  $ (4.01  \pm  0.54)  \cdot 10^4 $ \\
 $M^2_{H_2}$ &  $(-1.298 \pm 0.004)\cdot 10^5$ &
  $(4.1  \pm  3.2) \cdot 10^4 $\\
 $A_{ti t} $ & $ -446 \pm 14$   & $ -100 \pm 54 $  \\ \hline
 $\tan\beta$ & $9.9  \pm  0.9$ & --- \\ \hline
\end{tabular}
\end{center}
\end{table}
}

\begin{figure}[htb]
{\small $1/M_i$~[GeV$^{-1}]$ ~~~~~~~ ~~~~~~~~~ 
~~~~~~~ $M^2_{\tilde j}$~[$10^3$ GeV$^2$]}
\centering
\includegraphics*[height=3.5cm,width=4cm]{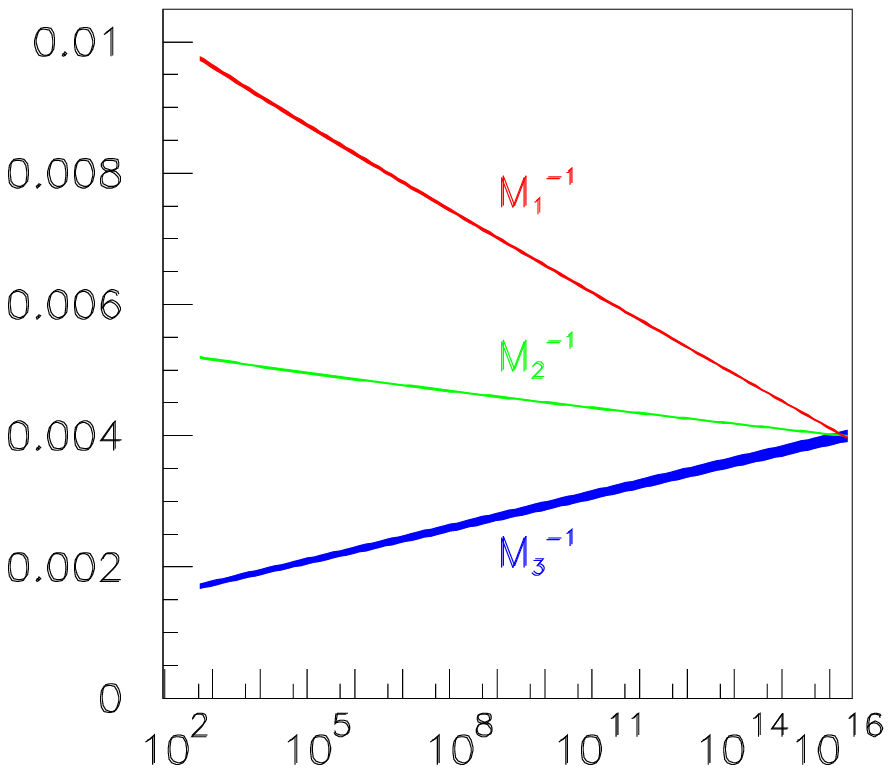} 
\includegraphics*[height=3.5cm,width=4cm]{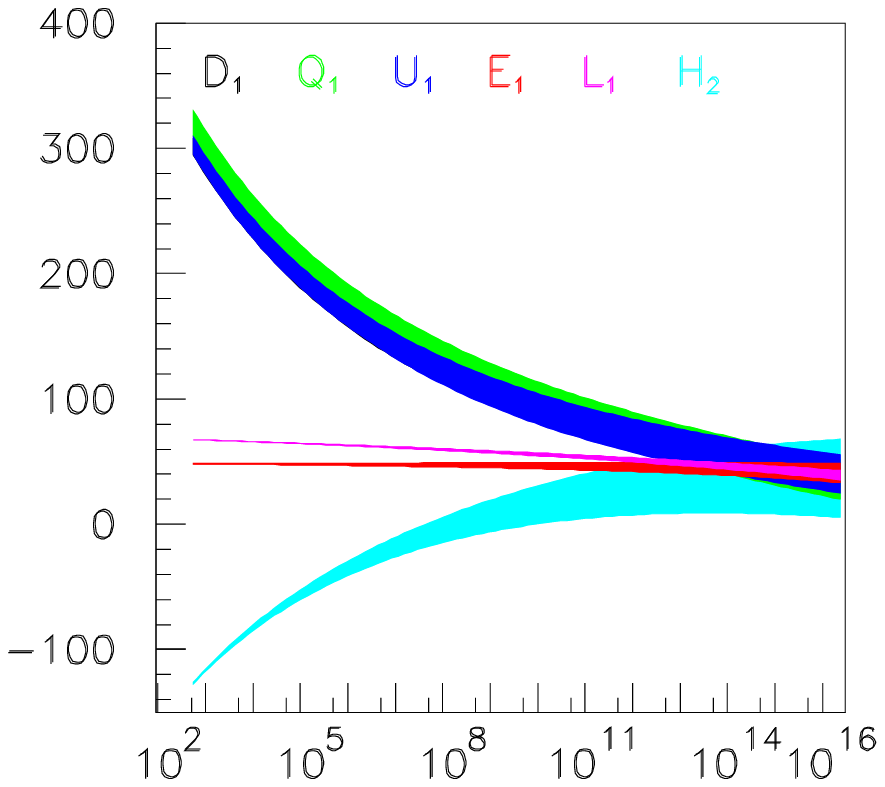}\\
$Q$~[GeV] ~~~~~~~~~~~  ~~~~~~~~~~~~~~~~ $Q$~[GeV]
\caption{mSUGRA: Evolution, from low to high scales, of 
gaugino mass parameters (left),  and first two generation sfermion mass
parameters and  
     the Higgs mass parameter $M^2_{H_2}$ (right). 
The widths of the bands indicate the 1$\sigma$ CL \cite{bpz-new}. }
\label{fig:bpz-sugra}
\end{figure}

Based on simulations and estimates of expected  precision,
the low-energy 'experimental' values  
are taken as the input values for
the evolution of the mass parameters in the bottom-up
approach to the GUT scale.  
The results for the evolution of the mass parameters to the GUT scale $M_U$
are shown in \fig{fig:bpz-sugra}.
The left panel  presents the evolution of the gaugino
parameters $M_i^{-1}$, while the right panel shows  
the extrapolation
of the slepton mass parameters squared of the first two generations. 
The accuracy
deteriorates for the squark mass parameters and for the Higgs mass parameter
$M^2_{H_2}$.
The origin of the differences between the errors for slepton, squark and
Higgs mass parameters can be traced back to the numerical 
size of the coefficients.  The quality of the test is apparent from
\tab{sugradata},  
where it is shown how well the reconstructed mass parameters at the
GUT scale reproduce the input values 
$M_{1/2}=250$ GeV and $M_0=200$ GeV. 

The above analysis has also been extended \cite{bpz-new} to a
left--right supersymmetric $SO(10)$ model in which the $SO(10)$
symmetry is assumed to be realized at a scale between the standard
$SU(5)$ scale $M_U \simeq 2 \cdot 10^{16}$, derived from the
unification of the gauge couplings, and the Planck scale $M_P \simeq
10^{19}$~GeV.  The right--handed neutrinos are assumed heavy, with
masses at intermediate scales between $O( 10^{10})$~GeV and
$O(10^{15})$~GeV, so that the observed light neutrino masses are
generated by the see-saw mechanism.  The evolution of the gaugino and
scalar mass parameters of the first two generations is not affected by
the left--right extension.  It is only different for the third
generation and for $M^2_{H_2}$ owing to the enhanced Yukawa coupling
in this case. The sensitivity to the intermediate $\nu_R$ scales is
rather weak because neutrino Yukawa couplings affect the evolution of
the sfermion mass parameters only mildly. Nevertheless, a rough
estimate of the intermediate scale follows from the evolution of the
mass parameters to the low experimental scale if universality holds at
the Grand Unification scale.

\subsection{Gauge mediated SUSY breaking}
In GMSB 
the scalar and the F components of a Standard--Model singlet superfield $S$
acquire vacuum expectation values $\langle S \rangle$ and 
$\langle F_S \rangle $ through interactions with
fields in the secluded sector, thus breaking supersymmetry. %
Vector-like messenger fields $M$, carrying non--zero 
$SU(3)\times SU(2) \times U(1)$ charges and coupling to $S$, transport
the supersymmetry breaking to the eigen--world.
The system is characterized by the mass $M_M \sim$ $\langle S \rangle $ 
of the messenger
fields and the mass 
scale $\Lambda = {\langle F_S \rangle} / {\langle S \rangle} $ setting
the size  
of the gaugino and scalar masses.  $M_M$ is expected to be in the
range of 10 to $10^6$ TeV and $\Lambda$ has to be smaller than $M_M$. 

The gaugino masses  
are generated by loops of scalar and fermionic messenger component
fields, while masses of the scalar fields in the
visible sector
are generated by 2-loop effects of gauge/gaugino and messenger
fields, and the $A$ parameters
are generated at 3-loop level and they are practically zero at $M_M$.
Scalar particles with identical Standard--Model 
charges squared have equal
masses at the messenger scale $M_M$, which is a  characterictic
feature of the GMSB model. 

\begin{figure}[htb]
\centering
\includegraphics*[height=4cm,width=8.2cm]{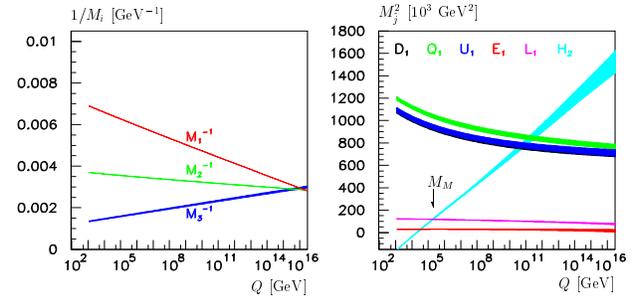} 
\caption{GMSB: Evolution, from low to high scales, of 
gaugino mass parameters (left),  and first two generation sfermion mass
parameters and  
     the Higgs mass parameter $M^2_{H_2}$ (right). 
The widths of the bands indicate the 1$\sigma$ CL \cite{bpz-new}.}
\label{fig:bpz-Gmsb}
\end{figure}

This scheme has been investigated for the point $\Lambda = 100$~TeV,
$M_M = 200$~TeV, $N_5=1$, $N_{10}=0$, $\tan\beta=15$ and $\mu>0$
corresponding to the Snowmass Point SPS\#8. 
The evolution
 of the gaugino and sfermion mass parameters
of the first two generations as well as the Higgs mass parameters, 
including 2-loop $\beta$--functions,
is presented in \fig{fig:bpz-Gmsb}. 

The gaugino masses in GMSB evolve nearly in the same way as in
mSUGRA. However, due to the influence of the $A$--parameters in the 2-loop
RGEs for the gaugino mass parameters, they 
do not meet at the same point as the gauge couplings in this
scheme. On the other hand 
the running of the scalar masses is quite different in both theories. 
The bands of the slepton $L$--doublet mass parameter $M^2_{\tilde L}$ and the
Higgs parameter $M^2_{H_2}$, which carry the same moduli of
standard--model charges, cross at the scale $M_M$. The crossing,
indicated by an arrow in the \fig{fig:bpz-Gmsb}, is a necessary
condition (in the minimal form) for the GMSB scenario to be realized. 
Moreover, at the messenger scale the ratios of scalar masses squared in
the simplest version of 
GMSB are determined solely by group factors and gauge couplings, being 
independent of the specific GMSB characteristics, i.~e. messenger 
multiplicities and $\Lambda$ mass scale.

The two scales $\Lambda$ and $M_M$, and the messenger multiplicity 
$N_M=N_5+3N_{10}$ can be extracted from the spectrum
of the gaugino and scalar particles.    For the point
analyzed in the example above, the following accuracy for the mass
parameters and the messenger
multiplicity has been found:
\begin{eqnarray}
\Lambda &=& (1.01 \pm 0.03) \cdot 10^2 \; \rm {TeV}\\
M_M &=&(1.92 \pm 0.24) \cdot 10^2 \; \rm {TeV} \\
N_M &=& 0.978 \pm 0.056
\end{eqnarray}

\subsection{String induced SUSY breaking}
Four--dimensional
strings naturally give rise to a minimal set of fields for inducing
supersymmetry breaking; they play the r\^ole of the fields in the hidden
sectors: the dilaton $S$ and the moduli $T_m$ chiral
superfields which are generically present in large classes of
4--dimensional heterotic string theories. In the analysis
only one moduli field $T$ has been considered.  
SUSY breaking, mediated by a goldstino field,  
originates in the vacuum expectation
values of $S$ and $T$ generated by genuinely non--perturbative
effects. The properties of the model depend on the composition
of the goldstino which is a mixture of the 
dilaton field $S$ and the moduli field $T$, 
\begin{eqnarray}
\tilde G &=& {S}\,  \sin \theta   + {T} \, \cos \theta  
\end{eqnarray}
Universality is generally broken in such a scenario by a set of non-universal
modular weights $n_j$ that determine the coupling of $T$ to the SUSY matter
fields $\Phi_j$. 
The gaugino and scalar mass parameters can be expressed to leading order
by the gravitino mass $m_{3/2}$, the vacuum values $\langle S \rangle$
and  $\langle T \rangle$, 
the mixing parameter
$\sin\theta$, the modular weights $n_j$ and the Green-Schwarz
parameter $\delta_{\rm GS}$. 
The relations between the universal gauge coupling
$\alpha(M_{\rm string})$ at the string scale $M_{\rm string}$ 
and the gauge couplings $\alpha_i(M_{\rm GUT})$ at the 
SU(5) unification scale $M_{\rm GUT}$:
\begin{equation}
\alpha^{-1}_i(\rm M_{\rm GUT}) = \alpha^{-1}({\rm M_{\rm string}}) + 
\Delta \alpha^{-1}_i[n_j]
\end{equation}
receive small deviations from universality at
the GUT scale which  are accounted for by string loop effects transporting the
couplings from the universal string scale to the GUT scale.
The gauge coupling at $M_{\rm string}$ is related to the dilaton field, 
$g_s^2 = 1/{\langle S \rangle}$.

A mixed dilaton/moduli
superstring scenario with dominating dilaton
field component and with different couplings of the
moduli field to the (L,R) sleptons, the (L,R) squarks and to the Higgs
fields, corresponding to O--I representation has been chosen for the
analysis \cite{bpz-new}, for which    
$\sin^2 \theta = 0.9$,  $n_{L_i} = -3$, $n_{E_i} = -1$, 
$n_{H_1} =n_{H_2}=-1$,
 $n_{Q_i} = 0$, $n_{D_i} = 1$, $n_{U_i} = -2$, 
and the gravitino mass 180~GeV.

\begin{figure}[htb]
{\small $1/M_i$~[GeV$^{-1}] \times 10^{-2}$ ~~~~~~~ ~~~~~~~~~ 
 $M^2_{\tilde j}$~[$10^3$ GeV$^2$]}
\centering
\includegraphics*[height=3.4cm,width=3.8cm]{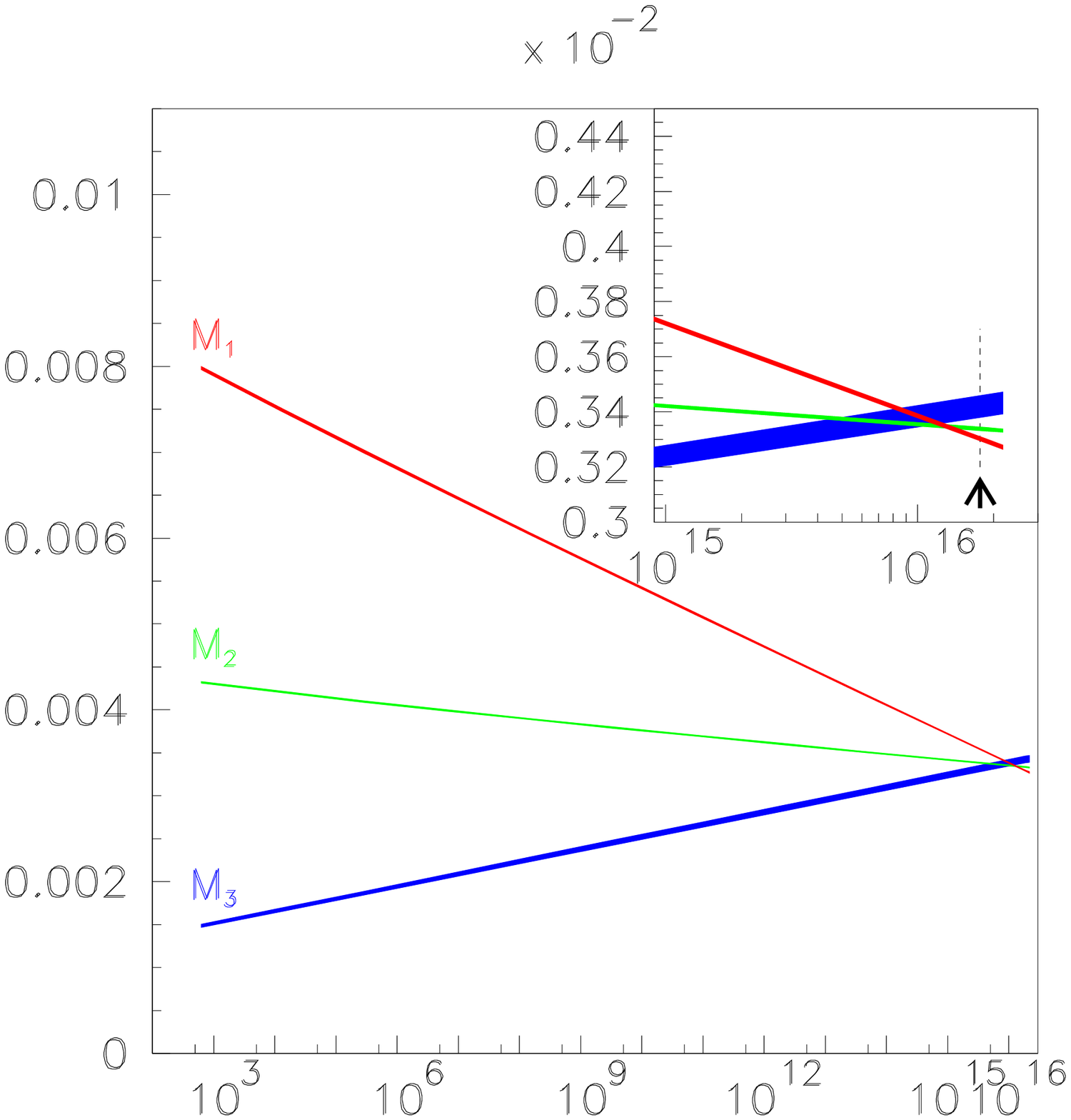} 
\includegraphics*[height=3.5cm,width=4cm]{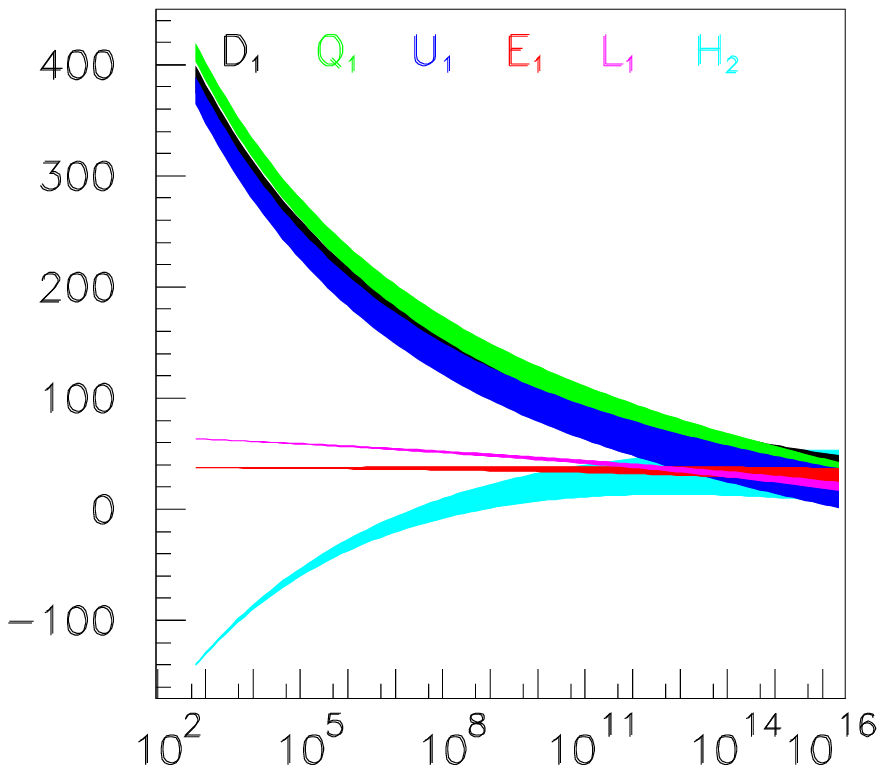}\\
$Q$~[GeV] ~~~~~~~~~~~  ~~~~~~~~~~~~~~~~ $Q$~[GeV]
\caption{String scenario: Evolution, from low to high scales, of 
gaugino mass parameters (left),  and first two generation sfermion mass
parameters and  
     the Higgs mass parameter $M^2_{H_2}$ (right). 
The widths of the bands indicate the 1$\sigma$ CL \cite{bpz-new}. }
\label{fig:bpz-String}
\end{figure}

The evolution of the
gaugino and scalar mass parameters is displayed in \fig{fig:bpz-String}.
The pattern of the trajectories is remarkably different from other
scenarios.  The breaking of universality in the gaugino
sector, induced by string threshold corrections, is shown in the
insert.  

{\small
\begin{table}
\caption[]{ Comparison of the experimentally reconstructed values with the
            ideal fundamental parameters in a specific example for a string 
            effective field theory. [All mass parameters are 
            in units of GeV.]
\label{tab:parameters_string2}}
\begin{center}
\begin{tabular}{c|c|c}
Parameter           & Ideal & Reconstructed \\ \hline\hline
$m_{3/2}$           &  180  &     179.9  $\pm$  0.4 \\
$\langle S \rangle$ &   2   &      1.998  $\pm$  0.006 \\
$\langle T \rangle$ &  14   &      14.6  $\pm$  0.2 \\
$\sin^2\theta$        & 0.9 &      0.899  $\pm$  0.002 \\
$g_s^2$             & 0.5   &      0.501  $\pm$  0.002 \\
$\delta_{GS}$       &   0   &      0.1  $\pm$  0.4 \\ \hline
$\tan \beta$        &  10   &      10.00  $\pm$  0.13 \\ 
\end{tabular}
\end{center}
\end{table}}

The reconstructed 
values the fundamental parameters 
of the string effective field theory are compared with the ideal values in 
Table~\ref{tab:parameters_string2}. Also the reproduction of 
moduli weights as 'integers' at the per-cent level provides a highly
non-trivial check of the string model \cite{bpz-new}.

\section{Summary}

Much progress has been achieved during the Extended ECFA/DESY
Workshop. It has been demonstrated that a high luminosity LC with
polarized beams, and with additional $e\gamma$, $\pp$ and
$e^-e^-$ modes, can provide high quality data for the precise 
determination of
low-energy SUSY Lagrangian parameters. In the bottom--up approach, 
through the evolution of the parameters from the electroweak scale,  
the regularities  in different scenarios at the high scales can be
unravelled if 
precision analyses of the supersymmetric particle  sector at
$e^+ e^-$ linear colliders are combined with analyses at the LHC.
In this way the basis 
of the SUSY breaking 
mechanism can be explored and the crucial elements of the fundamental 
supersymmetric theory can be reconstructed.

So far most  analyses were
based on lowest--order expressions. With higher order corrections now
available, one of the goals of the SUSY WG in the new ECFA Study
would be to refine the above program.   Many new theoretical calculations
and future
experimental analyses will be necessary. However, the prospect of exploring
elements of the ultimate unification of the interactions provides
a strong stimulus in this direction.   



\begin{thebibliography}{9}   

\bibitem{Aguilar-Saavedra:2001rg}
J.~A.~Aguilar-Saavedra {\it et al.}  [ECFA/DESY LC Physics Working Group
                  Collaboration],
``TESLA Technical Design Report Part III: Physics at an e+e- Linear
Collider,'' 
arXiv:hep-ph/0106315.


\bibitem{Snowmass}Understanding Matter, Energy, Space and Time: The
  Case for the $e^+e^-$ Linear Collider,
  http://sbhepnt.physics.sunysb.
edu/~grannis/ilcsc/lc$_-$icfa$_-$v6.3.pdf  

\bibitem{LHCrec} F.~E.~Paige,
arXiv:hep-ph/0211017.
J.~G.~Branson, D.~Denegri, I.~Hinchliffe, F.~Gianotti, F.~E.~Paige and
P.~Sphicas  (eds.)  [ATLAS and CMS Collaborations],
Eur.\ Phys.\ J.\ directC {\bf 4} (2002) N1.

\bibitem{JKsusy03}
J.~Kalinowski,
arXiv:hep-ph/0212388.

\bibitem{lhclc}
G. Weiglein, talk in Amsterdam. 
The web address of the LHC/LC Study Group:  
http://www.ippp.dur.ac.uk/ $\tilde{ }$ge\-org/lhclc/. 

\bibitem{meetings} The web addresses of the 
Extended ECFA-DESY meetings:\\  
Cracow http://fatcat.ifj.edu.pl/ecfadesy-krakow/\\ 
St.~Malo http://www-dapnia.cea.fr/ecfadesy-stmalo/ \\ 
Prague http://www-hep2.fzu.cz/ecfadesy/Talks/SUSY/\\ 
Amsterdam http://www.nikhef.nl/ecfa-desy/start.html  


\bibitem{power}
G.~Moortgat-Pick, talk in Amsterdam. The web address of the POWER
Study Group: 
http://www.ippp.dur.ac.uk/ $\tilde{ }$gu\-drid/power/.

\bibitem{sps} B.~C.~Allanach {\it et al.},
in {\it Proc. of the APS/DPF/DPB Summer Study on the Future of
Particle Physics (Snowmass 2001) } ed. N.~Graf; 
Eur.\ Phys.\ J.\ C {\bf 25} (2002) 113
[eConf {\bf C010630} (2001) P125]
[arXiv:hep-ph/0202233] LC-TH-2003-022.
The values of  benchmarks are listed on
http://www.ippp.dur.ac.uk/ $\tilde{ }$georg/sps

\bibitem{isajet} H.~Baer, F.~E.~Paige, S.~D.~Protopopescu and X.~Tata,
arXiv:hep-ph/0001086.


\bibitem{GMAKP1}N.~Ghodbane and H.~U.~Martyn,
in {\it Proc. of the APS/DPF/DPB Summer Study on the Future of
  Particle Physics (Snowmass 2001) } ed. N.~Graf, 
arXiv:hep-ph/0201233, 
LC-PHSM-2003-055, LC-TH-2001-079. 
B.~Allanach, S.~Kraml and W.~Porod,
arXiv:hep-ph/0207314.

\bibitem{AKP2}
B.~C.~Allanach, S.~Kraml and W.~Porod,
JHEP {\bf 0303} (2003) 016
[arXiv:hep-ph/0302102]  LC-TH-2003-017.





\bibitem{softsusy}B.~C.~Allanach,
Comput.\ Phys.\ Commun.\  {\bf 143} (2002) 305
[arXiv:hep-ph/0104145].

\bibitem{spheno}W.~Porod,
Comput.\ Phys.\ Commun.\  {\bf 153} (2003) 275
[arXiv:hep-ph/0301101] LC-TOOL-2003-042.


\bibitem{suspect}A.~Djouadi, J.~L.~Kneur and G.~Moultaka,
arXiv:hep-ph/0211331.


\bibitem{susygen} S.~Katsanevas and P.~Morawitz,
Comput.\ Phys.\ Commun.\  {\bf 112} (1998) 227
[arXiv:hep-ph/9711417].

\bibitem{pythia} T.~Sjostrand, L.~Lonnblad and S.~Mrenna,
arXiv:hep-ph/0108264.

\bibitem{postlep}M.~Battaglia {\it et al.},
in {\it Proc. of the APS/DPF/DPB Summer Study on the Future of
  Particle Physics (Snowmass 2001) } ed. N.~Graf, 
eConf {\bf C010630} (2001) P347
[arXiv:hep-ph/0112013].


\bibitem{jejuslep}
A.~Bartl {\it et al.}  [ECFA/DESY SUSY Collaboration],
arXiv:hep-ph/0301027, LC-TH-2003-039.
A.~Freitas {\it et al.}, [ECFA/DESY SUSY Collaboration],
arXiv:hep-ph/0211108. LC-PHSM-2003-019.



\bibitem{WMsusy02} W.~Majerotto,
arXiv:hep-ph/0209137.



\bibitem{freitas}
A.~Freitas, D.~J.~Miller and P.~M.~Zerwas,
Eur.\ Phys.\ J.\ C 21 (2001) 361
[arXiv:hep-ph/0106198] LC-TH-2001-011.
A.~Freitas and D.~J.~Miller,
in {\it Proc. of the APS/DPF/DPB Summer Study on the Future of
  Particle Physics 
(Snowmass 2001) } eds. R.~Davidson and C.~Quigg, [hep-ph/0111430]. 
A.~Freitas and A.~von Manteuffel,
arXiv:hep-ph/0211105.

\bibitem{0207364} J.~Guasch, W.~Hollik and J.~Sola,
JHEP {\bf 0210} (2002) 040
[arXiv:hep-ph/0207364].
J.~Guasch, W.~Hollik and J.~Sola,
arXiv:hep-ph/0307011, LC-TH-2003-033.
W.~Hollik and H.~Rzehak,
arXiv:hep-ph/0305328.

\bibitem{BRV}M.~Beccaria, F.~M.~Renard and C.~Verzegnassi,
arXiv:hep-ph/0203254, LC-TH-2002-005.
M.~Beccaria, M.~Melles, F.~M.~Renard and C.~Verzegnassi,
arXiv:hep-ph/0210283.

\bibitem{martyn-prague} H.U.~Martyn, LC-PHSM-2003-07, and talk in Prague.
\bibitem{nieto-ams} H. Nieto-Chaupis, LC-DET-2003-074, and talk in Amsterdam.
\bibitem{martyn-susy02} H.U.~Martyn,
arXiv:hep-ph/0302024.

\bibitem{Dima:01}
M.~Dima {\it et al.},
Phys.\ Rev.\ D 65 (2002) 071701.

\bibitem{0303110} 
E.~Boos, G.~Moortgat-Pick, H.~U.~Martyn, M.~Sachwitz and A.~Vologdin,
arXiv:hep-ph/0211040.
E.~Boos, H.U.~Martyn, G.~Moortgat-Pick,
  M.~Sachwitz, A.~Sherstnev and P.M.~Zerwas, 
arXiv:hep-ph/0303110, LC-PHSM-2003-018.

\bibitem{taupol}M.M.~Nojiri,
Phys.\ Rev.\ D {\bf 51} (1995) 6281
[arXiv:hep-ph/9412374].
M.M.~Nojiri, K.~Fujii and T.~Tsukamoto,
Phys.\ Rev.\ D {\bf 54} (1996) 6756
[arXiv:hep-ph/9606370].









\bibitem{FNS} A.~Finch, H.~Nowak and A.~Sopczak,
arXiv:hep-ph/0211140, LC-PHSM-2003-075.

\bibitem{0306181} S.~Heinemeyer, S.~Kraml, W.~Porod and G.~Weiglein,
arXiv:hep-ph/0306181, LC-TH-2003-052.



\bibitem{Bloechi:02}
C.~Bl\"ochinger, H.~Fraas, G.~Moortgat-Pick and W.~Porod,
Eur.\ Phys.\ J.\ C 24 (2002) 297
[arXiv:hep-ph/0201282] LC-TH-2003-031.
 
\bibitem{freitas-phd} 
A.~Freitas,
DESY-THESIS-2002-023, 
A.~Freitas, A.~von Manteuffel, in \cite{freitas}, 
A.~Freitas, A.~von Manteuffel, P.M.~Zerwas, in preparation.


\bibitem{BMW} A.~Brandenburg, M.~Maniatis and M.~M.~Weber,
arXiv:hep-ph/0207278.
M.~Maniatis,
DESY-THESIS-2002-007



\bibitem{Baer:01} H.~Baer, C.~Balazs, S.~Hesselbach, J.~K.~Mizukoshi
  and X.~Tata, 
Phys.\ Rev.\ D {\bf 63} (2001) 095008
[arXiv:hep-ph/0012205].





\bibitem{sleptonedm} A.~Bartl, W.~Majerotto, W.~Porod and D.~Wyler,
arXiv:hep-ph/0306050.




\bibitem{edmAf} D.~Chang, W.~Y.~Keung and A.~Pilaftsis,
Phys.\ Rev.\ Lett.\  {\bf 82} (1999) 900
[Erratum-ibid.\  {\bf 83} (1999) 3972]
[arXiv:hep-ph/9811202].
A.~Pilaftsis,
Phys.\ Lett.\ B {\bf 471} (1999) 174
[arXiv:hep-ph/9909485].


\bibitem{0207186} A.~Bartl, K.~Hidaka, T.~Kernreiter and W.~Porod,
Phys.\ Rev.\ D {\bf 66} (2002) 115009
[arXiv:hep-ph/0207186] LC-TH-2003-027.




\bibitem{0306281} A.~Bartl, S.~Hesselbach, K.~Hidaka, T.~Kernreiter
  and W.~Porod, 
arXiv:hep-ph/0306281, LC-TH-2003-04.



\bibitem{0202198} A.~Bartl, T.~Kernreiter and W.~Porod,
Phys.\ Lett.\ B {\bf 538} (2002) 59
[arXiv:hep-ph/0202198]  LC-TH-2003-028.


\bibitem{Anan:02} B.~Ananthanarayan, S.~D.~Rindani and A.~Stahl,
Eur.\ Phys.\ J.\ C {\bf 27} (2003) 33
[arXiv:hep-ph/0204233] LC-PHSM-2002-006.



\bibitem{Hagiwara}
K.~Hagiwara {\it et al.}  [Particle Data Group Collaboration],
Phys.\ Rev.\ D {\bf 66} (2002) 010001.

\bibitem{nuosc} Y.~Fukuda {\it et al.}  [Super-Kamiokande Collaboration],
Phys. Rev. Lett. {\bf 81}, 1562 (1998)
Q.~R.~Ahmad {\it et al.}  [SNO Collaboration],
Phys.\ Rev.\ Lett.\  {\bf 89} (2002) 011301
[arXiv:nucl-ex/0204008].
Q.~R.~Ahmad {\it et al.}  [SNO Collaboration],
Phys.\ Rev.\ Lett.\  {\bf 89}, 011302 (2002)
[arXiv:nucl-ex/0204009].
K.~Eguchi {\it et al.}  [KamLAND Collaboration],
Phys.\ Rev.\ Lett.\  {\bf 90} (2003) 021802
[arXiv:hep-ex/0212021].



\bibitem{bm}F.~Borzumati, A.~Masiero,
Phys.\ Rev.\ Lett.\  {\bf 57} (1986) 961.

\bibitem{lfv-lc} 
N.~Arkani-Hamed, J.~L.~Feng, L.~J.~Hall and H.~Cheng,
Phys.\ Rev.\ Lett.\ {\bf 77} (1996) 1937
[hep-ph/9603431]. 
~Hisano, M.~M.~Nojiri, Y.~Shimizu and M.~Tanaka,
Phys.\ Rev.\ D {\bf 60} (1999) 055008
[hep-ph/9808410].
M.~Guchait, J.~Kalinowski and P.~Roy,
Eur.\ Phys.\ J.\ C {\bf 21} (2001) 163
[arXiv:hep-ph/0103161] LC-TH-2001-073.
J.~Kalinowski,
Acta Phys.\ Polon.\ B {\bf 32} (2001) 3755.



\bibitem{0210326} W.~Porod, W.~Majerotto,
Phys.\ Rev.\ D {\bf 66} (2002) 015003 
[arXiv:hep-ph/0201284] LC-TH-2003-030.
W.~Porod and W.~Majerotto,
arXiv:hep-ph/0210326.


\bibitem{jk} J.~Kalinowski,
Acta Phys.\ Polon.\ B {\bf 33} (2002) 2613
[arXiv:hep-ph/0207051]  LC-TH-2003-040.

\bibitem{prz} E.~Perazzi, G.~Ridolfi and F.~Zwirner,
Nucl.\ Phys.\ B {\bf 574} (2000) 3
[arXiv:hep-ph/0001025].



\bibitem{checchia} P.~Checchia and E.~Piotto,
arXiv:hep-ph/0102208,  LC-TH-2001-015.


\bibitem{cdsz}
S.~Y.~Choi, A.~Djouadi, H.~S.~Song and P.~M.~Zerwas,
Eur.\ Phys.\ J.\ {\bf C8} (1999) 669
[hep-ph/9812236].

\bibitem{cdgksz} 
S.~Y.~Choi, M.~Guchait, J.~Kalinowski and P.~M.~Zerwas,
Phys.\ Lett.\ B {\bf 479} (2000) 235
[hep-ph/0001175]; 
S.~Y.~Choi, A.~Djouadi, M.~Guchait, J.~Kalinowski, H.~S.~Song and
P.~M.~Zerwas, 
Eur.\ Phys.\ J.\ C {\bf 14} (2000) 535
[hep-ph/0002033].



\bibitem{six} 
L.~Chau and W.~Keung,
Phys.\ Rev.\ Lett.\  {\bf 53} (1984) 1802;
H.~Fritzsch and J.~Plankl,
Phys.\ Rev.\ D {\bf 35} (1987) 1732.

\bibitem{ckmz} S.~Y.~Choi, J.~Kalinowski, G.~Moortgat-Pick and P.~M.~Zerwas,
Eur.\ Phys.\ J.\ C {\bf 22} (2001) 563
[arXiv:hep-ph/0108117] LC-TH-2003-024.


\bibitem{hollikgaug}
T.~Fritzsche and W.~Hollik,
Eur.\ Phys.\ J.\ C {\bf 24} (2002) 619
[arXiv:hep-ph/0203159].
T.~Blank and W.~Hollik,
arXiv:hep-ph/0011092, LC-TH-2000-054.

\bibitem{vienna}
W.~\"Oller, H.~Eberl, W.~Majerotto and C.~Weber,
arXiv:hep-ph/0304006.
H.~Eberl, M.~Kincel, W.~Majerotto and Y.~Yamada,
Phys.\ Rev.\ D {\bf 64} (2001) 115013
[arXiv:hep-ph/0104109].

\bibitem{0205257}
M.~A.~Diaz and D.~A.~Ross,
arXiv:hep-ph/0205257.
M.~A.~Diaz and D.~A.~Ross,
JHEP {\bf 0106} (2001) 001
[arXiv:hep-ph/0103309].

\bibitem{add} S.~Y.~Choi, J.~Kalinowski, G.~Moortgat-Pick and P.~M.~Zerwas,
Eur.\ Phys.\ J.\ C {\bf 23} (2002) 769
[arXiv:hep-ph/0202039]  LC-TH-2003-023.

\bibitem{KM} 
J.~L.~Kneur and G.~Moultaka,
Phys.\ Rev.\ D {\bf 59} (1999) 015005
[hep-ph/9807336];
V.~Barger, T.~Han, T.~Li and T.~Plehn,
Phys.\ Lett.\ B {\bf 475} (2000) 342
[hep-ph/9907425];
J.~L.~Kneur and G.~Moultaka,
Phys.\ Rev.\ D {\bf 61} (2000) 095003
[hep-ph/9907360].


\bibitem{hensel} C.~Hensel, 
DESY-THESIS-2002-047, 
and talk in Cracow.



\bibitem{0209108}T.~Mayer, C.~Blochinger, F.~Franke and H.~Fraas,
Eur.\ Phys.\ J.\ C {\bf 27} (2003) 135
[arXiv:hep-ph/0209108].


\bibitem{tevlinball} M. Ball, talk in Prague. \\
C.~Tevlin, talk in Amsterdam. 

\bibitem{AASB} F.~del Aguila and J.~A.~Aguilar-Saavedra,
Phys.\ Lett.\ B {\bf 386} (1996) 241
[hep-ph/9605418];
J.~A.~Aguilar-Saavedra and G.~C.~Branco,
Phys.\ Rev.\ D {\bf 62} (2000) 096009
[hep-ph/0007025].


\bibitem{CJ} C.~Jarlskog,
Phys.\ Rev.\ Lett.\  {\bf 55} (1985) 1039.

\bibitem{kittel} O. Kittel, talk in Prague. 
A.~Bartl, H.~Fraas, O.~Kittel and W.~Majerotto,
arXiv:hep-ph/0308143, LC-TH-2003-065, and   
arXiv:hep-ph/0308141.

\bibitem{gaissmaier} B.~Gaissmaier,
arXiv:hep-ph/0211407, and 
talk in Amsterdam.

\bibitem{0306272} 
J.~Kalinowski, Acta Phys.\ Polon.\ B {\bf 34} (2003) 3441
[arXiv:hep-ph/0306272, LC-TH-2003-038]. 


\bibitem{11a}
S.~T.~Petcov,
Phys.\ Lett.\ B {\bf 178} (1986) 57.
G.~Moortgat-Pick and H.~Fraas, 
Eur.\ Phys.\ J.\ C {\bf 25} (2002) 189
[arXiv:hep-ph/0204333].


\bibitem{R2}  
J.~Ellis, J.~M.~Fr\`ere, J.~S.~Hagelin, G.~L.~Kane and S.~T.~Petcov,
Phys.\ Lett.\ B {\bf 132} (1983) 436.


\bibitem{gluino} S.~Berge and M.~Klasen,
Phys.\ Rev.\ D {\bf 66} (2002) 115014
[arXiv:hep-ph/0208212].
S.~Berge and M.~Klasen,
arXiv:hep-ph/0303032.




\bibitem{hprv} M. Hirsch, W. Porod, J.C.~Rom\~ao, J.W.F.~Valle,
Phys.\ Rev.\ D {\bf 66} (2002) 095006
[arXiv:hep-ph/0207334]  LC-TH-2003-029.



\bibitem{BRpVnu-LC} A.~Bartl, M.~Hirsch, T.~Kernreiter, W.~Porod and
  J.~W.~Valle, 
arXiv:hep-ph/0306071.
W.~Porod, M.~Hirsch, J.~Rom\~ao and J.~W.~Valle,
Phys.\ Rev.\ D {\bf 63} (2001) 115004
[arXiv:hep-ph/0011248]  LC-TH-2003-026.

\bibitem{0307364} M.~Hirsch and W.~Porod,
arXiv:hep-ph/0307364.

\bibitem{bi-tri} A.~Bartl, M.~Hirsch, T.~Kernreiter, W.~Porod, 
J.W.F.~Valle,  
Slepton LSP Decays: Trilinear versus Bilinear  R-parity Breaking,
LC-TH-2003-047, ZU-TH 10/03, UWThPh-2003-13. 

\bibitem{ghosh} D.~K.~Ghosh and S.~Moretti,
Phys.\ Rev.\ D {\bf 66} (2002) 035004
[arXiv:hep-ph/0112288]      LC-TH-2003-063.



\bibitem{sourov2}M.~Chaichian, K.~Huitu, S.~Roy and Z.~H.~Yu,
Phys.\ Lett.\ B {\bf 518} (2001) 261
[arXiv:hep-ph/0107111].


\bibitem{Franke} F.~Franke and S.~Hesselbach,
Phys.\ Lett.\ B {\bf 526} (2002) 370
[arXiv:hep-ph/0111285].
S.~Hesselbach and F.~Franke,
arXiv:hep-ph/0210363,  LC-TH-2003-006.

\bibitem{hff}
S.~Hesselbach, F.~Franke and H.~Fraas,
arXiv:hep-ph/0003272.

\bibitem{Witten:nf}
E.~Witten,
Nucl.\ Phys.\ B {\bf 188} (1981) 513.



\bibitem{bpz}
G.~A.~Blair, W.~Porod and P.~M.~Zerwas,
Phys.\ Rev.\ D 63 (2001) 017703;
P.~M.~Zerwas {\it et al.}, hep-ph/0211076,   LC-TH-2003-020.
P.~M.~Zerwas {\it et al.}, [ECFA/DESY SUSY Collaboration],
arXiv:hep-ph/0211076, LC-TH-2003-020.


\bibitem{bpz-new}
G.~A.~Blair, W.~Porod and P.~M.~Zerwas,
Eur.\ Phys.\ J.\ C {\bf 27} (2003) 263
[arXiv:hep-ph/0210058]  LC-TH-2003-021.






\end{thebibliography}
\end{document}